%% file: templateArxiv.tex
\documentclass{article}

\usepackage{PRIMEarxiv}

\usepackage[utf8]{inputenc} 
\usepackage[T1]{fontenc}    
\usepackage{hyperref}       
\usepackage{url}            
\usepackage{booktabs}       
\usepackage{amsfonts}       
\usepackage{nicefrac}       
\usepackage{microtype}      
\usepackage{lipsum}
\usepackage{fancyhdr}       
\usepackage{graphicx}       
\usepackage{float}

\usepackage{listings}

\usepackage[frozencache=true,cachedir=minted-cache]{minted}

\usepackage{caption}
\usepackage{subcaption}

\usepackage{csquotes}
\usepackage[
  backend=biber,
  style=numeric,
  sorting=nty
  ]{biblatex}
\graphicspath{{media/}}     

\newcommand{\thetitle}{Energy-Efficiency Evaluation of OpenMP Loop Transformations and Runtime Constructs
}

\title{\thetitle
}

\author{
  Henrik Valter, Axel Karlsson, Miquel Pericàs \\
  Computer Science and Engineering \\
  Chalmers University of Technology \\
  Gothenburg, Sweden\\
  \texttt{valterh@chalmers.se, axeka@student.chalmers.se, miquelp@chalmers.se} \\
}

\begin{document}
\maketitle

\begin{abstract}

\input{include/abstract.tex}

\end{abstract}


\lstset{language=C++,
                basicstyle=\ttfamily,
                keywordstyle=\color{blue}\ttfamily,
                stringstyle=\color{red}\ttfamily,
                commentstyle=\color{green}\ttfamily,
                morecomment=[l][\color{magenta}]{\#}
}

\section{Introduction}
\input{include/introduction.tex}

\section{Background}
\input{include/background.tex}

\section{Benchmark Programs}
\input{include/benchmarkprograms.tex}

\section{Experimental Methodology}
\input{include/experimentalmethodology.tex}

\section{Results}
\input{include/results.tex}

\section{Directives and Programmer Recommendations}
\label{chap:directives}
\input{include/directives.tex}

\section{Conclusion}
\input{include/conclusion.tex}


\printbibliography

\newpage

\section*{Appendix}
\input{include/appendix.tex}

\end{document}

%% file: include/abstract.tex
OpenMP is the de facto API for parallel programming in HPC applications. These programs are often computed in data centers, where energy consumption is a major issue. Whereas previous work has focused almost entirely on performance, we here analyse aspects of OpenMP from an energy consumption perspective. This analysis is accomplished by executing novel microbenchmarks and common benchmark suites on data center nodes and measuring the energy consumption. Three main aspects are analysed: directive-generated loop tiling and unrolling, parallel for loops and explicit tasking, and the policy of handling blocked threads. 
For loop tiling and unrolling, we find that tiling can yield significant energy savings for some, mostly unoptimised programs, while directive-generated unrolling provides very minor improvement in the best case and degenerates performance majorly in the worst case. 
For the second aspect, we find that parallel for loops yield better results than explicit tasking loops in cases where both can be used. This becomes more prominent with more fine-grained workloads.
For the third, we find that significant energy savings can be made by not descheduling waiting threads, but instead having them spin, at the cost of a higher power consumption.
We also analyse how the choice of compiler affects the above questions by compiling programs with each of \emph{ICC}, \emph{Clang} and \emph{GCC}, and find that while neither is strictly better than the others, they can produce very different results for the same compiled programs.
As a final step, we combine the findings of all results and suggest novel compiler directives as well as general recommendations on how to reduce energy consumption in OpenMP programs.

%% file: include/introduction.tex
Up until approximately the 1990s, computers were rapidly made faster due to increasing clock frequencies and exploitation of instruction-level parallelism (ILP), with little concern for power consumption \cite{sjalander_power-efficient_2008}. Methods to increase performance using ILP generally didn't scale well and used significant amount of hardware resources for relatively little gain \cite{sjalander_power-efficient_2008}. This was however not a problem due to Moore's law and Dennard scaling being in full effect and made these costs justifiable. Moore's law states that the number of transistors in a dense integrated circuit doubles approximately every two years \cite{kaxiras_computer_2008}, while Dennard scaling states that power usage stays in proportion with area as chips get smaller \cite{sjalander_power-efficient_2008}.
However, in the mid to end 1990s, power became a major problem as increasing processor frequencies would become impossible to cool, and Dennard scaling was declared broken around 2005-2007 \cite{sjalander_power-efficient_2008}. To get around these problems and keep improving performance, the industry instead moved towards putting several cores on one chip, so thread-level parallelism (TLP) also could be exploited. This allowed further performance without relaying on additional costly ILP optimisations or increased clock frequency.

However, these new performance gains did not come free, which normally was the case for ILP optimisations, but instead required explicit effort by the programmer. To properly utilise these new multicore processors, code needed to be written in a way that took into account correctness issues, such as race conditions, starvation and deadlock, but also performance issues, like false sharing and minimising parallel overheads such as data synchronisation \cite{rauber_thread_2013}. Generally, writing parallel code required significantly more effort than serial, and automatically converting serial code to efficient parallel code also turned out to be a very challenging task \cite{lee_problem_2006}.

The difficulty of writing parallel code lead to the creation of parallel APIs such as OpenMP. It is a widely applied framework for fork-join model parallelism, especially in computer clusters running HPC workloads. These environments are in turn becoming increasingly more interesting due to cloud computing and offloading of heavy computations in areas such as machine learning inference \cite{brewer_inference_2020}. This motivates an analysis of OpenMP, especially in terms of performance, but also in terms of power and energy.

Power efficiency is especially important in data center contexts, where cooling makes up a significant cost and challenge \cite{bianchini_power_2004}\cite{kaxiras_computer_2008}. Increased heat dissipation raises the operating temperature, which in turn causes increased aging of hardware components causing degraded performance and shortened lifetime \cite{medeiros_mitigating_2021}.
In addition to the challenges of heat dissipation, high energy consumption is also linked with low reliability in HPC systems \cite{nandamuri_power_2014}. 
Beyond the context of data centers, energy consumption is also highly important for mobile devices due to the obvious issue of battery lifetime.

OpenMP is not a programming language in itself, but rather an API that is called using compiler directives in C, C++ and Fortran. The semantics of these directives is defined in OpenMP specifications. Recently, in November 2020, the specification for OpenMP 5.1 was introduced \cite{openmp_5.1_spec}. This specification introduced the \emph{tile} and \emph{unroll} directives, which are loop transformation techniques used for optimisation.
These optimisation techniques are described in the background, see section~\ref{sec:codetransforms}.

The most well-known directive is probably \emph{parallel for}, which executes iterations of a loop in parallel using a team of threads. There is also the \emph{task} directive, which packages a section of code to be possibly computed in parallel by another thread.

Another aspect of the runtime aspect of OpenMP (and parallel programming in general) is what a thread should do when blocked by another thread (such as on a barrier). There are in general two solutions: running an idle loop (spin-locking) or descheduling the thread. In OpenMP, such behavior can be controlled with an environment variable called the waiting policy.
 
\subsection{Our contribution}

In short, we answer the following research questions:

\begin{itemize}
    \item \textbf{RQ1}: How much impact do code transformations such as the recently introduced unrolling and tiling have on energy consumption?
    \item \textbf{RQ2}: How do the runtime parallelism constructs of OpenMP (tasks and parallel loops) compare in terms of energy?
    \item \textbf{RQ3}: How does the OpenMP waiting policy affect power and energy?
    \item \textbf{RQ4}: How do different implementations of OpenMP impact the research questions above?
    \item \textbf{RQ5}: How can the findings from the points above be combined into novel directives targeting energy efficiency?
\end{itemize}

\subsection{Limitations}

Although OpenMP supports Fortran, we will focus on C and C++ implementations only.
Besides being a framework for parallelism for multi-core computers, OpenMP also has support for offloading work to other devices such as GPUs \cite{openmp_5.2_spec}. There has also been some work on offloading work to FPGAs (Field-Programmable Gate Arrays) \cite{knaust_openmp_2019}. These are, however, out of scope of this project, which only considers shared-memory computers with homogeneous architectures.

When performing analysis on source code optimisations, we never go lower than assembly code, for example into compiler source code. This is mainly due to a lack of time and expertise in the area.

%% file: include/background.tex
This chapter covers the relevant background information for the project. First, we cover the aspects of OpenMP that are relevant to the research questions. Then, we describe the concepts of loop unrolling and tiling, how they can improve programs, and how they can be applied automatically by OpenMP. Finally we present related work concerning energy optimisation of OpenMP programs.

\subsection{OpenMP}

As described in the introduction, OpenMP is an API for shared-memory programming. It consists of compiler directives, runtime library functions and environment variables that simplifies parallel programming on shared-memory machines \cite{openmp_5.1_spec}.

\subsubsection{Runtime constructs}
\label{sec:background:runtimeconstructs}
When writing parallel programs with OpenMP there are three available directives that directly describes how the parallel parts of the program should behave, those being: \emph{parallel for}, \emph{tasks} and \emph{sections}. They are used to describe parallel behaviour for slightly different scenarios. Sections are, however, in our experience, rarely used in practise, so we do not consider it in our analysis.

The \emph{parallel for} directive is used to parallelise \emph{for} loops, and is the most used and straight forward of the three directives. The loop iterations are split up into chunks, distributed over the assigned number of threads, and then run in parallel. It is possible to change the size of these chunks and how they are distributed with input parameters passed to the directive. There exist four scheduling options that decide the distribution: \emph{static}, \emph{dynamic}, \emph{guided}, and \emph{auto}. The \emph{static} option distributes the chunks evenly over the threads at compile time, \emph{dynamic} instead dynamically distributes chunks at run time as threads complete their previous work, \emph{guided} works in a similar way as \emph{dynamic} but utilizing a dynamic chunk size, and lastly, \emph{auto} gives full control to the compiler to decide how to handle the scheduling (e.g. the GCC compiler just defaults to the \emph{static} setting when \emph{auto} is selected \cite{libgomp_code}) \cite{openmp_5.1_spec}. An example of a parallelized loop using the directive can be seen in listing \ref{lst:parallel_for}.

The \emph{task} directive is used to defines a scope of code that can be run concurrently by any thread. An example where this directive would be useful is in while loops, where the number of iterations is unknown and a \emph{task} can instead be created for each iteration. Another common example is recursive functions, where each recursive call can be its own \emph{task}. These two scenarios can be seen in listings \ref{lst:task_while} and \ref{lst:task_rec} respectively.

\begin{lstlisting}[caption=Parallelized for loop using the \emph{parallel for} directive,frame=tlrb,label={lst:parallel_for},language=C]{Name}
#pragma omp parallel for
for(int i = 0; i < N; i++) 
{
    foo(i); // Some work
}
\end{lstlisting}

\begin{lstlisting}[caption=Parallelized while loop using the \emph{task} directive.,frame=tlrb,label={lst:task_while},language=C]{Name}
#pragma omp parallel
#pragma omp single
while(...) 
{
#pragma omp task
    foo(i); // Some work
    i++;
}
\end{lstlisting}

\noindent\begin{minipage}{\textwidth}
\begin{lstlisting}[caption=Recursive function using the \emph{task} directive,frame=tlrb,label={lst:task_rec},language=C]{Name}
int main() {
  ...
  #pragma omp parallel
  #pragma omp single
  fib(N); // Executed in a parallel region
  ...
}

int fib(int n) 
{
  int i, j;
  if (n<2) {
    return n;
  }
  else {
    #pragma omp task shared(i)
    i=fib(n-1);
    #pragma omp task shared(j)
    j=fib(n-2);
    #pragma omp taskwait
    return i+j;
  }
}
\end{lstlisting}
\end{minipage}

\subsubsection{Waiting policy}
\label{sec:background:waitingpolicy}

The waiting policy, specified by the \emph{OMP\_WAIT\_POLICY}
\footnote{Information about the waiting policy in OpenMP can be found in \\ \href{https://www.openmp.org/spec-html/5.0/openmpse55.html}{https://www.openmp.org/spec-html/5.0/openmpse55.html}}
environment variable, provides a hint about the desired behaviour of waiting threads to the runtime system.
It can be set to \emph{active} or \emph{passive}. With the \emph{active} setting, threads should be active and consume processor cycles (such as spinning) while \emph{passive} threads should avoid consuming cycles by, for example, yielding the processor core.
If the value is not set, an implementation of OpenMP may have a third alternative as default behaviour, which is often a mix of the two.

\subsection{Code transformations}
\label{sec:codetransforms}

Code transformations are methods of changing or rearranging source code, often with the aim of reducing different kinds of performance overheads while being semantically equivalent.
This section covers the two approaches considered in this work: loop tiling and loop unrolling.

\subsubsection{Loop Unrolling}
\label{sec:background:unrolling}

Loop unrolling is a method that coalesces the logic of several loop iterations into one iteration with the goal of reducing overhead related to loop control logic such as counter updates, termination conditions and branching \cite{kukunas_chapter_12}. 

An example of a partially unrolled loop can be seen in listing \ref{lst:unrolled}, the unmodified version of the loop can be seen in listing \ref{lst:not_unrolled}. Here, the functionality of five iterations have been grouped together, effectively reducing the amount of control operations needed by the loop by 80\%.

\noindent\begin{minipage}{.45\textwidth}
\begin{lstlisting}[caption=Unmodified loop.,frame=tlrb,label={lst:not_unrolled},language=C]{Name}
int i;
int n[N];
for(i = 0; i < N; i++)
  n[i] = 10 * i;
\end{lstlisting}
\end{minipage}\hfill
\begin{minipage}{.45\textwidth}
\begin{lstlisting}[caption=Partially unrolled loop,frame=tlrb, label={lst:unrolled}]{Name}
int i;
int n[N];
for (i=0;i<N;i+=5) {
  n[i+0] = 10 * (i+0)
  n[i+1] = 10 * (i+1)
  n[i+2] = 10 * (i+2)
  n[i+3] = 10 * (i+3)
  n[i+4] = 10 * (i+4)
}
\end{lstlisting}
\end{minipage}

Note, however, that if the unrolling factor does not evenly divide the total number of iterations, this will lead to more iterations being performed than in the unmodified code, leading to incorrect behaviour. Additionally, when the total number of iterations is not known at compile time, assumptions about evenly divisibility can not be guaranteed. One way to solve this is seen in listing~\ref{lst:unrolled_with_check}, where the first loop is unrolled and the second handles the special case. 

\begin{lstlisting}[caption=Partially unrolled loop,frame=tlrb, label={lst:unrolled_with_check}]{Name}
int n[N];
int i=0;
for (;i<N-4;i+=5) {
  n[i+0] = 10 * (i+0);
  n[i+1] = 10 * (i+1);
  n[i+2] = 10 * (i+2);
  n[i+3] = 10 * (i+3);
  n[i+4] = 10 * (i+4);
}
for(;i<N;i++) {
  n[i] = 10 * i;
}
\end{lstlisting}

One should be careful when applying loop unrolling, since blindly unrolling loops does not always lead to better performance. A problem with excessive unrolling is that the L1 instruction cache misses can go up since the number of unique instruction has increased \cite{kukunas_chapter_12}, overall degrading performance due to cache eviction. Unrolling also has the undesirable effect of making the program less readable and larger in size \cite{anil_kumar_enhancing_2020}.

Directive-generated unrolling was introduced in OpenMP 5.1 \cite{openmp_5.1_spec}. 
Listing~\ref{lst:omp_unrolled} shows how a loop can be unrolled with similar results to listing~\ref{lst:unrolled_with_check}.

\noindent\begin{minipage}{\textwidth}
\begin{lstlisting}[caption=Unrolled loop by OpenMP,frame=tlrb,label={lst:omp_unrolled}]{Name}
int i;
int n[N];
#pragma omp unroll partial(5)
for(i = 0; i < N; i++) {
  n[i] = 10 * i
}
\end{lstlisting}
\end{minipage}

\subsubsection{Loop Tiling}
\label{sec:backgroundlooptiling}
Loop tiling, also known as \textit{loop blocking}, is a loop transformation that targets nested loops with the aim of improving data locality on data accesses. It does this by splitting each loop affected into two separate loops, the outer loop defines the size of blocks of data, while the inner iterates over said block. The reason for this loop rearrangement is to maximise the reuse of data before it becomes evicted from the cache, thus improving performance.

\vspace{1cm}

\begin{lstlisting}[caption=Basic square matrix multiplication,frame=tlrb,label={lst:basic_naive_matmul}]{Name}
// C = A * B
for (int row = 0; row < N; row++) {
  for (int col = 0; col < N; col++) {
    for (int k = 0; k < N; k++) {
      C[row*N + col] += A[row*N + k] * B[k*N + col];
    }
  }
}
\end{lstlisting}

\noindent\begin{minipage}{\textwidth}
\begin{lstlisting}[caption=Basic square matrix multiplication with loop tiling,frame=tlrb, label={lst:tiled_naive_matmul}]{Name}
// C = A * B
for (int r0 = 0; r0 < N; r0 += tile_size) {
  for (int c0 = 0; c0 < N; c0 += tile_size) {
    for (int k0 = 0; k0 < N; k0 += tile_size) {
      for (int r1 = r0; r1 < r0 + tile_size; r1++) {
        for (int c1 = c0; c1 < c0 + tile_size; c1++) {
          for (int k1 = k0; k1 < tile_size; k1++) {
            C[r1*N + c1] += A[r1*N + k1] * B[k1*N + c1];
          }
        }
      }
    }
  }
}
\end{lstlisting}
\end{minipage}

\noindent\begin{minipage}{\textwidth}
\begin{lstlisting}[caption=Basic square matrix multiplication with loop tiling with bound checks,frame=tlrb, label={lst:tiled_naive_matmul_with_checks}]{Name}
// C = A * B
for (int r0 = 0; r0 < N; r0 += tile_size) {
  int rmax = r0 + tile_size > N ? N : r0 + tile_size;
  for (int c0 = 0; c0 < N; c0 += tile_size) {
    int cmax = c0 + tile_size > N ? N : c0 + tile_size;
    for (int k0 = 0; k0 < N; k0 += tile_size) {
      int kmax = k0 + tile_size > N ? N : k0 + tile_size;
      for (int r1 = r0; r1 < rmax; r1++) {
        for (int c1 = c0; c1 < cmax; c1++) {
          for (int k1 = k0; k1 < kmax; k1++) {
            C[r1*N + c1] += A[r1*N + k1] * B[k1*N + c1];
          }
        }
      }
    }
  }
}
\end{lstlisting}
\end{minipage}

The tile size needs to be chosen in such a way that all the data needed for one tile iteration can fit into the memory hierarchy that is optimised for. If the tile size chosen is too small, then the overhead from the additional loops can instead start to dominate the execution time, undoing any performance gain. Therefore, it is of utmost importance to find an appropriate value of the tile size to balance these two factors.

An example of a loop tiled version of matrix multiplication can be seen in listing \ref{lst:tiled_naive_matmul}, with the unmodified version in listing \ref{lst:basic_naive_matmul}.
However, similar to unrolling, this assumes that the number of iterations are evenly divided by the tile size. The code in listing~\ref{lst:tiled_naive_matmul_with_checks} fixes this.

OpenMP introduced compiler-generated loop tiling in OpenMP 5.1 \cite{openmp_5.1_spec}. This makes it possible to perform loop tiling using the \emph{tile} directive. Listing~\ref{lst:omptiled_naive_matmul} shows how matrix multiplication can be loop tiled using OpenMP with a similar result to listing~\ref{lst:tiled_naive_matmul_with_checks}.

\noindent\begin{minipage}{\textwidth}
\begin{lstlisting}[caption=Basic square matrix multiplication with OpenMP loop tiling,frame=tlrb, label={lst:omptiled_naive_matmul}]{Name}
// C = A * B
#pragma omp tile (tile_size, tile_size, tile_size)
for (int row = 0; row < N; row++) {
  for (int col = 0; col < N; col++) {
    for (int k = 0; k < N; k++) {
      C[row*N + col] += A[row*N + k] * B[k*N + col];
    }
  }
}
\end{lstlisting}
\end{minipage}

\subsection{Related work}
\label{sec:related_work}

We begin with an example of loop unrolling being used to increase energy efficiency. In~\cite{oo_effect_2021}, the authors evaluate loop unrolling on an energy-efficient implementation of Strassen's algorithm. Strassen's algorithm is an algorithm for fast matrix multiplication, significantly faster than naive matrix multiplication for large matrices. Their method yields a performance gain of 98 \% and a reduction in energy consumption of 95 \%, which is mainly due to better utilisation of vector instructions.

To effectively experiment with energy consumption, we need a way to measure it. Goel and McKee \cite{goel_methodology_2016} present a method for measuring and modelling dynamic and static power in cores and uncores. Uncore is Intel's term for components close but not part of the CPU such as last-level cache, memory controller and interconnects. They find that while more cores generally yield higher performance, the optimal number of running threads may be lower due to stalling in serial sections of the program.

Two of our research questions, those regarding tasking versus parallel loops and the waiting policy, consider the runtime resources of OpenMP. In these contexts, the number of threads with which to execute parallel regions, often referred to as DCT (Dynamic Concurrency Throttling), is a major concern. In \cite{lorenzon_aurora_2019}, Lorenzon et al. perform DCT on OpenMP programs targeting performance, energy or energy-delay-product (EDP). The tool, called Aurora, performs a hill-climbing approach to selecting the optimal number of threads for a parallel OpenMP loop optimising performance, energy or EDP. It is made fully transparent to both programmer and end user by extending the \emph{libgomp} library (The OpenMP implementation for the GCC compiler), which is dynamically linked at runtime. The work improves EDP by 98 \% compared to the baseline (which uses as many threads as machine cores), 86 \% compared to using the OMP\_DYNAMIC (which selects number of threads based on previous machine workload during the last 15 minutes) and 91 \% compared to feedback-driven threading (\cite{suleman_feedback-driven_2008}).
However, the work targets only parallel loops (eg. \textit{\#pragma omp parallel for [...]}) and not sections or tasks.
Energy minimisation is also beneficial due to mitigating hardware ageing, as mentioned in the introduction. In \cite{medeiros_mitigating_2021}, the authors perform DCT specifically for this purpose. Their work is very closely related to the other DCT work mentioned above.

In~\cite{jordan_multi-objective_2012}, the authors introduce a novel auto-tuning framework. 
The framework tunes parameters such as unrolling factors, tile sizes and loop ordering to optimise a combination of performance, energy and power. It also selects the optimal number of threads for code regions.
The framework is based on the \emph{Insieme Compiler and Runtime} infrastructure \cite{inseime}, which provides source-to-source transformations for parallel C, C++, OpenMP, MPI and OpenCL programs. Internally, Inseime transforms the input source code into an internal representation called INSPIRE on which the transformations and optimisations occur \cite{zangerl_compiler_nodate}. As a final step the internal representation is translated back into the input language. Evaluations show a 70 \% performance improvement over solutions tuned for a specific number of threads \cite{jordan_multi-objective_2012}. 

Related to this work is OpenMPE, which is an extension of OpenMP for application-level optimisation of energy, power and performance \cite{alessi_application-level_2015}. Most importantly, OpenMPE introduces an \emph{objective} clause, which specifies how a region of code should be optimised in terms of performance, energy and power.
As a simple example, \emph{\#pragma omp ... objective(E)} will optimise a code region with the only objective of minimising energy consumption.
A more advanced example is \emph{\#pragma omp ... objective(0.4*E+0.6*T : P<200)} meaning that the optimisation is weighted 40 \% for energy, 60 \% for performance and the power must not exceed 200 W.
The authors have also implemented a version of OpenMPE.
In the first step of the compilation process the source code is transformed into INSPIRE representation, on which the objective directives are applied.
The INSPIRE code is then translated back into C, and then normally compiled using GCC.
Evaluation of OpenMPE shows energy savings of 15 \% across 9 use cases.

%% file: include/benchmarkprograms.tex
This chapter covers the benchmark programs used in the project. These programs fall into one of three categories. The first is \emph{microbenchmarks}, which are very small programs created to test very specific feature of the language, such as the overhead of starting a parallel region. These are written by us and allows well-controlled experimentation, but make no claim to represent a realistic program. 

The second category is \emph{own benchmarks}, which consists of two larger programs: matrix multiplication and a 2-dimensional stencil. These provide a slightly more realistic workload. They allow experimentation with different optimisations and algorithmic changes, providing stronger grounds for conclusions than the microbenchmarks. They should however be supplemented with standardised benchmark suites in order to be compared with other work.

The final category is the \emph{benchmark suites}: well-known benchmarks what are commonly used in papers and therefore much easier for other researchers to compare against, with no potential source of error from biased or poorly written own benchmarks. Unfortunately, these often have the drawback of being big and complex, making it difficult to determine the impact of aspects such as loop tiling, for example.

\subsection{Microbenchmarks}

This section covers the three microbenchmarks introduced for this project. As mentioned above, these are very small kernels meant for experimenting with very specific program aspects.
First is the \emph{parallel constructs microbenchmark}, the purpose of which is to experiment with the different ways of introducing parallelism to a program. It is especially important for the second research question, which considers tasking versus parallel loops.
The second benchmark is the so-called \emph{inactivity microbenchmark}, which is meant for evaluation of the waiting policy.
The third and final program is the \emph{unrolling microbenchmark}, which as the name implies evaluates loop unrolling.

\subsubsection{Parallel constructs microbenchmark}
\label{sec:theory:parallelismconstructsmicrobenchmark}

The purpose of this microbenchmark is to measure the overhead of the different parallel constructs, as well as the overhead of creating and managing parallel regions.

The program creates a number of tasks to complete in parallel. The main kernel, in its most basic form, can be seen in listing~\ref{lst:basic_sleeping_microbenchmark}. The tasks themselves simply stall a random amount of time.
Stalling in this case refers to running a busy-waiting loop for the required amount of time in contrast to sleeping, which yields the running thread. The stalling is implemented using the \emph{gettimeofday} function and can be seen in listing~\ref{lst:stalling}.

\noindent \begin{minipage}{\linewidth}
\begin{lstlisting}[caption=Basic sleeping microbenchmark,frame=tlrb, label={lst:basic_sleeping_microbenchmark}]{Name}
for (int i = 0; i < iterations; i++) {
    // New parallel region for every iteration
    #pragma omp parallel
    {
        #pragma omp [...]
        for (int t = 0; t < num_tasks; t++) {
            perform_task(taskarray[t]);
        }
    }
}
\end{lstlisting}
\end{minipage}

\noindent \begin{minipage}{\linewidth}
\begin{lstlisting}[caption=Implementation of busy-waiting for a number of microseconds,frame=tlrb, label={lst:stalling}]{Name}
void stall_us(double us) {
    struct timeval t0;
    gettimeofday(&t0, 0);
    struct timeval t1;
    double elapsed;
    do {
        gettimeofday(&t1, 0);
        elapsed=((t1.tv_sec-t0.tv_sec)*
            1000000+t1.tv_usec-t0.tv_usec);
    } while (elapsed < us);
}
\end{lstlisting}
\end{minipage}

The inner loop can be made parallel using one of many OpenMP directives. One is using a \emph{for} loop, which in this case would divide the loop iterations among the threads. Another is using a \emph{parallel for} loop, which in this case would create a nested parallel region. A third is the \emph{taskloop} directive. Example of a taskloop can be seen in listing~\ref{lst:sleeping_taskloop}.

We also try a manual variant of the taskloop construct where a single or multiple thread create explicit tasks for each iteration as seen in listings~\ref{lst:sleeping_s_mtasking} and~\ref{lst:sleeping_f_mtasking}.

\noindent \begin{minipage}{\linewidth}
\begin{lstlisting}[caption=Task generation with taskloop directive,frame=tlrb, label={lst:sleeping_taskloop}]{Name}
#pragma omp single
#pragma omp taskloop
for (int t = 0; t < num_tasks; t++) {
    perform_task(waittimes[t]);
}
\end{lstlisting}
\end{minipage}

\noindent \begin{minipage}{\linewidth}
\begin{lstlisting}[caption=Single-threaded task generation,frame=tlrb, label={lst:sleeping_s_mtasking}]{Name}
#pragma omp single
for (int t = 0; t < num_tasks; t++) {
    #pragma omp task
    perform_task(waittimes[t]);
}
\end{lstlisting}
\end{minipage}

\noindent \begin{minipage}{\linewidth}
\begin{lstlisting}[caption=Multi-threaded task generation,frame=tlrb, label={lst:sleeping_f_mtasking}]{Name}
#pragma omp for
for (int t = 0; t < num_tasks; t++) {
    #pragma omp task
    perform_task(waittimes[t]);
}
\end{lstlisting}
\end{minipage}

The iterations parameter defines how many times to perform all tasks, and a new parallel region is created for each iteration.
The granularity of each task is how many microseconds to stall. This is uniformly randomised between $0$ and a maximum task size parameter.
The final parameter is the number of tasks.
Since the task granularity is uniformly distributed between $0$ and $max\_task\_size$, the expected execution time is
$$T_{exec} = \frac{iterations \cdot num\_tasks \cdot max\_task\_size}{2 \cdot num\_threads}$$
without considering overheads.
This microbenchmark has the characteristics of an embarrassingly parallel workload, since no kind of dependencies between the tasks exists. 

\subsubsection{Inactivity microbenchmark}

The purpose of this microbenchmark is to evaluate the waiting policies of inactive threads.
The kernel consists of a team of threads where only one thread performs some dummy work while the other threads wait. This is repeated for a number of iterations to obtain a reasonable sample size.
The dummy work is the same as in the sleeping microbenchmark, meaning a busy-wait loop that stalls for the requested number of microseconds.

\noindent \begin{minipage}{\linewidth}
\begin{lstlisting}[caption=Kernel of inactivity microbenchmark,frame=tlrb, label={lst:inactivity_microbenchmark}]{Name}
#pragma omp parallel
{
  for (int i = 0; i < ITERATIONS; i++) {
    #pragma omp master
    stall_us(WAITTIME_US);
    #pragma omp barrier
  }
}
\end{lstlisting}
\end{minipage}

This can be executed with different values of the waiting time.
Intuitively, you would expect active scheduling to have a lower energy consumption for small task sizes due to the overheads of passive scheduling increasing execution time, while passive scheduling is better for large task sizes due to the power savings of not having threads spin.

\subsubsection{Unrolling microbenchmark}
\label{sec:background:unrolling_microbenchmark}
The purpose of this microbenchmark is to test the effects unrolling has on some simple for loops with various performance characteristics. For each scenario three different implementations will be tested, these are 1) no explicit unrolling, 2) OpenMP unrolling and 3) manual unrolling.

Four programs have been created for this microbenchmark, these are:
\begin{itemize}
  \item \texttt{SIMPLE\_MEM}, which loops over an array and initialises it with values (listing \ref{lst:SIMPLE_MEM}).
  \item \texttt{SIMPLE\_COMP}, which calculates the sum for some simple calculations (listing \ref{lst:SIMPLE_COMP}).
  \item \texttt{SIMPLE\_COMP\_DEPEND}, similar to the previous program but there are strict dependencies between each iteration (listing \ref{lst:SIMPLE_COMP_DEPEND}).
  \item \texttt{COMPLEX\_NESTED}, which contains an additional nested loop that does some calculation for each element in an array (listing \ref{lst:COMPLEX_NESTED}). Only the outermost loop is unrolled for this program.
\end{itemize}

\noindent\begin{minipage}{.45\textwidth}
\begin{lstlisting}[caption=The \texttt{SIMPLE\_MEM} program ,frame=tlrb,label={lst:SIMPLE_MEM},language=C]{Name}
for(int i=0; i<len; i++)
  A[i] = i;
\end{lstlisting}
\end{minipage}\hfill
\begin{minipage}{.45\textwidth}
\begin{lstlisting}[caption=The \texttt{SIMPLE\_COMP} program,frame=tlrb, label={lst:SIMPLE_COMP}]{Name}
float sum = 0;
for(int i=0; i<len; i++)
  sum += 1/(float)i;
\end{lstlisting}
\end{minipage}

\noindent\begin{minipage}{.45\textwidth}
\begin{lstlisting}[caption=The \texttt{SIMPLE\_COMP\_DEPEND} program ,frame=tlrb,label={lst:SIMPLE_COMP_DEPEND},language=C]{Name}
int sum = 0;
for(int i=0; i<len; i++)
  sum = (sum + i)%100;
\end{lstlisting}
\end{minipage}\hfill
\begin{minipage}{.45\textwidth}
\begin{lstlisting}[caption=The \texttt{COMPLEX\_NESTED} program,frame=tlrb, label={lst:COMPLEX_NESTED}]{Name}
for(int i=0; i<len; i++){
  A[i] = 0;
  for(int j=0; j<len; j++)
    A[i] += i*j;
}
\end{lstlisting}
\end{minipage}

The different unrolling implementation for each program have been created in the fashion described in section \ref{sec:background:unrolling}. The only exception is the manual unrolling implementations for \texttt{SIMPLE\_COMP} and \texttt{COMPLEX\_NESTED}, where some additional manual optimisations have been applied. The optimised versions can be seen in listings \ref{lst:SIMPLE_COMP_OPT} and \ref{lst:COMPLEX_NESTED_OPT} respectively.

For \texttt{SIMPLE\_COMP}, several different sum variables are used for the different offsets that are then added together after the loop is completed. This because it removes the implicit dependencies between the calculations, and can potentially improve performance. 

For \texttt{COMPLEX\_NESTED}, the inner nested loops are combined into a single nested loop (often called \emph{loop jamming}) when the outer loop is unrolled.

\noindent\begin{minipage}{.45\textwidth}
\begin{lstlisting}[caption={The \texttt{SIMPLE\_COMP} program, manually unrolled by a factor of two, with additional optimisations.} ,frame=tlrb,label={lst:SIMPLE_COMP_OPT},language=C]{Name}
float sum0, sum1;
for(int i=0; i<len; i+=2)
{
  sum0 += 1/(float)(i+0);
  sum1 += 1/(float)(i+1);
}
float sum = sum0+sum1;
\end{lstlisting}
\end{minipage}\hfill
\begin{minipage}{.45\textwidth}
\begin{lstlisting}[caption={The \texttt{COMPLEX\_NESTED} program, manually unrolled by a factor of two, with additional optimisations.},frame=tlrb, label={lst:COMPLEX_NESTED_OPT}]{Name}
for(int i=0; i<len; i+=2){
  A[(i+0)] = 0;
  A[(i+1)] = 0;
  for(int j=0; j<len; j++)
    {
    A[(i+0)] += (i+0)*j;
    A[(i+1)] += (i+1)*j;
    } 
}
\end{lstlisting}
\end{minipage}

\subsection{Own benchmarks}
\label{sec:theory:ownbenchmarks}

This section covers our own benchmarks, which are more realistic than our microbenchmarks. The ones we have made are particularly useful for experimenting with unrolling and tiling.

\subsubsection{Matrix multiplication}
\label{sec:methodology:matmul_and_strassen}

As seen in listing~\ref{lst:basic_naive_matmul}, basic matrix multiplication can be performed by iterating over the rows and columns of the target matrix and then for each element calculate the dot product of the row of the left matrix and column of the right matrix. 

However, it is also possible to change the order of the loops and achieve a significant speedup due to locality, as seen in listing~\ref{lst:alternate_matmul}.

\noindent \begin{minipage}{\linewidth}
\begin{lstlisting}[caption=Optimised code for matrix multiplication where the inner loops are switched.,frame=tlrb, label={lst:alternate_matmul}]{Name}
for (int row = 0; row < N; row++) {
  for (int k = 0; k < N; k++) {
    for (int col = 0; col < N; col++) {
      C[row*N + col] += A[row*N + k] * B[k*N + col];
    }
  }
}
\end{lstlisting}
\end{minipage}

We refer to the first as \emph{naive} matrix multiplication and the second as \emph{reordered} matrix multiplication. Due to its better memory access pattern we expect a smaller improvement from using loop tiling.

Both loop tiling and loop unrolling can be applied to matrix multiplication due to having nested loops. Tiling is applied using OpenMP directives as seen in listing~\ref{lst:omptiled_naive_matmul}, but also manually as seen in listing~\ref{lst:tiled_naive_matmul}. Example code for loop unrolling using OpenMP can be seen in listing~\ref{lst:omp_unrolled_matmul} and the corresponding manual unrolling code can be seen in listing~\ref{lst:manually_unrolled_matmul}. Only the innermost loop is unrolled in both cases.

\noindent \begin{minipage}{\linewidth}
\begin{lstlisting}[caption=Loop unrolled matrix multiplication using OpenMP.,frame=tlrb, label={lst:omp_unrolled_matmul}]{Name}
for (int row = 0; row < N; row++) {
  for (int col = 0; col < N; col++) {
    #pragma omp unroll partial(UNROLL_FACTOR)
    for (int k = 0; k < N; k++) {
      C[row*N + col] += A[row*N + k] * B[k*N + col];
    }
  }
}
\end{lstlisting}
\end{minipage}

\noindent \begin{minipage}{\linewidth}
\begin{lstlisting}[caption=Loop unrolled matrix multiplication of using manual unrolling of factor 4.,frame=tlrb, label={lst:manually_unrolled_matmul}]{Name}
for (int r = 0; r < N; r++) {
  for (int c = 0; c < N; c++) {
    int k = 0;
    for (; k < N - 3; k += 4) {
      C[r*N + c] += A[r*N + k] * B[k*N + c];
      C[r*N + c] += A[r*N + (k+1)] * B[(k+1)*N + c];
      C[r*N + c] += A[r*N + (k+2)] * B[(k+2)*N + c];
      C[r*N + c] += A[r*N + (k+3)] * B[(k+3)*N + c];
    }
    for (; k < N; k++) {
      C[r*N + c] += A[r*N + k] * B[k*N + c];
    }
  }
}
\end{lstlisting}
\end{minipage}

\subsubsection{2D stencil}
\label{sec:methodology2dstencil}

Our own benchmark \emph{2D stencil} is a simple program that takes a matrix as an argument and, for each cell, computes the average of all its surrounding neighbours.
For simplicity, the program skips computing the average of outermost cells since they would be a special case otherwise. Listing~\ref{lst:2dstencil_basic} shows an outline of the basic code.
The purpose of introducing this benchmark is to have a program less computationally and memory intensive than matrix multiplication, while still being a relatively realistic pattern. More importantly, tiling can be applied very easily.

\noindent \begin{minipage}{\linewidth}
\begin{lstlisting}[caption=Basic code outline for the 2D stencil program. Note that no calculation is performed for the edges.,frame=tlrb, label={lst:2dstencil_basic}]{Name}
for (int i = 1; i < N - 1; i++)
  for (int j = 1; j < N - 1; j++)
      matrix[i*N + j] = (
          matrix[(i-1)*N+(j-1)] +
          matrix[(i-1)*N + j] +
          matrix[(i-1)*N + (j+1)] +
          matrix[(i)*N + (j-1)] +
          matrix[(i)*N + j] +
          matrix[(i)*N + (j+1)] +
          matrix[(i+1)*N + (j-1)] +
          matrix[(i+1)*N + j] +
          matrix[(i+1)*N + (j+1)]
      ) / 9.0;
\end{lstlisting}
\end{minipage}

It should also be mentioned that there are possible race conditions in this program. As one thread operates on row $i$, another can operate on row $i+1$ leading to race conditions as the second thread will read from row $i$ which the first thread writes to. This is, however, not an issue we will consider.

\subsection{Common benchmark suites}
\label{sec:theory:benchmarks}

This section covers the used benchmark suites: BOTS, NPB, and PARSEC.
All of these are used for the analysis of the waiting policy by executing their respective programs with the different waiting policy options, compilers and numbers of threads. They are also used for a more realistic analysis of the code transformations, where we profile each program and try to apply loop unrolling and tiling in a more realistic setting.

BOTS (Barcelona OpenMP Task Suite) is a benchmark suite targeting exploitation of the irregular parallelism offered by OpenMP tasking\footnote{The BOTS benchmarks can be found in \href{https://github.com/bsc-pm/bots}{https://github.com/bsc-pm/bots}}. From BOTS we use all programs: \emph{alignment}, \emph{fft}, \emph{floorplan}, \emph{nqueens}, \emph{sparselu}, \emph{strassen} and \emph{uts}.
Two of these, \emph{alignment} and \emph{sparselu}, have the option to create tasks in either a single-threaded or multi-threaded manner, similar to how the parallelism constructs microbenchmark does as seen in listings~\ref{lst:sleeping_s_mtasking} and~\ref{lst:sleeping_f_mtasking}. We therefore use these as part of our analysis of parallelism constructs.

The NAS Parallel Benchmark Suite (NPB) was developed by NASA in 1991 to evaluate the performance of parallel supercomputers.
The programs are derived from computational fluid dynamics applications \cite{noauthor_nas_nodate}.
Most of these programs are, however, implemented in Fortran, which is not in the scope of this thesis. Therefore, we instead use an unofficial implementation ported to C \footnote{NAS Parallel Benchmarks ported to C can be found in \href{https://github.com/benchmark-subsetting/NPB3.0-omp-C}{https://github.com/benchmark-subsetting/NPB3.0-omp-C}}.
From NPB we use the following programs: \emph{BT}, \emph{CG}, \emph{EP}, \emph{FT}, \emph{IS}, \emph{LU} and \emph{MG}. There is also the SP program, but we decided to not use it due to issues with compilation and execution.
The problem sizes with which to run NPB are known as classes
\footnote{NPB classes are described in \href{https://www.nas.nasa.gov/software/npb\_problem\_sizes.html}{https://www.nas.nasa.gov/software/npb\_problem\_sizes.html}}.
The smallest problem size (S) is very small, making parallel computations not worth the overhead. Instead, we use the larger A class.

The final benchmark suite we used was PARSEC, (Princeton Application Repository for Shared-Memory Computers). PARSEC is a benchmark suite designed to test multiprocessors for parallel workloads ranging from many different domains such as computer vision and financial analysis \cite{bienia_parsec_2008}.
The programs each have several implementations utilising different parallelism models, such as pthreads and OpenMP. But since our main focus was OpenMP, instead an extension of the benchmark called PARSECSs \footnote{The Git repository for PARSECSs can be found here: https://pm.bsc.es/gitlab/benchmarks/parsec-ompss}, which includes two OpenMP versions, a \emph{parallel for} and \emph{tasking} implementation, was used. We use these versions in our analysis of the OpenMP parallelism constructs.

Not all of the original 12 PARSEC benchmarks was ported to use both the OpenMP implementations, and some of the benchmarks we did not get to work or had other technical issues with. In the end the two following benchmarks was used for the final evaluation: \emph{Blackscholes} and \emph{Fluidanimate}.
PARSEC comes with many different input sizes with execution times ranging from almost instantaneous to several minutes. In our analysis we used the largest input set, called \emph{Native}.

%% file: include/experimentalmethodology.tex
This chapter describes details about the experimental setup and procedure of the project, especially how measurement of energy consumption is performed. It also includes an overview of the overall workflow used, as well as potential sources of error.

\subsection{Parameter search space}

We systematically evaluate loop unrolling, loop tiling, tasking and parallel for loops with different compiler optimisations, number of threads and waiting policies.

Loop unrolling and tiling are evaluated both by applying them to our own benchmarks and the common benchmark suites BOTS, NAS and PARSEC. In the benchmark suites, we analyse the program and attempt to apply unrolling and tiling where it is possible and evaluate the results.
These programs are heavily optimised and often have manually unrolled and tiled loops. These optimisations can be removed and replaced with both unrolled and tiled loops using OpenMP and with the optimisation removed. We expect however that the original version will be better since it has optimisations that OpenMP or the compiler might not be able to replicate.

\emph{Tasking} and \emph{parallel for} are two distinct ways of achieving parallelism with OpenMP. However, in many cases, it is possible to solve a problem using either one of them. Therefor we will run some of the applications using two different implementations, one implemented with tasks and the other with parallel for, to compare if there is an advantage of using one over the other.

Waiting policies is evaluated mainly by executing the common benchmark suites and our microbenchmarks with the different settings.

\subsection{Hardware and software details}

Experiments were performed on nodes on the Tetralith HPC cluster. These nodes have 16-core dual socket Intel Xeon Gold 6130 processors (32 cores in total), 96 GiB RAM and run Linux version 3.10.0. Allocated nodes can have different disk sizes, but we limit ourselves to only the smallest disk size of 240 GB (SSD).
The Tetralith nodes have support to model energy consumption using RAPL, described further in section~\ref{sec:method:rapl}.

Three compilers were used to compile C code with OpenMP:
the Intel compiler \emph{ICC} version 18.0.1, which was available on Tetralith,
the GNU C compiler \emph{GCC} version 10.1.0, compiled from source, and the LLVM C compiler \emph{Clang} version 13.0.0, also compiled from source. Clang 13.0.0 is the only compiler of the three which has support for the newly introduced tiling and unrolling OpenMP directives.

\subsection{Energy consumption measurements}
\label{sec:method:rapl}

Many modern processors provide means to obtain the power consumption of the processor through model-based power estimations~\cite{goel_methodology_2016}. In Intel processors, since the Sandy Bridge architecture, this is done using the tool Running Average Power Limit (RAPL), which provides data on the accumulated energy consumption. The tool has an power estimation accuracy of 1 \%, with a standard deviation of 1.1 \% \cite{david_rapl_2010}. Through RAPL it is also possible to constrain the maximum power consumption allowed by various subsystems, but it was decided that exploring this feature of RAPL was out of scope for this paper and will not be used in experimentation.

On Tetralith, RAPL reports energy consumption as a running total, which resets after reaching a threshold. To measure the energy consumption between two points in time is therefore done by reading this value at both points in time, and then subtracting the initial energy value from the energy value at the second point in time. It could happen that the counter reaches the threshold during this time and resets, so this overflow must also be taken into account. The energy counters are stored as files, which means that accessing them is done by reading them as any other file. The energy consumption of each 16-core socket and their main memory is also reported separately, but we do not consider this in our analysis and simply add these together.

\subsection{Evaluation workflow}
\label{sec:experimentalprocedure}

In order to integrate energy measurements into the benchmarks as seamlessly as possible, we implement all of the energy reading logic in a separate C file. This allows us to measure the energy consumption of only parts of a program, instead of measuring the energy for all of it.
Listing~\ref{lst:basic_energy_measurements} shows how this can be done with the example of matrix multiplication. The \emph{poll\_before} function creates a data structure with the readings before the program kernel. \emph{poll\_after} reads the energy counters again, compensates for potential overflow, and outputs the energy consumption to a JSON file.

Performing no readings during the execution has the benefit of not interrupting the program during execution to read the counter values. It has however a major disadvantage in that the peak power cannot be measured. Average power can however be calculated easily as the energy consumption divided by the execution time. \textbf{All reported power values in this report consider average power}.

\noindent \begin{minipage}{\linewidth}
\begin{lstlisting}[caption=Methodology of energy measurements with a matrix multiplication example,frame=tlrb, label={lst:basic_energy_measurements}]{Name}
int main() {
    matrix_t* A, B, C;
    rand_matrices(A, B);
    measurements_t* m = poll_before();
    compute_matmul(C, A, B);
    poll_after(m, "measurements_outputs.json");
    free_matrices(A, B, C);
}
\end{lstlisting}
\end{minipage}

Thus far, the approach does not take idle power into account, i.e., the power consumed while the machine is inactive.
This can give misleading results, especially when the program only uses a small fraction of the machine such as only running with one thread.
This is assumed to be fairly constant for a given system, and can therefore be measured once and the compensated energy given by

$$E_{compensated} = E_{total} - P_{static} \cdot T_{exec}$$

where $E_{total}$ is the total energy consumed, $P_{static}$ is the static energy consumption of the system, and $T_{exec}$ is the execution time of the program kernel.

However, static power can differ between nodes on the data center.
Therefore, it is measured before running a set of benchmarks on the allocated node, simply by sleeping for a number of seconds and measuring the used consumed energy.

\begin{figure}[h]
    \centering
    \includegraphics[width=0.9\textwidth]{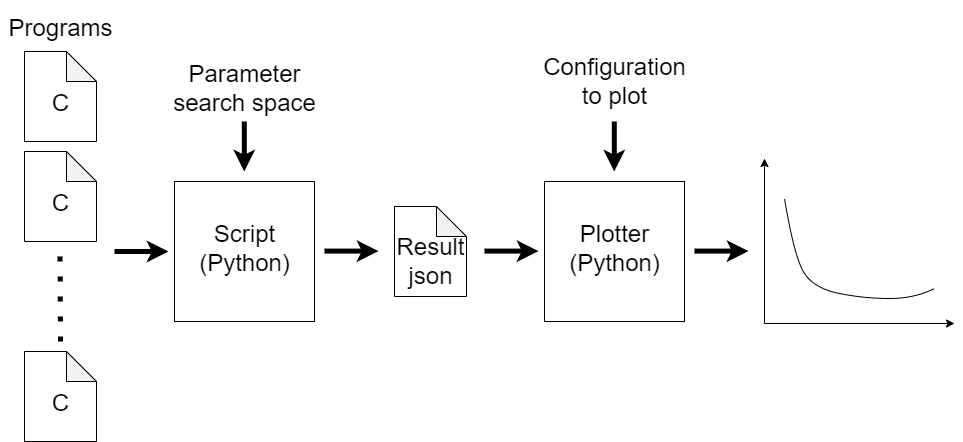}
    \caption{The workflow we developed and used for running benchmarks and plotting their result. A python script is used that takes some benchmarks written in C and also a set of runtime parameters, such as waiting policy and number of threads. The script then runs these benchmarks with all parameters combinations and saves the results to a json file. Another python script can then read this file and make some plot after a specified configuration.}
    \label{fig:workflow}
\end{figure}

We use Python scripts to compile and execute a range of configurations, saving the outputs to a results file \emph{results.json}. We then use another set of scripts to read these results and plot figures. This process was developed so large batch jobs could be executed with no further input from the user, which let us easily and methodically test large amounts of benchmarks and  configurations. The whole process of our workflow when testing the benchmarks can be seen in figure~\ref{fig:workflow}.

When compiling code, optimisation level \emph{O3} is always used unless specified otherwise. Furthermore, for every configuration a program is executed between 3 and 10 times, and the measurements (execution time, energy, and power) are reported as the geometric mean of these executions.

\subsection{Potential sources of error}

On Tetralith nodes, processor frequency is controlled by system administrators and is not possible to adjust by us.
The nodes run the intel\_pstate driver on the scaling governor \emph{powersave}, which selects p states based on current CPU utilisation \footnote{Information about intel\_pstate can be found in\\ \href{https://www.kernel.org/doc/html/v5.12/admin-guide/pm/intel\_pstate.html}{https://www.kernel.org/doc/html/v5.12/admin-guide/pm/intel\_pstate.html}}. While it perhaps would be better to run with the highest performance all the time, this is a more realistic setting.

C states are also controlled by system administrators and are configured to aggressively limit the power consumption.
In \cite{ournani_taming_2020}, it is found that enabled C-states highly impacts variations in energy of benchmarks, so it would have been better to disable them. Again, however, enabled C-states provides more realistic data since most data centers will have them enabled.

%% file: include/results.tex
This chapter presents the results of the experiments that we ran for the different benchmarks.
First, we cover our own matrix multiplication and 2D stencil programs in sections~\ref{sec:results:matmul} and~\ref{sec:results:stencil}.
Next, we present the microbenchmark results in sections~\ref{sec:results:parconstructs_microbench},~\ref{sec:results:inactivity_microbench} and~\ref{sec:results:loop_unrolling_microbench}.
After that, we cover the common benchmark suites in sections~\ref{sec:results:bots},~\ref{sec:results:nas}, and~\ref{sec:results:parsec}. Finally, we conclude the chapter with a compilation of the most interesting data in section~\ref{sec:results:summary} and answer some of the research questions based on that.

In general, we present metrics (energy, execution time and power) either in absolute numbers or relative to some baseline. Use of relative numbers is useful when the differences between configurations is very small, which is often the case for example in unrolling. 

When plotting data, we often show only the energy consumption and not execution time. This is because the two quantities are almost always proportional to one another, which makes showing both redundant. As this project focuses on energy consumption, we choose to focus on that. Execution time and power are however often reported in summarising tables along with energy.

\subsection{Matrix multiplication}
\label{sec:results:matmul}

This section evaluates loop transformations on matrix multiplication. The input matrices are 1024x1024 single-precision floating point numbers. This size is chosen because the execution time is about a second, which is short enough to run many configurations in a short time but long enough for the overheads of parallelism management to be minor.
All experiments use the \emph{active} waiting policy and the \emph{static} loop scheduling.

\subsection{Loop tiling}

As described in section~\ref{sec:backgroundlooptiling}, loop tiling transforms for loops to improve data locality and reduce overhead.
We apply three versions of tiling. The first is using the OpenMP tiling directive, as described in listing~\ref{lst:omptiled_naive_matmul}. The second is a tiling version similar to listing~\ref{lst:tiled_naive_matmul}, where we do not assume that the number of iterations are divisible by the tile size. The final version, similar to listing~\ref{lst:tiled_naive_matmul_with_checks}, is a manually tiled version where we do make this assumption.
The tested tile sizes are 1, 2, 4, 8, 16, 32 and 64. Size 8 proved to be the best, which is used in the presented results.

\begin{figure}[h]
  \centering
  \includegraphics[width=\textwidth]{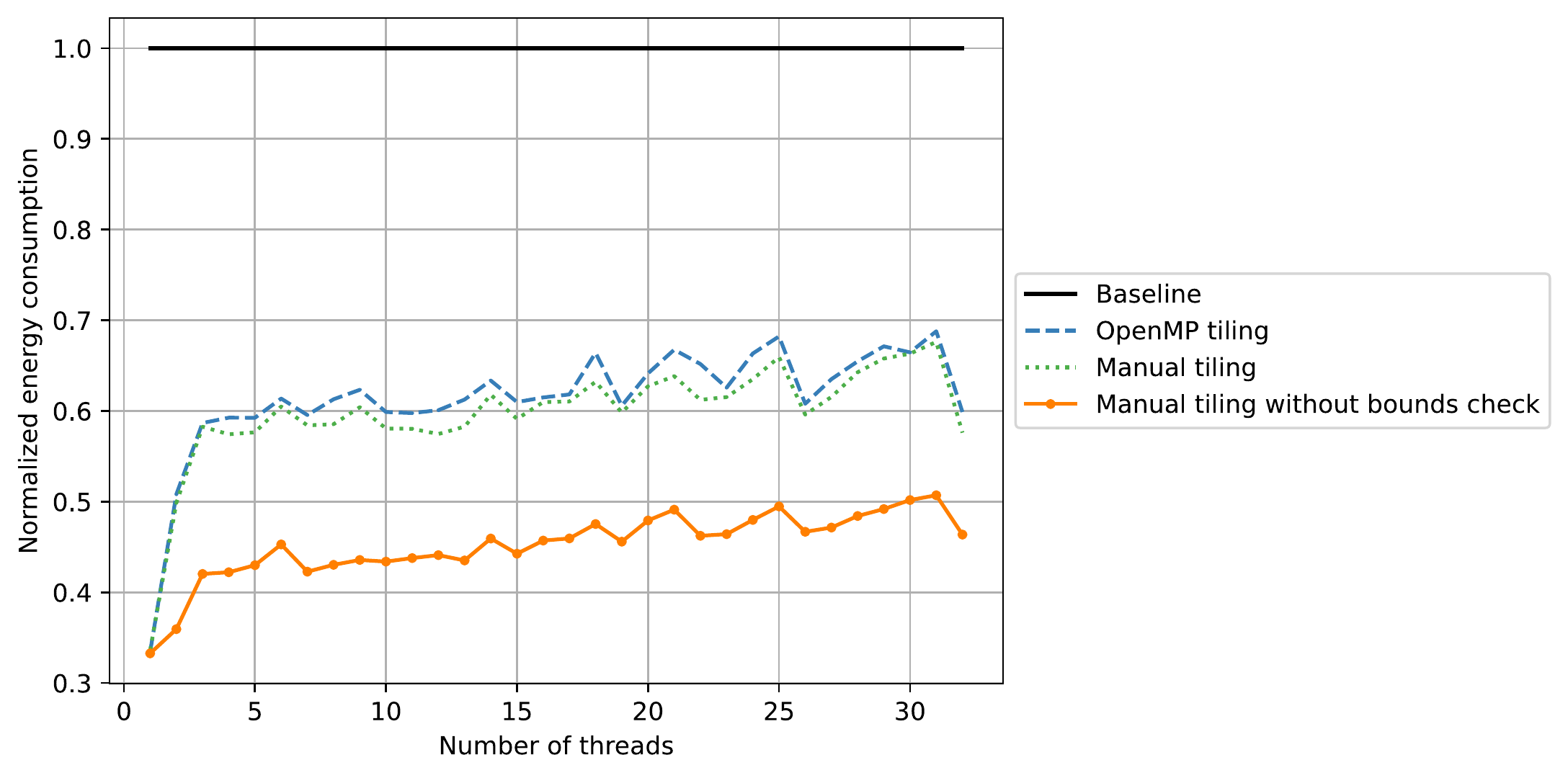}
  \caption{Relative energy consumption for the naive matrix multiplication with tiling.}
  \label{fig:matmul_clang_tiling_naive_energy}
\end{figure}

We first present the results for the Clang compiler only.
In figure~\ref{fig:matmul_clang_tiling_naive_energy} we show the energy consumption of the tiled versions relative to the baseline program when using the naive matrix multiplication algorithm. 
We see a clear reduction in energy when applying loop tiling, which seems fairly consistent for most numbers of threads but is especially effective for very low numbers of threads (1 or 2). We also see that both manual versions outperform the OpenMP version, but the second version much more so.

\begin{figure}[h]
  \centering
  \includegraphics[width=\textwidth]{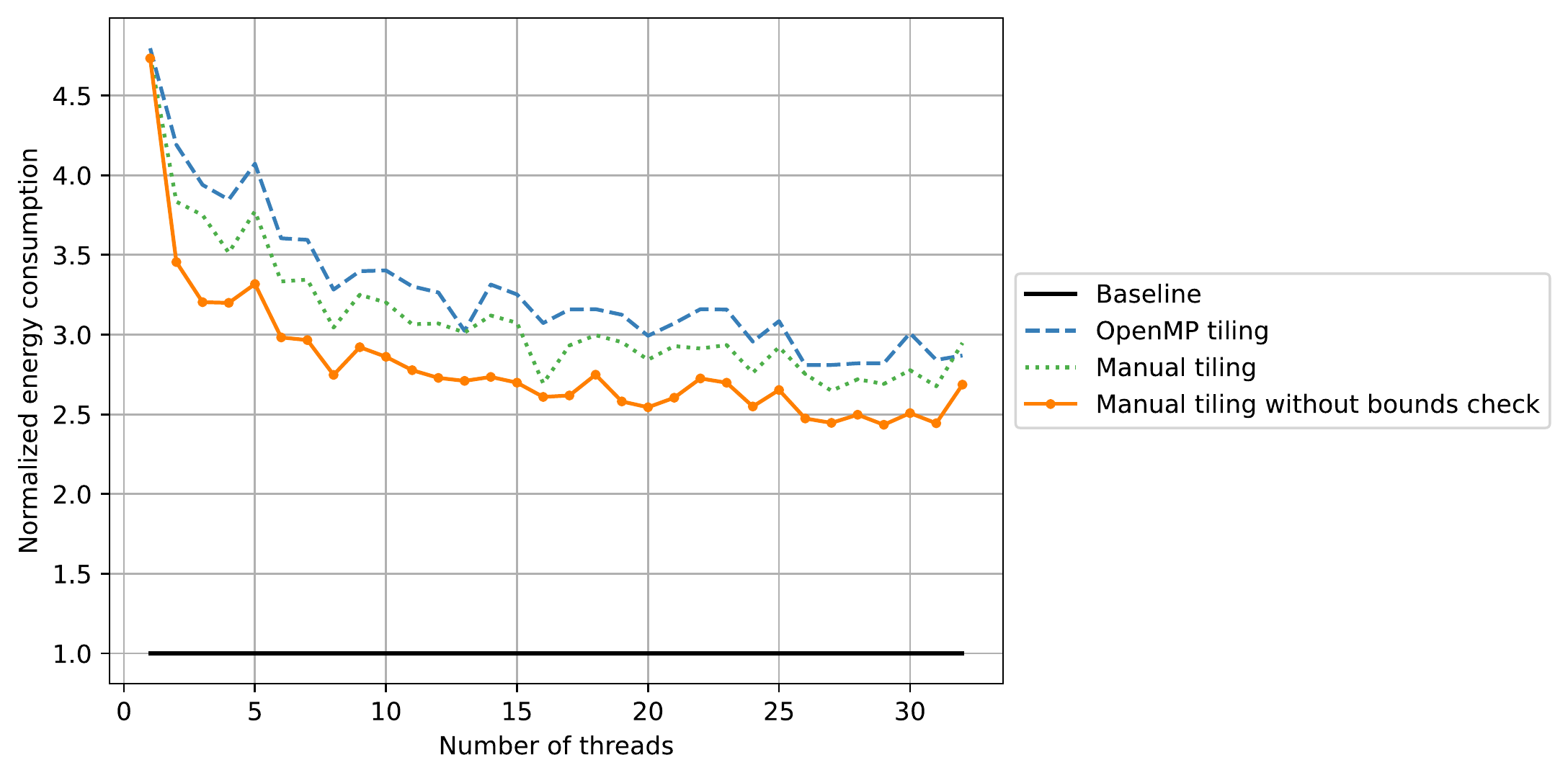} \caption{Relative energy consumption for the reordered matrix multiplication with tiling.}
  \label{fig:matmul_clang_tiling_reordered_energy}
\end{figure}

Moving on to the reordered algorithm described in section~\ref{sec:methodology:matmul_and_strassen}, we see the energy reduction results in figure~\ref{fig:matmul_clang_tiling_reordered_energy}.
While we do not expect to see any major speedup from tiling with the reordered algorithm, we see a major performance downgrade. This is probably not due to the tiling itself, but rather that the tiling prevents other, more significant compiler optimisations.

To investigate the cause of this performance downgrade, we use the \emph{perf} tool to analyse L1-D cache miss rate with and without loop tiling. We use 16 threads and the \emph{active} waiting policy. Results can be seen in table~\ref{tab:tiling_perf_hunt}. For the naive algorithm, we see that the load misses go down significantly while the total loads increase somewhat when applying tiling, which is the expected result since tiling increases locality. With the reordered algorithm however, both the load misses and total loads are significantly higher when applying tiling. This could be due to a compiler optimisation that is applied successfully without tiling, but not with it.

\begin{table}[h]
\caption{L1-D cache behaviour for 16-threaded matrix multiplication with and without tiling.}
\begin{tabular}{|l|l|l|l|}
\hline
                         & L1-D load misses & L1-D loads    & L1-D load miss rate \\ \hline
Naive, baseline          & 1,080,290,162    & 2,184,350,663 & 49.46 \%           \\ \hline
Naive, OpenMP tiling     & 304,085,476      & 2,439,204,188 & 12.47 \%           \\ \hline
Naive, Manual tiling     & 307,412,605      & 2,439,204,151 & 12.60 \%           \\ \hline
Reordered, baseline      & 67,327,997       & 573,736,879   & 11.73 \%           \\ \hline
Reordered, OpenMP tiling & 191,507,084      & 3,397,602,579 & 5.64 \%            \\ \hline
Reordered, Manual tiling & 195,533,659      & 3,399,716,260 & 5.75 \%            \\ \hline
\end{tabular}
\label{tab:tiling_perf_hunt}
\end{table}

Thus far we have analysed results produced by Clang only.
The results for GCC are very similar to those of Clang, with tiling being highly beneficial for the naive algorithm and highly detrimental to the reordered algorithm. Plots can be found in figures~\ref{fig:matmul_gcc_tiling_naive_energy} and~\ref{fig:matmul_gcc_tiling_reordered_energy} in the appendix.
For ICC, however, tiling is highly detrimental in all cases except for the reordered algorithm where the loop iterations are assumed to be divisible by the tile size. In that specific case, the tiling provides a noticeable energy reduction that seems to be fairly constant across different numbers of threads. Plots can be found in figures~\ref{fig:matmul_icc_tiling_naive_energy} and~\ref{fig:matmul_icc_tiling_reordered_energy}, also in the appendix. For ICC, the naive algorithm without tiling performs about as well as the reordered algorithms for Clang and GCC without tiling. 

Table~\ref{tab:matmul_tiling_summary} shows the average improvements across all the number of threads for each compiler. Manual v1 refers to code such as listing~\ref{lst:tiled_naive_matmul_with_checks}, where it is not assumed that the tile size evenly divides the iterations, while manual v2 refers to code such as listing~\ref{lst:tiled_naive_matmul_with_checks}, where this is assumed.

We note that for ICC, the naive algorithm with no tiling performs about as well as the reordered algorithms for Clang and GCC without tiling. This indicates that ICC is capable of changing loop ordering and turning the naive algorithm into the reordered one by itself, which in this case is a drastic improvement. However, when we try to apply tiling, it fails to apply this optimisation and the performance deteriorates to being comparable to those of the other compilers.

To summarise, there are essentially two groups of results here: either everything gets much better with tiling, as with Clang and GCC on the naive version, or it gets much worse, as with Clang and GCC on the reordered version. ICC is the odd one out, as it gets significantly worse with all versions of tiling except the second tiling version for the reordered algorithm. While ICC seems to perform the best overall, the single best result is actually using GCC with the baseline reordered version.

\begin{table}[h]
\begin{center}
\caption{Summary of the matrix multiplication tiling results across all numbers of threads. The reported execution time, power and energy are relative to the unmodified program.}
\label{tab:matmul_tiling_summary}
\begin{tabular}{|c|c|c|ccc|ccc|}
\hline
Compiler & Version & Tiling & \multicolumn{3}{c|}{Absolute} & \multicolumn{3}{c|}{Relative} \\ \hline
 &  &  & \multicolumn{1}{c|}{T {[}s{]}} & \multicolumn{1}{c|}{P {[}W{]}} & E {[}J{]} & \multicolumn{1}{c|}{T} & \multicolumn{1}{c|}{\textbf{P}} & \textbf{E} \\ \hline
clang & Naive & Baseline & \multicolumn{1}{c|}{0.290} & \multicolumn{1}{c|}{172.3} & 49.90 & \multicolumn{1}{c|}{1} & \multicolumn{1}{c|}{1} & 1 \\ \hline
clang & Naive & OpenMP & \multicolumn{1}{c|}{0.183} & \multicolumn{1}{c|}{166.7} & 30.47 & \multicolumn{1}{c|}{0.631} & \multicolumn{1}{c|}{0.967} & \textbf{0.610} \\ \hline
clang & Naive & Manual v1 & \multicolumn{1}{c|}{0.181} & \multicolumn{1}{c|}{163.8} & 29.66 & \multicolumn{1}{c|}{0.625} & \multicolumn{1}{c|}{0.950} & \textbf{0.594} \\ \hline
clang & Naive & Manual v2 & \multicolumn{1}{c|}{0.138} & \multicolumn{1}{c|}{163.2} & 22.47 & \multicolumn{1}{c|}{0.475} & \multicolumn{1}{c|}{0.947} & \textbf{0.450} \\ \hline
gcc & Naive & Baseline & \multicolumn{1}{c|}{0.276} & \multicolumn{1}{c|}{172.8} & 47.71 & \multicolumn{1}{c|}{1} & \multicolumn{1}{c|}{1} & 1 \\ \hline
gcc & Naive & Manual v1 & \multicolumn{1}{c|}{0.129} & \multicolumn{1}{c|}{172.3} & 22.31 & \multicolumn{1}{c|}{0.469} & \multicolumn{1}{c|}{0.998} & \textbf{0.468} \\ \hline
gcc & Naive & Manual v2 & \multicolumn{1}{c|}{0.059} & \multicolumn{1}{c|}{131.2} & 7.764 & \multicolumn{1}{c|}{0.214} & \multicolumn{1}{c|}{0.760} & \textbf{0.163} \\ \hline
icc & Naive & Baseline & \multicolumn{1}{c|}{0.043} & \multicolumn{1}{c|}{119.6} & 5.140 & \multicolumn{1}{c|}{1} & \multicolumn{1}{c|}{1} & 1 \\ \hline
icc & Naive & Manual v1 & \multicolumn{1}{c|}{0.109} & \multicolumn{1}{c|}{167.6} & 18.30 & \multicolumn{1}{c|}{2.541} & \multicolumn{1}{c|}{1.401} & \textbf{3.560} \\ \hline
icc & Naive & Manual v2 & \multicolumn{1}{c|}{0.067} & \multicolumn{1}{c|}{125.9} & 8.44 & \multicolumn{1}{c|}{1.560} & \multicolumn{1}{c|}{1.052} & \textbf{1.642} \\ \hline
clang & Reordered & Baseline & \multicolumn{1}{c|}{0.050} & \multicolumn{1}{c|}{121.2} & 6.070 & \multicolumn{1}{c|}{1} & \multicolumn{1}{c|}{1} & 1 \\ \hline
clang & Reordered & OpenMP & \multicolumn{1}{c|}{0.133} & \multicolumn{1}{c|}{148.8} & 19.78 & \multicolumn{1}{c|}{2.654} & \multicolumn{1}{c|}{1.228} & \textbf{3.260} \\ \hline
clang & Reordered & Manual v1 & \multicolumn{1}{c|}{0.127} & \multicolumn{1}{c|}{147.0} & 18.66 & \multicolumn{1}{c|}{2.535} & \multicolumn{1}{c|}{1.213} & \textbf{3.075} \\ \hline
clang & Reordered & Manual v2 & \multicolumn{1}{c|}{0.117} & \multicolumn{1}{c|}{144.1} & 16.88 & \multicolumn{1}{c|}{2.340} & \multicolumn{1}{c|}{1.189} & \textbf{2.782} \\ \hline
gcc & Reordered & Baseline & \multicolumn{1}{c|}{0.034} & \multicolumn{1}{c|}{126.8} & 4.37 & \multicolumn{1}{c|}{1} & \multicolumn{1}{c|}{1} & 1 \\ \hline
gcc & Reordered & Manual v1 & \multicolumn{1}{c|}{0.064} & \multicolumn{1}{c|}{143.3} & 9.13 & \multicolumn{1}{c|}{1.846} & \multicolumn{1}{c|}{1.130} & \textbf{2.086} \\ \hline
gcc & Reordered & Manual v2 & \multicolumn{1}{c|}{0.100} & \multicolumn{1}{c|}{141.5} & 14.12 & \multicolumn{1}{c|}{2.891} & \multicolumn{1}{c|}{1.116} & \textbf{3.227} \\ \hline
icc & Reordered & Baseline & \multicolumn{1}{c|}{0.043} & \multicolumn{1}{c|}{119.1} & 5.15 & \multicolumn{1}{c|}{1} & \multicolumn{1}{c|}{1} & 1 \\ \hline
icc & Reordered & Manual v1 & \multicolumn{1}{c|}{0.081} & \multicolumn{1}{c|}{142.7} & 11.56 & \multicolumn{1}{c|}{1.871} & \multicolumn{1}{c|}{1.199} & \textbf{2.243} \\ \hline
icc & Reordered & Manual v2 & \multicolumn{1}{c|}{0.041} & \multicolumn{1}{c|}{108.2} & 4.44 & \multicolumn{1}{c|}{0.949} & \multicolumn{1}{c|}{0.909} & \textbf{0.862} \\ \hline
\end{tabular}
\end{center}
\end{table}

\subsubsection{Loop unrolling}

Partial loop unrolling is applied to the innermost loop in the matrix multiplication kernel.
Three versions of unrolling are used: with OpenMP, manual assuming the loop iterations are divisible by the unroll factor, and manual where this is not assumed.
The tested unroll factors are 1, 2, 4, 8, 16, 32 and 64, where 32 turns out to be the best and is used for the following demonstrated results.

\begin{figure}[h]
  \centering
  \includegraphics[width=\textwidth]{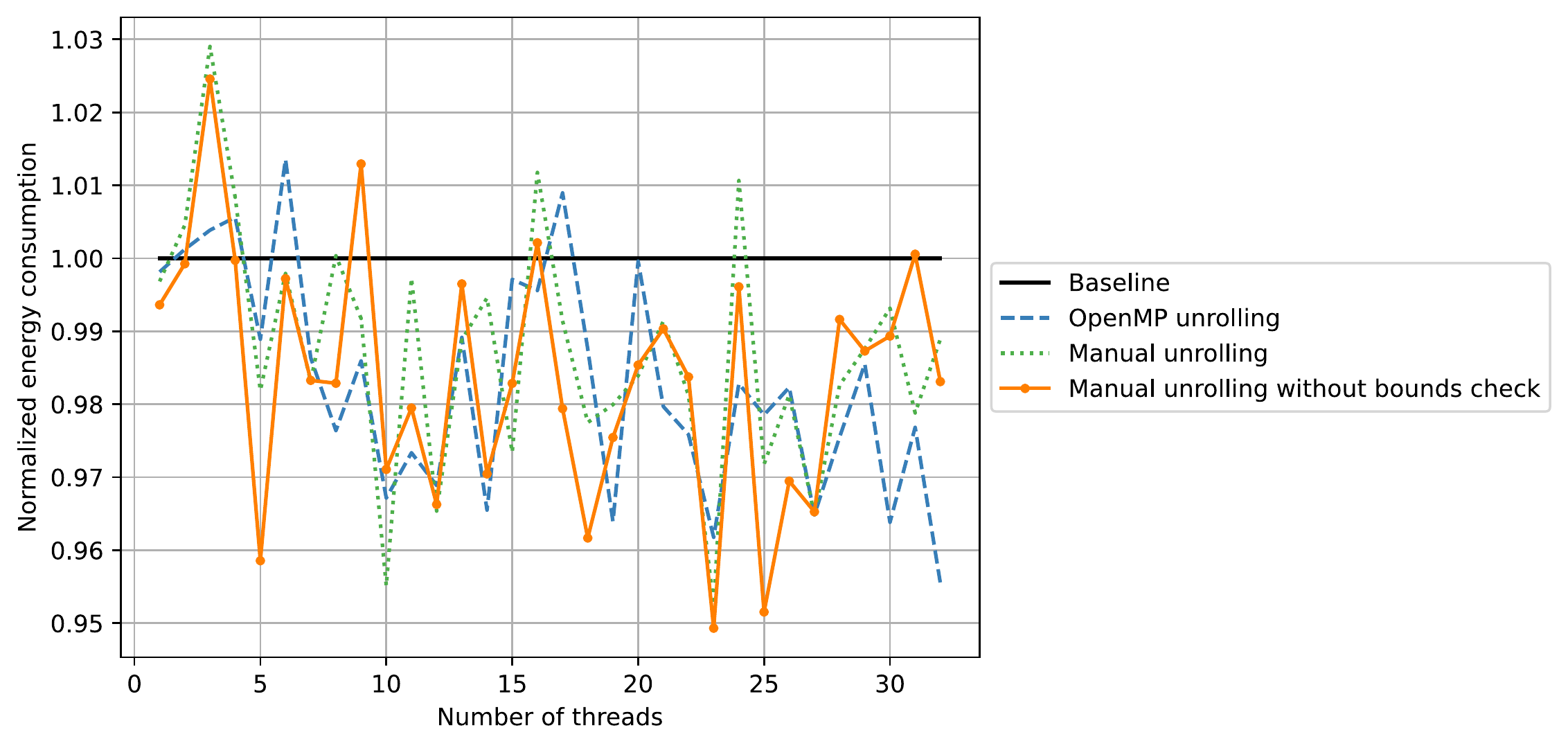}
  \caption{Relative energy consumption for the naive matrix multiplication with unrolling.}
  \label{fig:matmul_clang_unroll_naive_energy}
\end{figure}

Once again we begin with the results for Clang only. In figure~\ref{fig:matmul_clang_unroll_naive_energy} we see the relative energy consumption for unrolling the naive matrix multiplication algorithm. We see that all versions of unrolling seem to get better results when more threads are used, while being worse than the baseline when running with a single thread. At a glance, it looks like neither unrolling method is  definitely better than the others. We note that the differences between the versions are very small, only a few percent.

\begin{figure}[h]
  \centering
  \includegraphics[width=\textwidth]{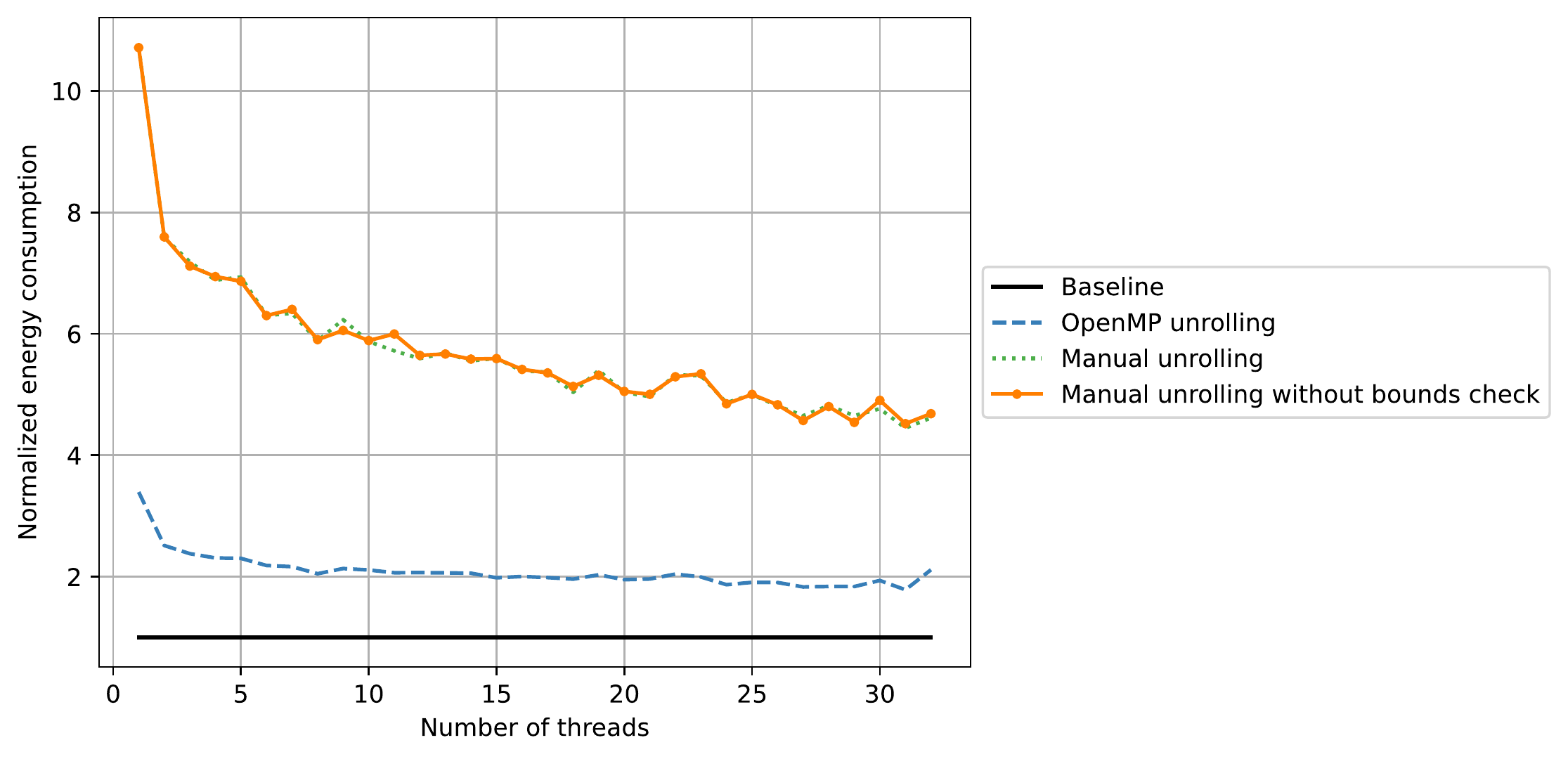}
  \caption{Relative energy consumption for the reordered matrix multiplication with unrolling.}
  \label{fig:matmul_clang_unroll_reordered_energy}
\end{figure}

Figure~\ref{fig:matmul_clang_unroll_reordered_energy} shows the same results using the reordered algorithm. Now, all versions of unrolling are strictly worse than not applying unrolling. We note that the increase in energy consumption is much larger than the decrease for the naive version. As for what causes this, we believe it is the same phenomenon as with loop tiling: the unrolling prevents other, more significant optimisations from being applied.

For GCC, the results are very similar to those of Clang; unrolling the naive versions provides a few percent lower energy, while unrolling the reordered version roughly doubles the energy. Plots can be found in the appendix, in figures~\ref{fig:matmul_gcc_unroll_naive_energy} and~\ref{fig:matmul_gcc_unroll_reordered_energy}.
ICC is, as usual, the odd one out. Unrolling the naive algorithm makes the results several times worse regardless of unrolling method. For the reordered algorithm we however see that the second unrolling method, where we assume that the unrolling factor evenly divides the loop iterations, the results become better than the baseline.

Table~\ref{tab:matmul_unrolling_summary} shows the average results for all numbers of threads. As with loop tiling, there are two groups of results. In the first group, we have minor energy reductions of about 1.5 \% for Clang and one GCC configuration. All other results are terrible, at least doubling energy consumption and generally increasing power about 30 \%.
We see that results from unrolling the naive version with ICC roughly matches the results for Clang and GCC, indicating that the same phenomenon has happened for loop unrolling as happened with loop tiling: ICC successfully performed loop reordering in the baseline version, but failed to do so when a loop transformation is applied.
To conclude, when applying loop unrolling to a program such as this, the programmer can expect either a very minor improvement or a massive deterioration of execution time and energy consumption likely caused by other optimisations not being applied successfully.

\begin{table}[h]
\begin{center}
\caption{Summary of the matrix multiplication unrolling results across all numbers of threads. The relative numbers are relative to the baseline with no unrolling for the same compiler.}
\label{tab:matmul_unrolling_summary}
\begin{tabular}{|c|c|c|ccc|ccc|}
\hline
Compiler & Version & Unrolling & \multicolumn{3}{c|}{Absolute} & \multicolumn{3}{c|}{Relative} \\ \hline
 &  &  & \multicolumn{1}{c|}{T {[}s{]}} & \multicolumn{1}{c|}{P {[}W{]}} & E {[}J{]} & \multicolumn{1}{c|}{T} & \multicolumn{1}{c|}{P} & \textbf{E} \\ \hline
clang & Naive & Baseline & \multicolumn{1}{c|}{0.289} & \multicolumn{1}{c|}{169.5} & 48.93 & \multicolumn{1}{c|}{1} & \multicolumn{1}{c|}{1} & \textbf{1} \\ \hline
clang & Naive & OpenMP & \multicolumn{1}{c|}{0.288} & \multicolumn{1}{c|}{167.0} & 48.13 & \multicolumn{1}{c|}{0.998} & \multicolumn{1}{c|}{0.985} & \textbf{0.984} \\ \hline
clang & Naive & Manual v1 & \multicolumn{1}{c|}{0.288} & \multicolumn{1}{c|}{167.3} & 48.13 & \multicolumn{1}{c|}{0.996} & \multicolumn{1}{c|}{0.987} & \textbf{0.984} \\ \hline
clang & Naive & Manual v2 & \multicolumn{1}{c|}{0.289} & \multicolumn{1}{c|}{167.3} & 48.27 & \multicolumn{1}{c|}{0.999} & \multicolumn{1}{c|}{0.987} & \textbf{0.986} \\ \hline
\hline
gcc & Naive & Baseline & \multicolumn{1}{c|}{0.280} & \multicolumn{1}{c|}{176.4} & 49.39 & \multicolumn{1}{c|}{1} & \multicolumn{1}{c|}{1} & \textbf{1} \\ \hline
gcc & Naive & Manual v1 & \multicolumn{1}{c|}{0.279} & \multicolumn{1}{c|}{170.7} & 47.65 & \multicolumn{1}{c|}{0.997} & \multicolumn{1}{c|}{0.968} & \textbf{0.965} \\ \hline
gcc & Naive & Manual v2 & \multicolumn{1}{c|}{0.278} & \multicolumn{1}{c|}{170.9} & 47.48 & \multicolumn{1}{c|}{0.992} & \multicolumn{1}{c|}{0.969} & \textbf{0.961} \\ \hline
\hline
icc & Naive & Baseline & \multicolumn{1}{c|}{0.043} & \multicolumn{1}{c|}{119.1} & 5.170 & \multicolumn{1}{c|}{1} & \multicolumn{1}{c|}{1} & \textbf{1} \\ \hline
icc & Naive & Manual v1 & \multicolumn{1}{c|}{0.284} & \multicolumn{1}{c|}{168.6} & 47.82 & \multicolumn{1}{c|}{6.537} & \multicolumn{1}{c|}{1.416} & \textbf{9.253} \\ \hline
icc & Naive & Manual v2 & \multicolumn{1}{c|}{0.284} & \multicolumn{1}{c|}{168.9} & 47.96 & \multicolumn{1}{c|}{6.545} & \multicolumn{1}{c|}{1.418} & \textbf{9.279} \\ \hline
\hline
clang & Reordered & Baseline & \multicolumn{1}{c|}{0.050} & \multicolumn{1}{c|}{119.5} & 5.960 & \multicolumn{1}{c|}{1} & \multicolumn{1}{c|}{1} & \textbf{1} \\ \hline
clang & Reordered & OpenMP & \multicolumn{1}{c|}{0.089} & \multicolumn{1}{c|}{135.7} & 12.12 & \multicolumn{1}{c|}{1.791} & \multicolumn{1}{c|}{1.135} & \textbf{2.034} \\ \hline
clang & Reordered & Manual v1 & \multicolumn{1}{c|}{0.089} & \multicolumn{1}{c|}{136.0} & 12.13 & \multicolumn{1}{c|}{1.788} & \multicolumn{1}{c|}{1.139} & \textbf{2.036} \\ \hline
clang & Reordered & Manual v2 & \multicolumn{1}{c|}{0.090} & \multicolumn{1}{c|}{135.8} & 12.18 & \multicolumn{1}{c|}{1.799} & \multicolumn{1}{c|}{1.136} & \textbf{2.044} \\ \hline
\hline
gcc & Reordered & Baseline & \multicolumn{1}{c|}{0.034} & \multicolumn{1}{c|}{131.8} & 4.520 & \multicolumn{1}{c|}{1} & \multicolumn{1}{c|}{1} & \textbf{1} \\ \hline
gcc & Reordered & Manual v1 & \multicolumn{1}{c|}{0.073} & \multicolumn{1}{c|}{143.7} & 10.50 & \multicolumn{1}{c|}{2.133} & \multicolumn{1}{c|}{1.090} & \textbf{2.325} \\ \hline
gcc & Reordered & Manual v2 & \multicolumn{1}{c|}{0.072} & \multicolumn{1}{c|}{144.4} & 10.40 & \multicolumn{1}{c|}{2.101} & \multicolumn{1}{c|}{1.095} & \textbf{2.302} \\ \hline
\hline
icc & Reordered & Baseline & \multicolumn{1}{c|}{0.043} & \multicolumn{1}{c|}{118.8} & 5.130 & \multicolumn{1}{c|}{1} & \multicolumn{1}{c|}{1} & \textbf{1} \\ \hline
icc & Reordered & Manual v1 & \multicolumn{1}{c|}{0.085} & \multicolumn{1}{c|}{139.3} & 11.81 & \multicolumn{1}{c|}{1.964} & \multicolumn{1}{c|}{1.172} & \textbf{2.303} \\ \hline
icc & Reordered & Manual v2 & \multicolumn{1}{c|}{0.039} & \multicolumn{1}{c|}{115.8} & 4.470 & \multicolumn{1}{c|}{0.894} & \multicolumn{1}{c|}{0.975} & \textbf{0.871} \\ \hline
\end{tabular}
\end{center}
\end{table}

\subsection{2D stencil}
\label{sec:results:stencil}

This section evaluates loop transformations on the 2D stencil described in section~\ref{sec:methodology2dstencil}. The input matrix consists of 2050x2050 single-precision floating-point numbers. This size is chosen for the same reason as the matrix multiplication input size; it is short enough to run many configurations in a reasonable time but long enough to not be majorly affected by external events such as the OS scheduling background processes.

\subsubsection{Loop tiling}

Like with matrix multiplication, we apply three versions of tiling: OpenMP, manual with bounds checks, and manual without.
We test sizes 1, 2, 4, 8, 16, 32 and 64, and proceed with size 8 which yields the best results.

\begin{figure}[h]
  \centering
  \includegraphics[width=\textwidth]{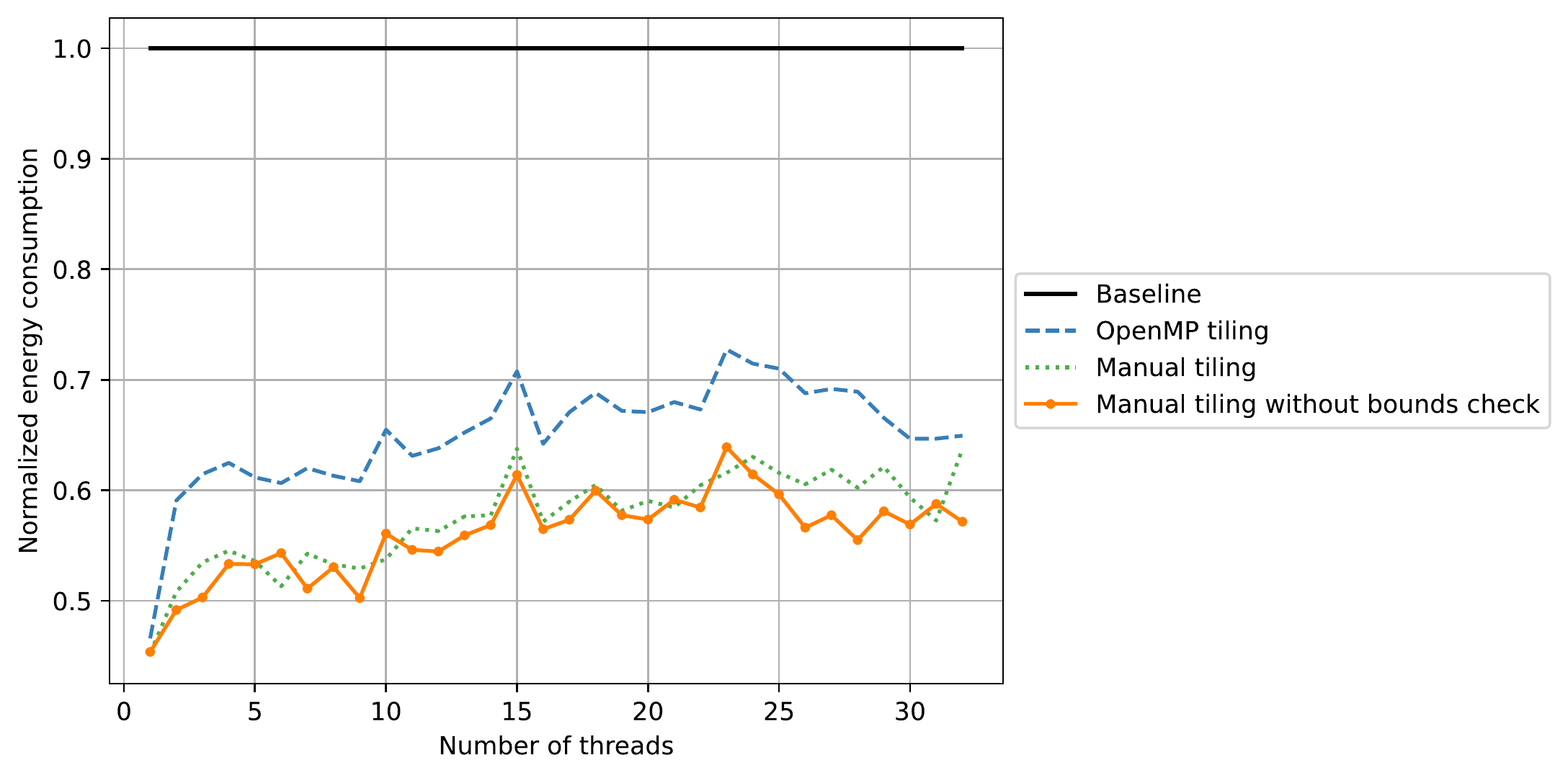}
  \caption{Energy consumption for the 2D stencil program with different versions of tiling, compiled with Clang.}
  \label{fig:stencil_clang_energy_tiling}
\end{figure}

Like before, we begin by focusing on the results for Clang. In figure~\ref{fig:stencil_clang_energy_tiling} we see the energy results relative to the baseline program without tiling. Like with matrix multiplication, we observe that OpenMP-induced tiling performs the worst, followed by both versions of manual tiling. Likewise, tiling becomes relatively less impactful as more threads are used. 
The corresponding plots for GCC and ICC are seen in the appendix, in figures~\ref{fig:stencil_gcc_energy_tiling} and~\ref{fig:stencil_icc_energy_tiling}. The results for GCC are very similar to those of Clang, with both versions of manual tiling performing about the same. The first version of tiling with ICC performs about the same as well.
The odd one out is ICC with the version of tiling where the loop iterations are assumed to be divisible by the tile size, which performs significantly better than anything else. We are unsure what causes this, but it is most likely due to allowing for other compiler optimisations.

To further analyze the impact of tiling, we use the tool \emph{perf} to perform an analysis on the cache behavior, the results of which can be seen in table~\ref{tab:stencil_tiling_cache_behavior}. Interestingly, the amount of misses actually increase with tiling, but the total amount of accesses decrease significantly. We are unsure what causes this.

\begin{table}[h]
\begin{center}
\caption{L1-D cache behavior for 16-threaded 2d stencil with and without tiling.}
\label{tab:stencil_tiling_cache_behavior}
\begin{tabular}{|l|l|l|l|}
\hline
                         & L1-D load misses & L1-D loads    & L1-D load miss rate \\ \hline
No tiling                & 17,861,341       & 1,754,756,062 & 1.02 \%            \\ \hline
OpenMP                   & 20,809,110       & 1,228,713,710 & 1.69 \%            \\ \hline
Manual v1                & 20,886,749       & 1,123,857,242 & 1.86 \%            \\ \hline
Manual v2                & 21,028,672       & 1,123,852,658 & 1.87 \%            \\ \hline
\end{tabular}
\end{center}
\end{table}

We conclude this section with table~\ref{tab:stencil_tiling_summary}, where we show the average result across all numbers of threads. In contrast to matrix multiplication, Clang outperforms GCC here. The table also confirms that manual tiling is better than that of OpenMP in this case.
We also note that ICC has the worst baseline results, but has also the best single result with the second version of unrolling.

\begin{table}[h]
\begin{center}
\caption{Summary of the 2d stencil tiling results across all numbers of threads. The relative numbers are relative to the baseline with no tiling for the same compiler.}
\label{tab:stencil_tiling_summary}
\begin{tabular}{|c|c|ccc|ccc|}
\hline
Compiler & Tiling & \multicolumn{3}{c|}{Absolute} & \multicolumn{3}{c|}{Relative} \\ \hline
 &  & \multicolumn{1}{c|}{T {[}s{]}} & \multicolumn{1}{c|}{P {[}W{]}} & E {[}J{]} & \multicolumn{1}{c|}{T} & \multicolumn{1}{c|}{P} & \textbf{E} \\ \hline
clang & Baseline & \multicolumn{1}{c|}{0.222} & \multicolumn{1}{c|}{126.1} & 27.98 & \multicolumn{1}{c|}{1} & \multicolumn{1}{c|}{1} & \textbf{1} \\ \hline
clang & OpenMP & \multicolumn{1}{c|}{0.143} & \multicolumn{1}{c|}{126.9} & 18.16 & \multicolumn{1}{c|}{0.645} & \multicolumn{1}{c|}{1.006} & \textbf{0.649} \\ \hline
clang & Manual v1 & \multicolumn{1}{c|}{0.128} & \multicolumn{1}{c|}{125.4} & 16.04 & \multicolumn{1}{c|}{0.576} & \multicolumn{1}{c|}{0.994} & \textbf{0.573} \\ \hline
clang & Manual v2 & \multicolumn{1}{c|}{0.124} & \multicolumn{1}{c|}{125.8} & 15.63 & \multicolumn{1}{c|}{0.560} & \multicolumn{1}{c|}{0.997} & \textbf{0.559} \\ \hline
gcc & Baseline & \multicolumn{1}{c|}{0.238} & \multicolumn{1}{c|}{123.9} & 29.45 & \multicolumn{1}{c|}{1} & \multicolumn{1}{c|}{1} & \textbf{1} \\ \hline
gcc & Manual v1 & \multicolumn{1}{c|}{0.144} & \multicolumn{1}{c|}{125.9} & 18.09 & \multicolumn{1}{c|}{0.604} & \multicolumn{1}{c|}{1.017} & \textbf{0.614} \\ \hline
gcc & Manual v2 & \multicolumn{1}{c|}{0.144} & \multicolumn{1}{c|}{125.9} & 18.08 & \multicolumn{1}{c|}{0.604} & \multicolumn{1}{c|}{1.017} & \textbf{0.614} \\ \hline
icc & Baseline & \multicolumn{1}{c|}{0.259} & \multicolumn{1}{c|}{132.2} & 34.29 & \multicolumn{1}{c|}{1} & \multicolumn{1}{c|}{1} & \textbf{1} \\ \hline
icc & Manual v1 & \multicolumn{1}{c|}{0.148} & \multicolumn{1}{c|}{134.0} & 19.89 & \multicolumn{1}{c|}{0.572} & \multicolumn{1}{c|}{1.014} & \textbf{0.580} \\ \hline
icc & Manual v2 & \multicolumn{1}{c|}{0.087} & \multicolumn{1}{c|}{125.3} & 10.94 & \multicolumn{1}{c|}{0.337} & \multicolumn{1}{c|}{0.948} & \textbf{0.319} \\ \hline
\end{tabular}
\end{center}
\end{table}

\subsubsection{Loop unrolling}

As with unrolling in matrix multiplication, we unroll the innermost loop with three variants: OpenMP-generated unrolling, manual unrolling that does assume that the total number of iterations is divisible by the unrolling factor, and manual unrolling that does not.
Similar to loop tiling, we determine the best unroll factor empirically by testing factors 1, 2, 4, 8, 16, 32 and 64. 32 turns out to be the best, which is the factor we use for the following presented data.

\begin{figure}[h]
  \centering
  \includegraphics[width=0.75\textwidth]{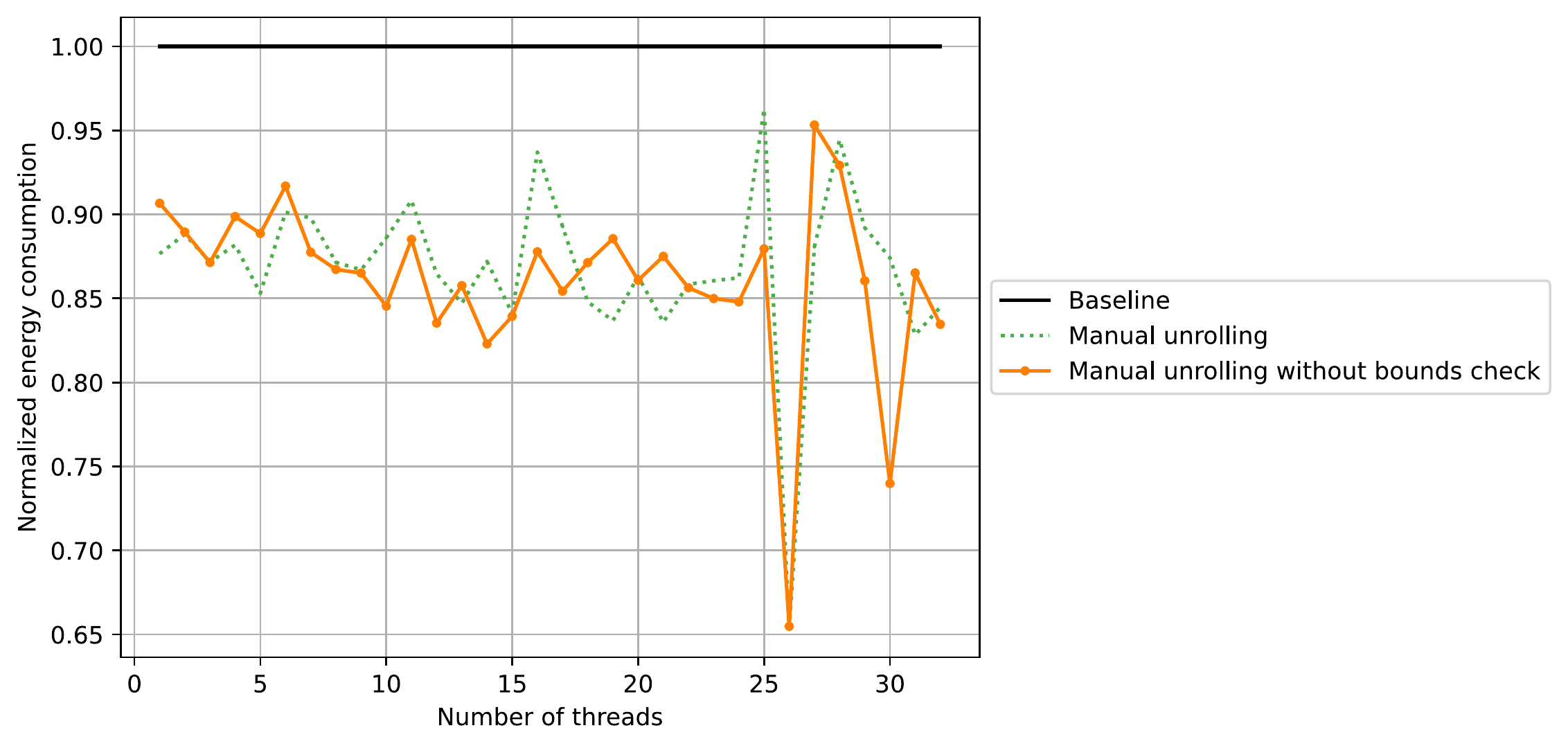}
  \caption{Relative energy consumption from applying unrolling to the 2D stencil program, compiled with GCC.}
  \label{fig:stencil_gcc_energy_unroll}
\end{figure}

\begin{figure}[h]
  \centering
  \includegraphics[width=0.75\textwidth]{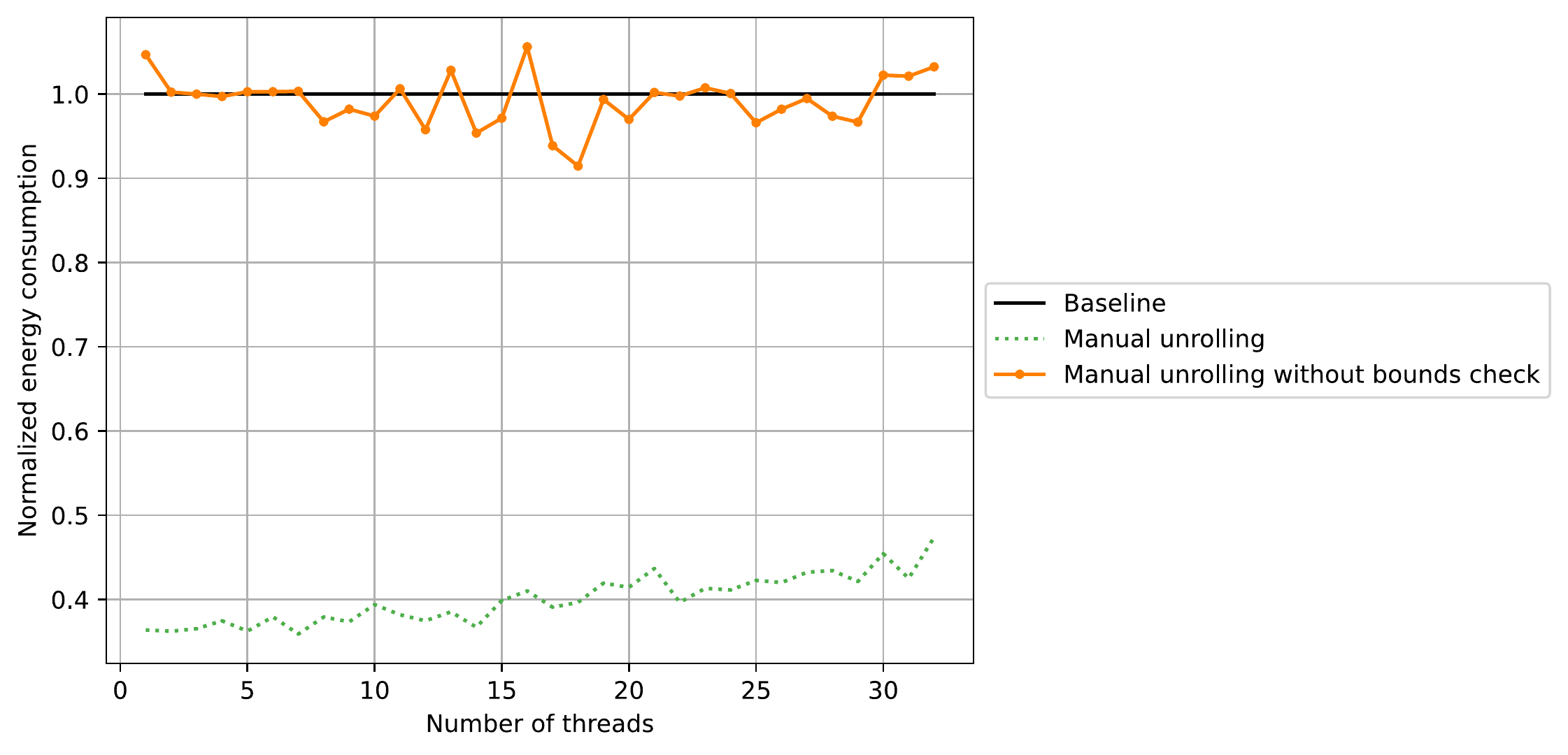}
  \caption{Relative energy consumption from applying unrolling to the 2D stencil program, compiled with ICC.}
  \label{fig:stencil_icc_energy_unroll}
\end{figure}

For Clang, there is almost no difference with either of the unrolling methods. The results can be seen in figure~\ref{fig:stencil_clang_energy_unroll} in the appendix. However, for both GCC and ICC there are some interesting results, as can be seen in figures~\ref{fig:stencil_gcc_energy_unroll} and~\ref{fig:stencil_icc_energy_unroll}. For GCC, both variants of unrolling perform around 10 \% better no matter the number of threads, with some variance around 25 threads.

For ICC, however, we see a huge improvement of 60 \% in one of the unrolling cases. This is a strange result, you would expect the unrolling variant without the bounds check to be faster, but it seems fairly equal to the baseline. Note that a similar result was observed for ICC with tiling: the second tiling variant performed much better than the first one.

Table~\ref{tab:stencil_unrolling_summary} summarises the results of unrolling the 2D stencil program. We can first of all confirm that unrolling has a small but positive effect on energy of about a percent, due to lower power consumption. GCC sees an improvement of 13-14 \% due to both shorter execution time and lower power. With this improvement, it becomes better than any version using Clang despite having a worse baseline.
ICC has the worst baseline, but the unrolling version without assuming the factor evenly divides the iterations provides a huge improvement that outclasses all other configurations. We are unsure what causes this.

\begin{table}[h]
\begin{center}
\caption{Summary of the 2D stencil unrolling results across all numbers of threads. The relative numbers are relative to the baseline, without unrolling for the same compiler.}
\label{tab:stencil_unrolling_summary}
\begin{tabular}{|c|c|ccc|ccc|}
\hline
Compiler & Unrolling & \multicolumn{3}{c|}{Absolute} & \multicolumn{3}{c|}{Relative} \\ \hline
 &  & \multicolumn{1}{c|}{T {[}s{]}} & \multicolumn{1}{c|}{P {[}W{]}} & E {[}J{]} & \multicolumn{1}{c|}{T} & \multicolumn{1}{c|}{P} & \textbf{E} \\ \hline
clang & Baseline & \multicolumn{1}{c|}{0.221} & \multicolumn{1}{c|}{120.7} & 26.66 & \multicolumn{1}{c|}{1} & \multicolumn{1}{c|}{1} & \textbf{1} \\ \hline
clang & OpenMP & \multicolumn{1}{c|}{0.221} & \multicolumn{1}{c|}{120.2} & 26.61 & \multicolumn{1}{c|}{1.002} & \multicolumn{1}{c|}{0.996} & \textbf{0.998} \\ \hline
clang & Manual v1 & \multicolumn{1}{c|}{0.222} & \multicolumn{1}{c|}{118.9} & 26.40 & \multicolumn{1}{c|}{1.005} & \multicolumn{1}{c|}{0.986} & \textbf{0.990} \\ \hline
clang & Manual v2 & \multicolumn{1}{c|}{0.222} & \multicolumn{1}{c|}{118.6} & 26.35 & \multicolumn{1}{c|}{1.005} & \multicolumn{1}{c|}{0.983} & \textbf{0.988} \\ \hline
gcc & Baseline & \multicolumn{1}{c|}{0.242} & \multicolumn{1}{c|}{123.7} & 29.98 & \multicolumn{1}{c|}{1} & \multicolumn{1}{c|}{1} & \textbf{1} \\ \hline
gcc & Manual v1 & \multicolumn{1}{c|}{0.214} & \multicolumn{1}{c|}{121.5} & 26.00 & \multicolumn{1}{c|}{0.883} & \multicolumn{1}{c|}{0.982} & \textbf{0.867} \\ \hline
gcc & Manual v2 & \multicolumn{1}{c|}{0.213} & \multicolumn{1}{c|}{121.1} & 25.77 & \multicolumn{1}{c|}{0.878} & \multicolumn{1}{c|}{0.979} & \textbf{0.860} \\ \hline
icc & Baseline & \multicolumn{1}{c|}{0.259} & \multicolumn{1}{c|}{132.2} & 34.29 & \multicolumn{1}{c|}{1} & \multicolumn{1}{c|}{1} & \textbf{1} \\ \hline
icc & Manual v1 & \multicolumn{1}{c|}{0.109} & \multicolumn{1}{c|}{125.1} & 13.67 & \multicolumn{1}{c|}{0.421} & \multicolumn{1}{c|}{0.947} & \textbf{0.399} \\ \hline
icc & Manual v2 & \multicolumn{1}{c|}{0.260} & \multicolumn{1}{c|}{130.5} & 33.99 & \multicolumn{1}{c|}{1.004} & \multicolumn{1}{c|}{0.987} & \textbf{0.991} \\ \hline
\end{tabular}
\end{center}
\end{table}

\subsection{Parallelism constructs microbenchmark}
\label{sec:results:parconstructs_microbench}

As described in section~\ref{sec:theory:parallelismconstructsmicrobenchmark}, the purpose of this microbenchmark is to experiment with the different ways of applying parallelism to a loop, and to measure and compare the overheads of doing so. It can also be used to evaluate the different waiting policies in this context as well.
The program is based around creating a number of tasks, with said tasks simply consisting of busy-waiting for a number of microseconds. The granularity of these is uniformly randomised from $0$ to $max\_task\_size$ to make the workload somewhat uneven. During our experiments, we use 20, 40, 60, 80, 100 and 200 $\mu$s for $max\_task\_size$.

We begin the analysis by focusing on the ways of using tasking, of which we use three methods: using a single thread to create all tasks, using multiple threads to create tasks, and finally by using the taskloop construct. The single-threaded variant is a fairly common construct, where one thread creates all tasks and other threads execute the tasks, similarly to listing~\ref{lst:task_while}.
The idea behind the multi-threaded variant is to have most threads create and execute mostly their own tasks, and then perform work-stealing to efficiently balance the workload. The final variant is to use the \emph{taskloop} clause in OpenMP, which is essentially syntactic sugar for a single-threaded task generation with optional parameters, which we do not consider.

Looking at the results for Clang only, table~\ref{tab:parconst_tasking} shows the results for some different values of task granularity. There are several interesting observations to make here. To start, we see that the multi-threaded variant actually consumes quite little power compared to both the other variants, which is not very intuitive. This makes it perform better in terms of energy for all levels of task granularity, even the larger ones where execution time is almost equal. We also see that active waiting outperforms passive waiting in most of the cases, except for the smallest granularity for the single-threaded variant. The reason for active waiting being better is probably due to the low amount of time threads spend idle in this microbenchmark.

\begin{table}[h]
\begin{center}
\caption{Characteristics of the tasking methods for the parallelism constructs microbenchmark using Clang. Measurements are reported as the geometric mean of said measurement for all numbers of threads.}
\label{tab:parconst_tasking}
\begin{tabular}{|c|c|ccc|ccc|ccc|}
\hline
            &         & \multicolumn{3}{c|}{Single-threaded}                                  & \multicolumn{3}{c|}{Multi-threaded}                                   & \multicolumn{3}{c|}{Taskloop}                                         \\ \hline
Granularity & Waiting & \multicolumn{1}{c|}{T {[}s{]}}    & \multicolumn{1}{c|}{P {[}W{]}}   & \textbf{E {[}J{]}}     & \multicolumn{1}{c|}{T {[}s{]}}    & \multicolumn{1}{c|}{P {[}W{]}}   & \textbf{E {[}J{]}}     & \multicolumn{1}{c|}{T {[}s{]}}    & \multicolumn{1}{c|}{P {[}W{]}}   & \textbf{E {[}J{]}}     \\ \hline
20          & Active  & \multicolumn{1}{c|}{0.31} & \multicolumn{1}{c|}{156} & \textbf{48.54} & \multicolumn{1}{c|}{0.1}  & \multicolumn{1}{c|}{120} & \textbf{12.44} & \multicolumn{1}{c|}{0.11} & \multicolumn{1}{c|}{132} & \textbf{14.09} \\ \hline
20          & Passive & \multicolumn{1}{c|}{0.27} & \multicolumn{1}{c|}{154} & \textbf{42.48} & \multicolumn{1}{c|}{0.12} & \multicolumn{1}{c|}{122} & \textbf{15.11} & \multicolumn{1}{c|}{0.11} & \multicolumn{1}{c|}{129} & \textbf{14.7}  \\ \hline
60          & Active  & \multicolumn{1}{c|}{0.12} & \multicolumn{1}{c|}{138} & \textbf{17.15} & \multicolumn{1}{c|}{0.1}  & \multicolumn{1}{c|}{114} & \textbf{11.69} & \multicolumn{1}{c|}{0.1}  & \multicolumn{1}{c|}{127} & \textbf{13.08} \\ \hline
60          & Passive & \multicolumn{1}{c|}{0.12} & \multicolumn{1}{c|}{132} & \textbf{15.84} & \multicolumn{1}{c|}{0.11} & \multicolumn{1}{c|}{113} & \textbf{12.44} & \multicolumn{1}{c|}{0.11} & \multicolumn{1}{c|}{122} & \textbf{13.15} \\ \hline
100         & Active  & \multicolumn{1}{c|}{0.1}  & \multicolumn{1}{c|}{126} & \textbf{13.11} & \multicolumn{1}{c|}{0.1}  & \multicolumn{1}{c|}{112} & \textbf{11.52} & \multicolumn{1}{c|}{0.1}  & \multicolumn{1}{c|}{124} & \textbf{12.82} \\ \hline
100         & Passive & \multicolumn{1}{c|}{0.11} & \multicolumn{1}{c|}{119} & \textbf{12.97} & \multicolumn{1}{c|}{0.11} & \multicolumn{1}{c|}{109} & \textbf{11.71} & \multicolumn{1}{c|}{0.11} & \multicolumn{1}{c|}{117} & \textbf{12.66} \\ \hline
200         & Active  & \multicolumn{1}{c|}{0.1}  & \multicolumn{1}{c|}{121} & \textbf{12.36} & \multicolumn{1}{c|}{0.1}  & \multicolumn{1}{c|}{114} & \textbf{11.89} & \multicolumn{1}{c|}{0.1}  & \multicolumn{1}{c|}{128} & \textbf{13.33} \\ \hline
200         & Passive & \multicolumn{1}{c|}{0.11} & \multicolumn{1}{c|}{112} & \textbf{11.82} & \multicolumn{1}{c|}{0.11} & \multicolumn{1}{c|}{106} & \textbf{11.53} & \multicolumn{1}{c|}{0.11} & \multicolumn{1}{c|}{121} & \textbf{12.98} \\ \hline
\end{tabular}
\end{center}
\end{table}

\begin{figure}[h]
    \centering
    \includegraphics[width=\textwidth]{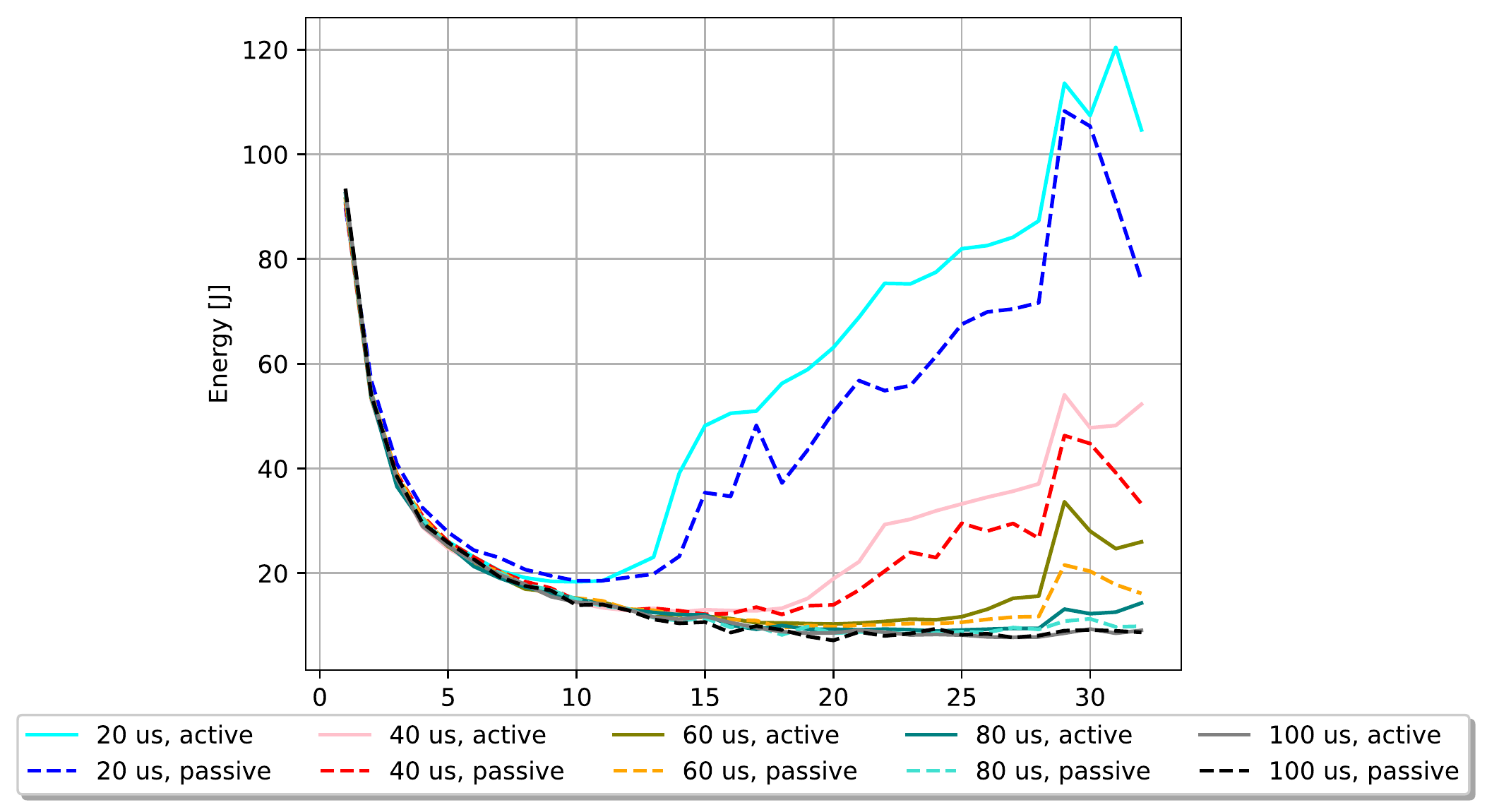}
    \caption{Energy consumption for the single-threaded task generation in the parallel constructs microbenchmark for Clang. The method scales well up to some number of threads, based on the granularity of tasks.}
    \label{fig:parconst_singlethreaded}
\end{figure}

Indeed, we see that using single-threaded task generation seems to be very inefficient for very small workloads. Figure~\ref{fig:parconst_singlethreaded} shows the energy consumption of this method for different levels of task granularity, with both active and passive waiting. This confirms that this method of task generation is inefficient for small tasks, and that it is worse the smaller tasks are. Furthermore, we see that passive waiting performs better than active waiting in this case, which is likely due to passive waiting being better when many threads are mostly inactive. It should be mention that this method scales fairly well up to about 8 threads even for the tiniest tasks, so it would likely be reasonably effective in any realistic setting where tasks are typically of a larger of a larger scale than tens of microseconds.

We select the multi-threaded task generation to be the best for now, and proceed to compare it to using a \emph{parallel for} loop. Since the size of each task is uneven, the time to compute each loop iteration can vary. This means that it makes sense to experiment with both the default \emph{static} scheduling, which statically divide the loop iterations among the team of threads, and the \emph{dynamic} scheduling, which dynamically schedules loop iterations among the threads. The optional chunk size parameter is unused, making it default to 1.

Table~\ref{tab:parconst_parfor_vs_tasking} shows an overview of the results for the \emph{parallel for} method using both static and dynamic scheduling, as well as the tasking method for reference. Once again there are several interesting observations to make. We see that the tasking construct consistently outperforms the loop method by a factor of 9.7 \% in terms of energy as long as the active waiting policy is used. If passive waiting is used, however, the parallel loop variant performs on 13.8 \% better in terms of energy due to the lower power consumption. Conversely, the execution time is actually around 15 \% higher. Furthermore, if dynamic scheduling is used, the parallel loop variant consistently outperforms tasking regardless of the waiting policy. With active waiting, the power consumption is roughly the same but the execution time is better, while with passive waiting both execution time and power consumption is better for the parallel loop than the tasking method. Comparing the best configurations, those being tasking with active waiting and parallel loops with passive waiting and dynamic scheduling, for each task granularity gives 9.7 \% better energy consumption in favor of parallel loops.

\begin{table}[h]
\begin{center}
\caption{Results for the parallel for versus tasking methods, with Clang. Measurements are reported as the geometric mean of said measurement for all numbers of threads.}
\label{tab:parconst_parfor_vs_tasking}
\begin{tabular}{|c|c|ccc|ccc|ccc|}
\hline
            &         & \multicolumn{3}{c|}{Tasking}                                          & \multicolumn{3}{c|}{ParFor, static}                                   & \multicolumn{3}{c|}{ParFor, dynamic}                                  \\ \hline
Granularity & Waiting & \multicolumn{1}{c|}{T {[}s{]}}    & \multicolumn{1}{c|}{P {[}W{]}}   & \textbf{E {[}J{]}}     & \multicolumn{1}{c|}{T {[}s{]}}    & \multicolumn{1}{c|}{P {[}W{]}}   & \textbf{E {[}J{]}}     & \multicolumn{1}{c|}{T {[}s{]}}    & \multicolumn{1}{c|}{P {[}W{]}}   & \textbf{E {[}J{]}}     \\ \hline
20          & Active  & \multicolumn{1}{c|}{0.1}  & \multicolumn{1}{c|}{120} & \textbf{12.44} & \multicolumn{1}{c|}{0.11} & \multicolumn{1}{c|}{123} & 12.97          & \multicolumn{1}{c|}{0.1}  & \multicolumn{1}{c|}{121} & 11.79          \\ \hline
20          & Passive & \multicolumn{1}{c|}{0.12} & \multicolumn{1}{c|}{122} & 15.11          & \multicolumn{1}{c|}{0.13} & \multicolumn{1}{c|}{91}  & \textbf{11.55} & \multicolumn{1}{c|}{0.11} & \multicolumn{1}{c|}{104} & \textbf{11.55} \\ \hline
60          & Active  & \multicolumn{1}{c|}{0.1}  & \multicolumn{1}{c|}{114} & \textbf{11.69} & \multicolumn{1}{c|}{0.11} & \multicolumn{1}{c|}{125} & 13.27          & \multicolumn{1}{c|}{0.1}  & \multicolumn{1}{c|}{115} & 11.32          \\ \hline
60          & Passive & \multicolumn{1}{c|}{0.11} & \multicolumn{1}{c|}{113} & 12.44          & \multicolumn{1}{c|}{0.12} & \multicolumn{1}{c|}{91}  & \textbf{10.58} & \multicolumn{1}{c|}{0.1}  & \multicolumn{1}{c|}{101} & \textbf{10.51} \\ \hline
100         & Active  & \multicolumn{1}{c|}{0.1}  & \multicolumn{1}{c|}{112} & \textbf{11.52} & \multicolumn{1}{c|}{0.11} & \multicolumn{1}{c|}{124} & 13.4           & \multicolumn{1}{c|}{0.1}  & \multicolumn{1}{c|}{119} & 11.81          \\ \hline
100         & Passive & \multicolumn{1}{c|}{0.11} & \multicolumn{1}{c|}{109} & 11.71          & \multicolumn{1}{c|}{0.12} & \multicolumn{1}{c|}{91}  & \textbf{10.72} & \multicolumn{1}{c|}{0.1}  & \multicolumn{1}{c|}{100} & \textbf{10.4}  \\ \hline
200         & Active  & \multicolumn{1}{c|}{0.1}  & \multicolumn{1}{c|}{114} & 11.89          & \multicolumn{1}{c|}{0.11} & \multicolumn{1}{c|}{122} & 13.27          & \multicolumn{1}{c|}{0.1}  & \multicolumn{1}{c|}{118} & 11.9           \\ \hline
200         & Passive & \multicolumn{1}{c|}{0.11} & \multicolumn{1}{c|}{106} & \textbf{11.53} & \multicolumn{1}{c|}{0.12} & \multicolumn{1}{c|}{88}  & \textbf{10.46} & \multicolumn{1}{c|}{0.1}  & \multicolumn{1}{c|}{100} & \textbf{10.41} \\ \hline
\end{tabular}
\end{center}
\end{table}

The corresponding results for ICC are found in the appendix, in tables~\ref{tab:parconst_tasking_icc} and~\ref{tab:parconst_parfor_vs_tasking_icc}, due to their similarity to Clang.
To summarize for ICC, we again see that single-threaded task generation with small task granularity performs very poorly, but that this quickly becomes less of a concern for tasks around 200 $\mu$s. Comparing tasking versus parallel looping, we see that tasking consistently performs the worst provided that the passive waiting policy is used. We conclude that for ICC, an active waiting policy should be used when using tasking methods. However, when there is a choice between tasking and parallel loops, the latter should be used and the waiting policy should be passive.

\begin{figure}[h]
    \centering
    \includegraphics[width=\textwidth]{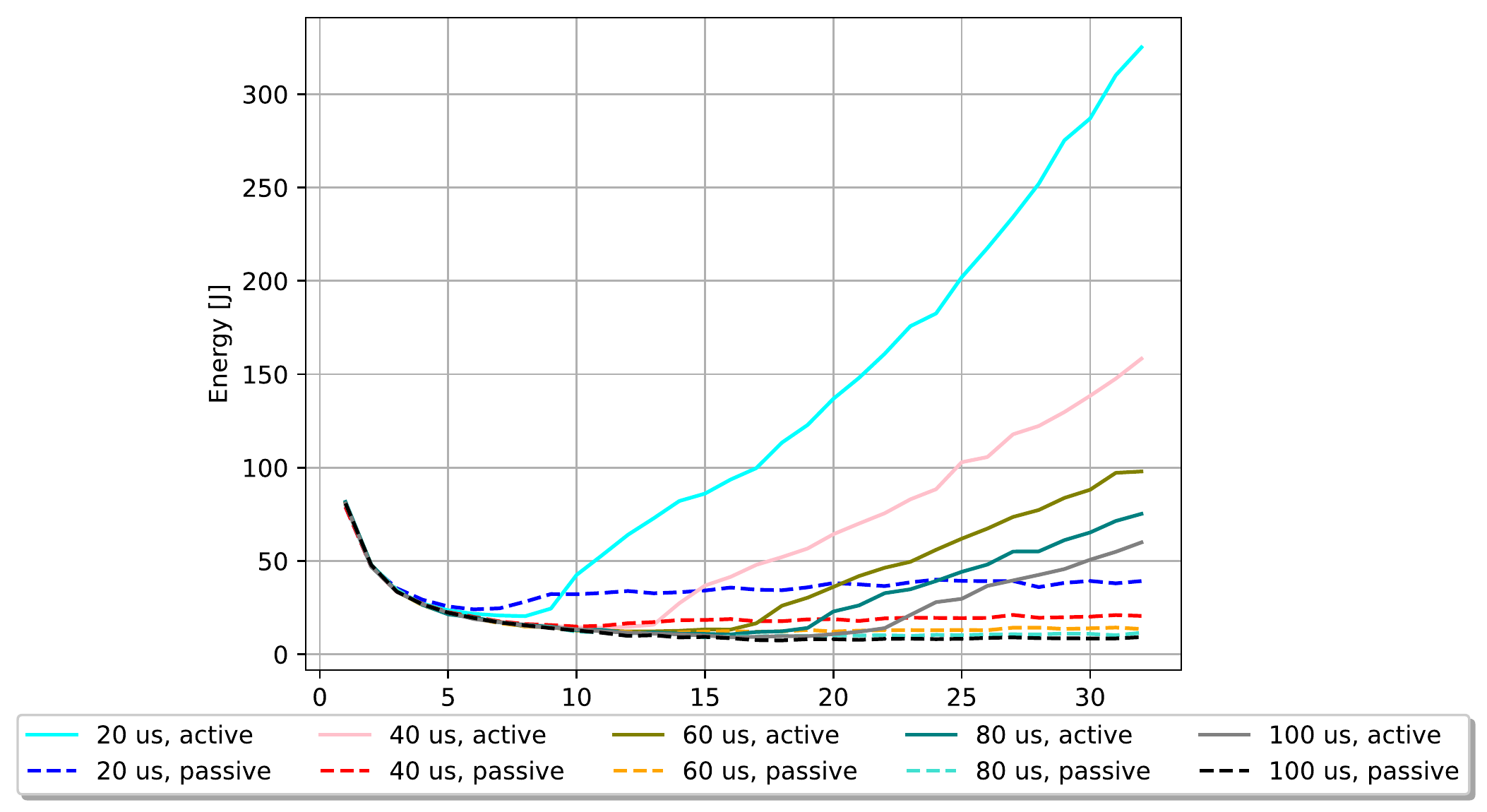}
    \caption{Energy consumption for the single-threaded task generation in the parallel constructs microbenchmark for GCC. The method scales well up to some number of threads, based on the granularity of tasks.}
    \label{fig:parconst_singlethreaded_gcc}
\end{figure}

The results for GCC are seen in tables~\ref{tab:parconst_tasking_gcc} and~\ref{tab:parconst_parfor_vs_tasking_gcc}. These results show some key differences compared to Clang and ICC. To start, we see that passive waiting consistently performs better than active waiting. Furthermore, comparing the tasking variants, we see that the taskloop is the best. This is therefore what is compared to the parallel for loop variants. In that table, we observe that the parallel for with static scheduling and the taskloop both contend for being the best, with the dynamically scheduled parallel for consistently performing the worst in terms of energy. It does however perform the best in terms of execution time. We conclude that taskloops and parallel for loops are about equal for GCC, but that manually creating tasks performs badly, be it single-threaded or multi-threaded.
Figure~\ref{fig:parconst_singlethreaded_gcc} shows the energy results for the single-threaded version. Compared to the corresponding result for Clang, the active waiting performs significantly worse than passive. It also seems to scale linearly with the amount of threads, while the plots for passive waiting flattens and becomes constant around 8 threads. This further suggests that passive waiting should be preferred for GCC.

\begin{table}[h]
\begin{center}
\caption{Characteristics of the tasking methods for the parallelism constructs microbenchmark using GCC.}
\label{tab:parconst_tasking_gcc}
\begin{tabular}{|c|c|ccc|ccc|ccc|}
\hline
            &         & \multicolumn{3}{c|}{Single-threaded}                                  & \multicolumn{3}{c|}{Multi-threaded}                                   & \multicolumn{3}{c|}{Taskloop}                                        \\ \hline
Granularity & Waiting & \multicolumn{1}{c|}{T {[}s{]}}    & \multicolumn{1}{c|}{P {[}W{]}}   & E {[}J{]}     & \multicolumn{1}{c|}{T {[}s{]}}    & \multicolumn{1}{c|}{P {[}W{]}}   & E {[}J{]}     & \multicolumn{1}{c|}{T {[}s{]}}    & \multicolumn{1}{c|}{P {[}W{]}}   & E {[}J{]}             \\ \hline
20          & Active  & \multicolumn{1}{c|}{0.65} & \multicolumn{1}{c|}{138} & 90.27          & \multicolumn{1}{c|}{0.3}  & \multicolumn{1}{c|}{138} & 41.26          & \multicolumn{1}{c|}{0.09} & \multicolumn{1}{c|}{126} & 11.7          \\ \hline
20          & Passive & \multicolumn{1}{c|}{0.61} & \multicolumn{1}{c|}{58}  & \textbf{35.54} & \multicolumn{1}{c|}{0.22} & \multicolumn{1}{c|}{135} & \textbf{29.52} & \multicolumn{1}{c|}{0.12} & \multicolumn{1}{c|}{79}  & \textbf{9.33} \\ \hline
60          & Active  & \multicolumn{1}{c|}{0.23} & \multicolumn{1}{c|}{137} & 31.85          & \multicolumn{1}{c|}{0.14} & \multicolumn{1}{c|}{135} & 18.98          & \multicolumn{1}{c|}{0.09} & \multicolumn{1}{c|}{127} & 11.73         \\ \hline
60          & Passive & \multicolumn{1}{c|}{0.23} & \multicolumn{1}{c|}{68}  & \textbf{15.6}  & \multicolumn{1}{c|}{0.12} & \multicolumn{1}{c|}{127} & \textbf{15.37} & \multicolumn{1}{c|}{0.1}  & \multicolumn{1}{c|}{87}  & \textbf{8.62} \\ \hline
100         & Active  & \multicolumn{1}{c|}{0.15} & \multicolumn{1}{c|}{135} & 20.78          & \multicolumn{1}{c|}{0.12} & \multicolumn{1}{c|}{132} & 15.53          & \multicolumn{1}{c|}{0.09} & \multicolumn{1}{c|}{126} & 11.54         \\ \hline
100         & Passive & \multicolumn{1}{c|}{0.15} & \multicolumn{1}{c|}{80}  & \textbf{11.96} & \multicolumn{1}{c|}{0.11} & \multicolumn{1}{c|}{125} & \textbf{13.32} & \multicolumn{1}{c|}{0.1}  & \multicolumn{1}{c|}{88}  & \textbf{8.82} \\ \hline
200         & Active  & \multicolumn{1}{c|}{0.1}  & \multicolumn{1}{c|}{130} & 12.74          & \multicolumn{1}{c|}{0.1}  & \multicolumn{1}{c|}{128} & 13.15          & \multicolumn{1}{c|}{0.09} & \multicolumn{1}{c|}{124} & 11.66         \\ \hline
200         & Passive & \multicolumn{1}{c|}{0.1}  & \multicolumn{1}{c|}{107} & \textbf{10.2}  & \multicolumn{1}{c|}{0.1}  & \multicolumn{1}{c|}{119} & \textbf{11.63} & \multicolumn{1}{c|}{0.1}  & \multicolumn{1}{c|}{89}  & \textbf{8.95} \\ \hline
\end{tabular}
\end{center}
\end{table}

\begin{table}[h]
\begin{center}
\caption{Results for the parallel for versus tasking methods, with GCC. In contrast to Clang and ICC, where the multi-threaded variant is the best tasking method, here we use the taskloop variant to compare with the parallel for variants.}
\label{tab:parconst_parfor_vs_tasking_gcc}
\begin{tabular}{|c|c|ccc|ccc|ccc|}
\hline
            &         & \multicolumn{3}{c|}{Tasking}                                         & \multicolumn{3}{c|}{ParFor, static}                                  & \multicolumn{3}{c|}{ParFor, dynamic}                                 \\ \hline
Granularity & Waiting & \multicolumn{1}{c|}{T {[}s{]}}    & \multicolumn{1}{c|}{P {[}W{]}}   & E {[}J{]}     & \multicolumn{1}{c|}{T {[}s{]}}    & \multicolumn{1}{c|}{P {[}W{]}}   & E {[}J{]}     & \multicolumn{1}{c|}{T {[}s{]}}    & \multicolumn{1}{c|}{P {[}W{]}}   & E {[}J{]}             \\ \hline
20          & Active  & \multicolumn{1}{c|}{0.09} & \multicolumn{1}{c|}{126} & 11.7          & \multicolumn{1}{c|}{0.09} & \multicolumn{1}{c|}{120} & 10.55         & \multicolumn{1}{c|}{0.08} & \multicolumn{1}{c|}{117} & 9.38          \\ \hline
20          & Passive & \multicolumn{1}{c|}{0.12} & \multicolumn{1}{c|}{79}  & \textbf{9.33} & \multicolumn{1}{c|}{0.12} & \multicolumn{1}{c|}{79}  & \textbf{9.9}  & \multicolumn{1}{c|}{0.09} & \multicolumn{1}{c|}{106} & \textbf{9.9}  \\ \hline
60          & Active  & \multicolumn{1}{c|}{0.09} & \multicolumn{1}{c|}{127} & 11.73         & \multicolumn{1}{c|}{0.09} & \multicolumn{1}{c|}{125} & 11.21         & \multicolumn{1}{c|}{0.08} & \multicolumn{1}{c|}{116} & 9.28          \\ \hline
60          & Passive & \multicolumn{1}{c|}{0.1}  & \multicolumn{1}{c|}{87}  & \textbf{8.62} & \multicolumn{1}{c|}{0.1}  & \multicolumn{1}{c|}{89}  & \textbf{8.89} & \multicolumn{1}{c|}{0.08} & \multicolumn{1}{c|}{108} & \textbf{9.07} \\ \hline
100         & Active  & \multicolumn{1}{c|}{0.09} & \multicolumn{1}{c|}{126} & 11.54         & \multicolumn{1}{c|}{0.09} & \multicolumn{1}{c|}{123} & 11.15         & \multicolumn{1}{c|}{0.08} & \multicolumn{1}{c|}{120} & 9.91          \\ \hline
100         & Passive & \multicolumn{1}{c|}{0.1}  & \multicolumn{1}{c|}{88}  & \textbf{8.82} & \multicolumn{1}{c|}{0.1}  & \multicolumn{1}{c|}{85}  & \textbf{8.76} & \multicolumn{1}{c|}{0.08} & \multicolumn{1}{c|}{113} & \textbf{9.59} \\ \hline
200         & Active  & \multicolumn{1}{c|}{0.09} & \multicolumn{1}{c|}{124} & 11.66         & \multicolumn{1}{c|}{0.09} & \multicolumn{1}{c|}{123} & 11.34         & \multicolumn{1}{c|}{0.08} & \multicolumn{1}{c|}{121} & 9.99          \\ \hline
200         & Passive & \multicolumn{1}{c|}{0.1}  & \multicolumn{1}{c|}{89}  & \textbf{8.95} & \multicolumn{1}{c|}{0.1}  & \multicolumn{1}{c|}{89}  & \textbf{9.23} & \multicolumn{1}{c|}{0.09} & \multicolumn{1}{c|}{110} & \textbf{9.43} \\ \hline
\end{tabular}
\end{center}
\end{table}

\subsection{Inactivity microbenchmark}
\label{sec:results:inactivity_microbench}

As mentioned previously, the purpose of this microbenchmark is to evaluate the effect of waiting policies of inactive threads.
This is done by creating a parallel section where only one thread performs work and all other threads wait. The work consists of simply busy-waiting for a number of microseconds. By varying how long this waiting time is we can change the granularity of the benchmark as a whole.
In our experiments we vary it between 10 and 100 $\mu$s, and set the number of iterations to $\frac{1000000}{waiting\ time}$. This means that the optimal execution time is about 1 second.
Upon analysis we notice that default and active waiting perform nearly identical for this microbenchmark regardless of compiler. Therefore, we present results for active and passive waiting only.

\begin{table}[h]
\begin{center}
\caption{Overview of the inactivity microbenchmark results across all numbers of threads. The smallest and largest task granularity of 10 and 100 are included, as well as the granularity where passive waiting becomes better than active for each compiler. These energy values are marked in bold.}
\label{tab:inactivity_overview}
\begin{tabular}{|c|c|ccc|ccc|}
\hline
Compiler & Granularity {[}us{]} & \multicolumn{3}{c|}{Active}                                                  & \multicolumn{3}{c|}{Passive}                                                 \\ \hline
         &                      & \multicolumn{1}{c|}{T{[}s{]}} & \multicolumn{1}{c|}{P{[}W{]}} & E{[}J{]}     & \multicolumn{1}{c|}{T{[}s{]}} & \multicolumn{1}{c|}{P{[}W{]}} & E{[}J{]}     \\ \hline
\hline
Clang    & 10                   & \multicolumn{1}{c|}{1.11}     & \multicolumn{1}{c|}{300}      & 334          & \multicolumn{1}{c|}{4.42}     & \multicolumn{1}{c|}{199}      & 878          \\ \hline
Clang    & 30                   & \multicolumn{1}{c|}{1.05}     & \multicolumn{1}{c|}{303}      & \textbf{317} & \multicolumn{1}{c|}{1.72}     & \multicolumn{1}{c|}{209}      & \textbf{360} \\ \hline
Clang    & 40                   & \multicolumn{1}{c|}{1.04}     & \multicolumn{1}{c|}{303}      & \textbf{314} & \multicolumn{1}{c|}{1.49}     & \multicolumn{1}{c|}{206}      & \textbf{306} \\ \hline
Clang    & 100                  & \multicolumn{1}{c|}{1.02}     & \multicolumn{1}{c|}{304}      & 311          & \multicolumn{1}{c|}{1.19}     & \multicolumn{1}{c|}{199}      & 235          \\ \hline
\hline
GCC      & 10                   & \multicolumn{1}{c|}{1.06}     & \multicolumn{1}{c|}{295}      & 312          & \multicolumn{1}{c|}{2.89}     & \multicolumn{1}{c|}{206}      & 595          \\ \hline
GCC      & 40                   & \multicolumn{1}{c|}{1.02}     & \multicolumn{1}{c|}{293}      & \textbf{298} & \multicolumn{1}{c|}{1.58}     & \multicolumn{1}{c|}{202}      & \textbf{320} \\ \hline
GCC      & 60                   & \multicolumn{1}{c|}{1.01}     & \multicolumn{1}{c|}{292}      & \textbf{295} & \multicolumn{1}{c|}{1.4}      & \multicolumn{1}{c|}{201}      & \textbf{282} \\ \hline
GCC      & 70                   & \multicolumn{1}{c|}{1.01}     & \multicolumn{1}{c|}{292}      & 295          & \multicolumn{1}{c|}{1.35}     & \multicolumn{1}{c|}{200}      & 271          \\ \hline
GCC      & 100                  & \multicolumn{1}{c|}{1.01}     & \multicolumn{1}{c|}{292}      & 294          & \multicolumn{1}{c|}{1.25}     & \multicolumn{1}{c|}{200}      & 250          \\ \hline
\hline
ICC      & 10                   & \multicolumn{1}{c|}{1.08}     & \multicolumn{1}{c|}{296}      & 321          & \multicolumn{1}{c|}{4.34}     & \multicolumn{1}{c|}{204}      & 884          \\ \hline
ICC      & 40                   & \multicolumn{1}{c|}{1.03}     & \multicolumn{1}{c|}{295}      & \textbf{303} & \multicolumn{1}{c|}{1.49}     & \multicolumn{1}{c|}{206}      & \textbf{307} \\ \hline
ICC      & 50                   & \multicolumn{1}{c|}{1.02}     & \multicolumn{1}{c|}{295}      & \textbf{302} & \multicolumn{1}{c|}{1.37}     & \multicolumn{1}{c|}{205}      & \textbf{281} \\ \hline
ICC      & 100                  & \multicolumn{1}{c|}{1.02}     & \multicolumn{1}{c|}{295}      & 300          & \multicolumn{1}{c|}{1.18}     & \multicolumn{1}{c|}{201}      & 237          \\ \hline
\end{tabular}
\end{center}
\end{table}

Some prominent results are seen in table~\ref{tab:inactivity_overview}. We see that for the lowest granularity, active waiting outperforms passive in terms of both execution time and energy. For the highest granularity however, passive waiting outperforms active due to the lower consumption. The table also includes the cutoff points, where passive waiting starts to outperform active in terms of energy. For Clang, we see that it is between 30-40 $\mu$s, while it is around 40-70 $\mu$s for GCC and 40-50 $\mu$s for ICC.

\begin{figure}[h]
    \centering
    \includegraphics[width=0.7\textwidth]{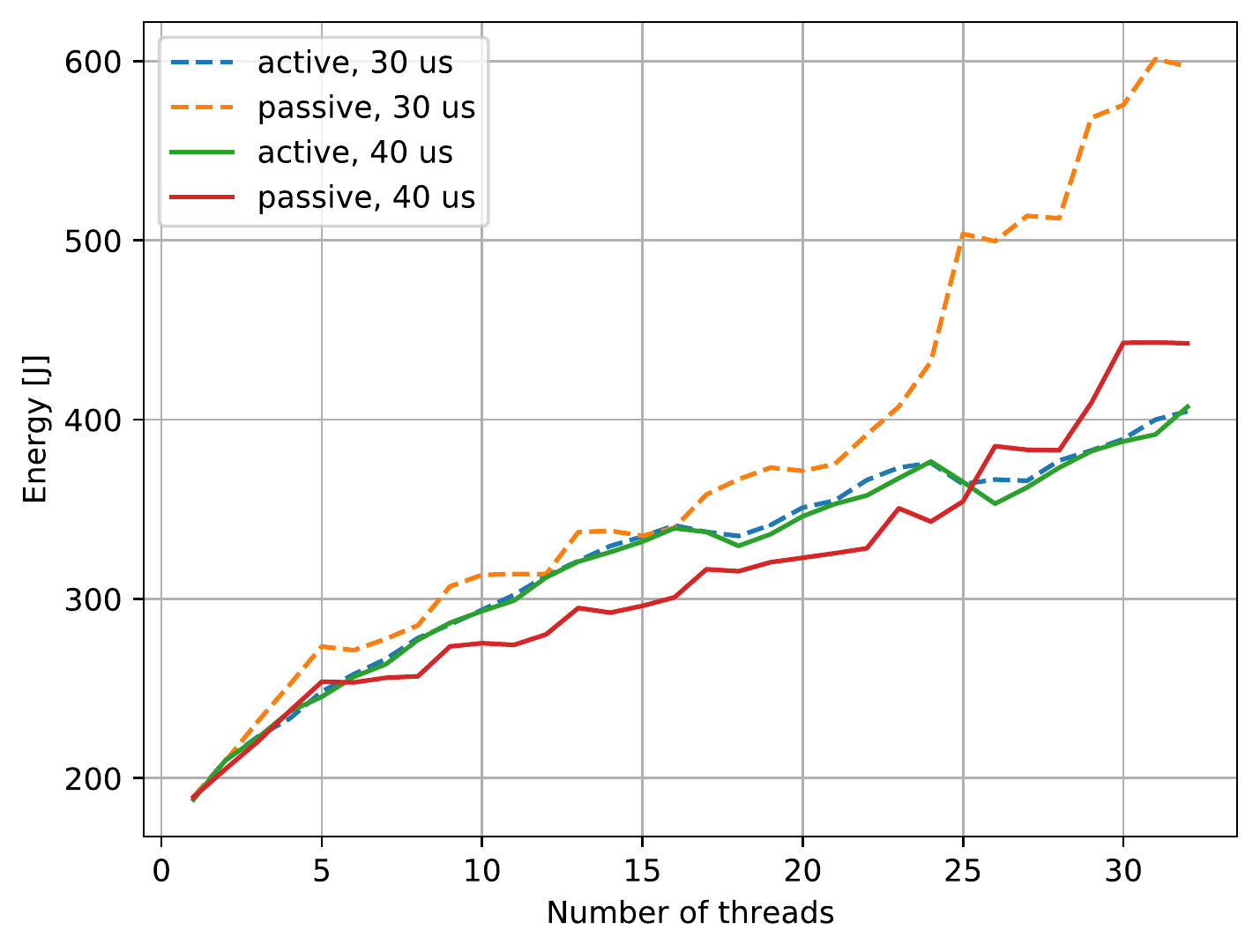}
    \caption{Energy consumption at the granularity where passive waiting overtakes active waiting for Clang.}
    \label{fig:inactivity_clang_energy}
\end{figure}

\begin{figure}[h]
    \centering
    \includegraphics[width=0.7\textwidth]{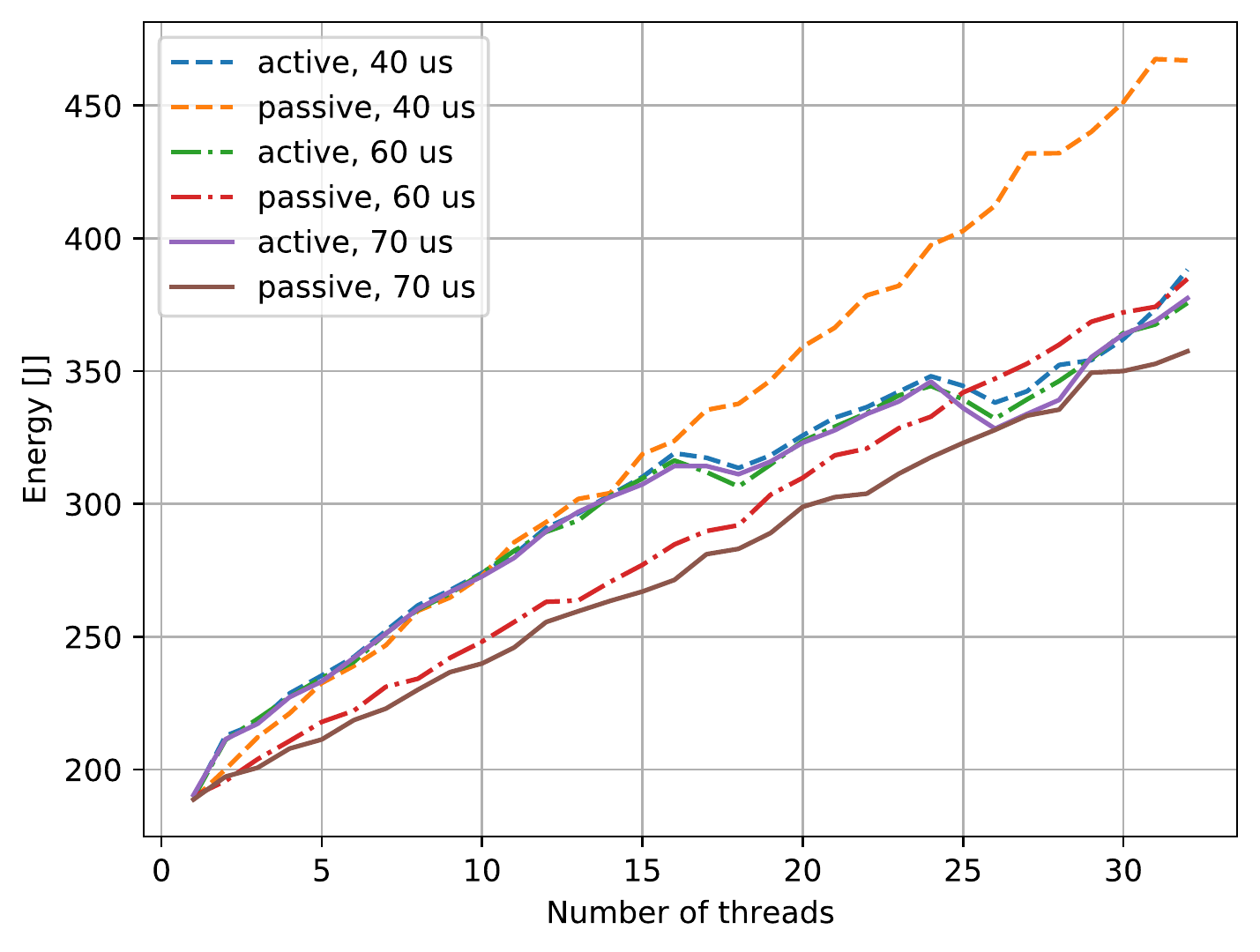}
    \caption{Energy consumption at the granularity where passive waiting overtakes active waiting for GCC.}
    \label{fig:inactivity_gcc_energy}
\end{figure}

In figure~\ref{fig:inactivity_clang_energy} we see the energy graph for Clang for these granularities. We see that for 30 $\mu$s, the active waiting consistently has lower energy consumption than the passive. Meanwhile, at 40 $\mu$s, passive waiting has a lower energy consumption for all but the highest number of threads. The corresponding plot for GCC can be seen in figure~\ref{fig:inactivity_gcc_energy} where we see a similar behavior but the cutoff point is at a higher granularity, as seen in the table before. For GCC, we see that the energy increase with passive waiting is a lot more linear than for Clang and ICC. Up to about 16 threads, the cutoff is about 40 $\mu$s, but for the highest numbers of threads it is around 60-70 $\mu$s. The plot for ICC can be found in figure~\ref{fig:inactivity_icc_energy} of the appendix due to its similarity to Clang.

To conclude, the results show that using the active waiting policy is better than the passive for most task sizes, but especially small ones. The cutoff point where passive scheduling is better is higher for GCC than for Clang and ICC. With very small task granularity passive waiting becomes significantly worse, very likely due to the runtime overheads of thread scheduling and descheduling. In these cases, while neither compiler performs well, GCC is better than both Clang and ICC. As a final note, it is worth mentioning that this microbenchmark favors passive waiting due to more threads giving no performance benefits at all, which is not reasonable for real programs. This means that active waiting will likely perform even better in practice due to reducing the execution time.

\subsection{Loop unrolling microbenchmark}
\label{sec:results:loop_unrolling_microbench}

This section evaluates the effect loop unrolling has on the four simple programs described in section \ref{sec:background:unrolling_microbenchmark}. All results were obtained using \emph{Clang}, since it is the only compiler that supports the new OpenMP \texttt{unroll} directive, without any parallelisation and with the 03 optimisation level enabled. An unrolling factor of eight was chosen for both the OpenMP and manual unrolling methods.

Measures were taken so the execution time for each program would be roughly five seconds for the default implementation. This was done by measuring the execution time and energy after running the program 4096 times with some appropriate value for the \texttt{len} variable.

\begin{figure}[h]
    \centering
    \includegraphics[width=0.7\textwidth]{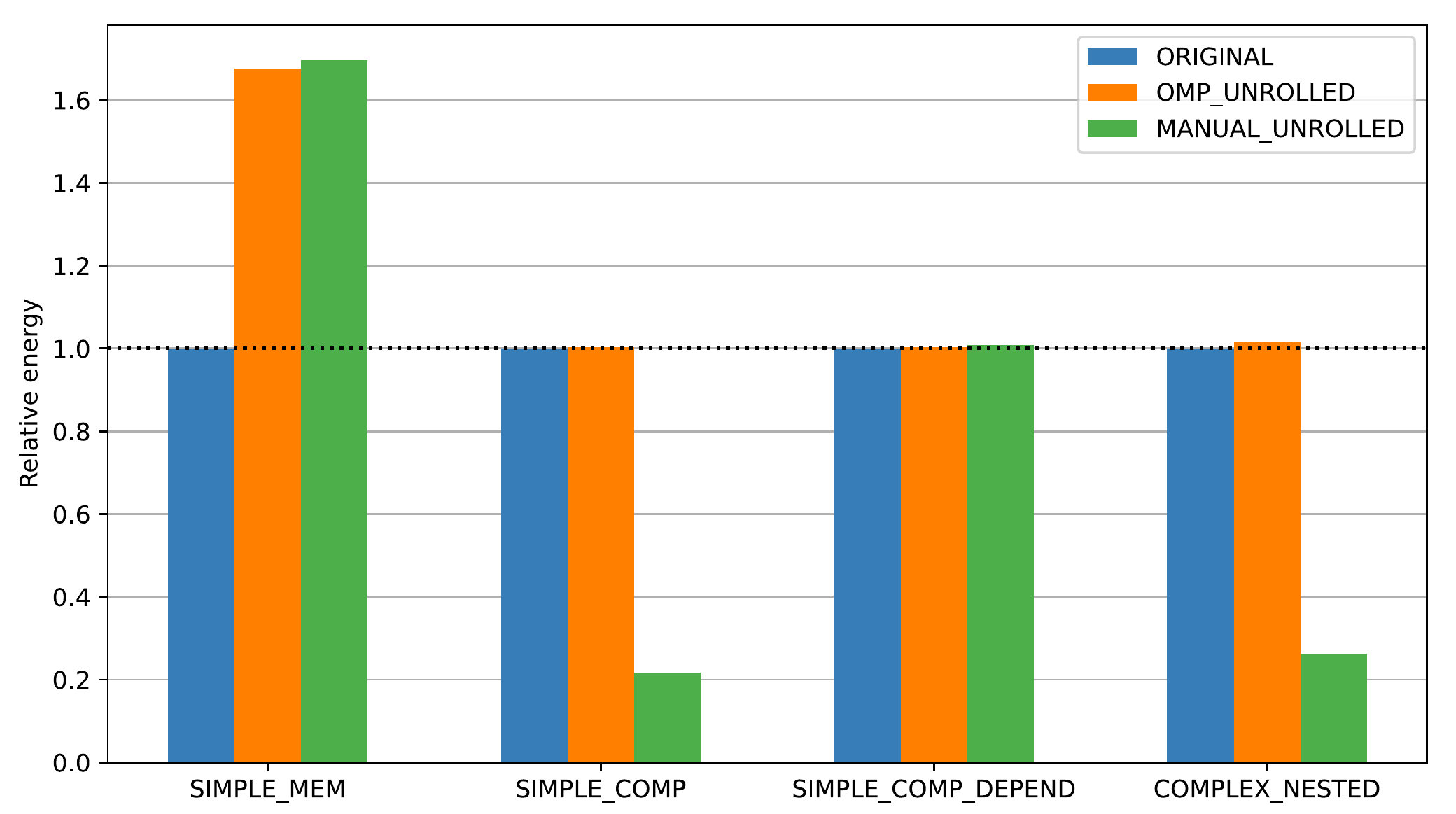}
    \caption{Relative energy used for the different unrolling implementations compared to no explicit unrolling for the unrolling microbenchmark programs.}
    \label{fig:unrolling_ress}
\end{figure}

The results for the tests can be seen in figure \ref{fig:unrolling_ress}. The figure shows the relative energy used by each implementation compared to the default implementation without any explicit unrolling for the four programs. What is immediate apparent from the results is that unrolling seems to either have no impact at all, or a very significant one, when looking at these four programs. The results from each program will be discussed in further detail below.

The \texttt{SIMPLE\_MEM} program got significantly worse for both unrolling implementations, using about 70 \% more energy. While all three implementations utilise vector instructions to do the calculations and memory assignment, the original implementation, which was automatically unrolled by a factor of two by the compiler, seem to make use of the instructions more efficiently. Interestingly, when running the OpenMP unrolling version again and matching the generated unrolling factor of two, it still performed worse than the original, but now by only 17 \%. 

The \texttt{SIMPLE\_COMP} program didn't see any performance change for the OpenMP unrolling, but saw a significant improvement for the manual unrolling implementation, using only 21 \% of the energy compared to the original implementation. The reason for this big difference is because in the manually unrolled version the calculations has been separated into eight independent parts, making it so they can be run in parallel using SIMD instructions. When naively unrolling the program manually without this optimisation, it performs on par with the original implementation.

The \texttt{SIMPLE\_COMP\_DEPEND} program has no meaningful difference in performance between the implementations. None of the implementations make us of vector instructions, unlike the two previous programs discussed, and is probably the reason why there has been no dramatic performance difference.

The implementations of the \texttt{COMPLEX\_NESTED} program performs very similarly to the \texttt{SIMPLE\_COMP} program, with OpenMP unrolling having no difference and manual unrolling significantly reducing energy, down to only 25 \%. The reason here is similarly also because more efficient usage of vector instructions. When doing the \emph{loop jamming} optimisation, the work in the nested loops gets clumped together into one loop, which the compiler then can parallelise with vector instructions.

In summary, unrolling for these very simple programs tends to either have an insignificant impact on the overall performance, or alternatively a huge impact that can be either positive or negative. What determines if there is a big difference is how well vector instructions can be utilized. If the unrolling disturbs highly efficient automatic optimisations made by the compiler, like for the \texttt{SIMPLE\_MEM} program, then performance can drop significantly. If instead the unrolling exposes opportunities for ILP, like for the \texttt{SIMPLE\_COMP} and \texttt{COMPLEX\_NESTED} programs, then huge performance gains can be achieved.
 
\subsection{Barcelona OpenMP Task Suite (BOTS)}
\label{sec:results:bots}

This section considers the BOTS (Barcelona OpenMP Task Suite) benchmark suite, presented previously in section~\ref{sec:theory:benchmarks}.
We first analyse the choice of waiting policy, which is covered in section~\ref{sec:results:bots_waitpolicy}.
Two of the benchmark programs also have the option to use either single-threaded and multi-threaded task generation, which we have previously analysed with the parallel constructs microbenchmark. This is treated in section~\ref{sec:results:bots_single_vs_multi_tasks}. 
Finally, we examine the source code for each program and apply loop tiling and unrolling in performance-critical areas to analyse the effects of these transformations in a realistic setting. Unfortunately, we find no programs in which we can apply loop tiling, so therefore only loop unrolling is considered, and is covered in section~\ref{sec:results:bots_unrolling}.

\subsubsection{Waiting policy}
\label{sec:results:bots_waitpolicy}

In these experiments we investigate the impact of the waiting policy on the BOTS benchmarks. We run the \emph{alignment}, \emph{fft} \emph{fib}, \emph{health}, \emph{sort}, \emph{sparselu}, \emph{strassen} and \emph{uts} benchmarks with the waiting policy set to active, passive or default.

\begin{figure}[h]
    \centering
    \includegraphics[width=0.8\textwidth]{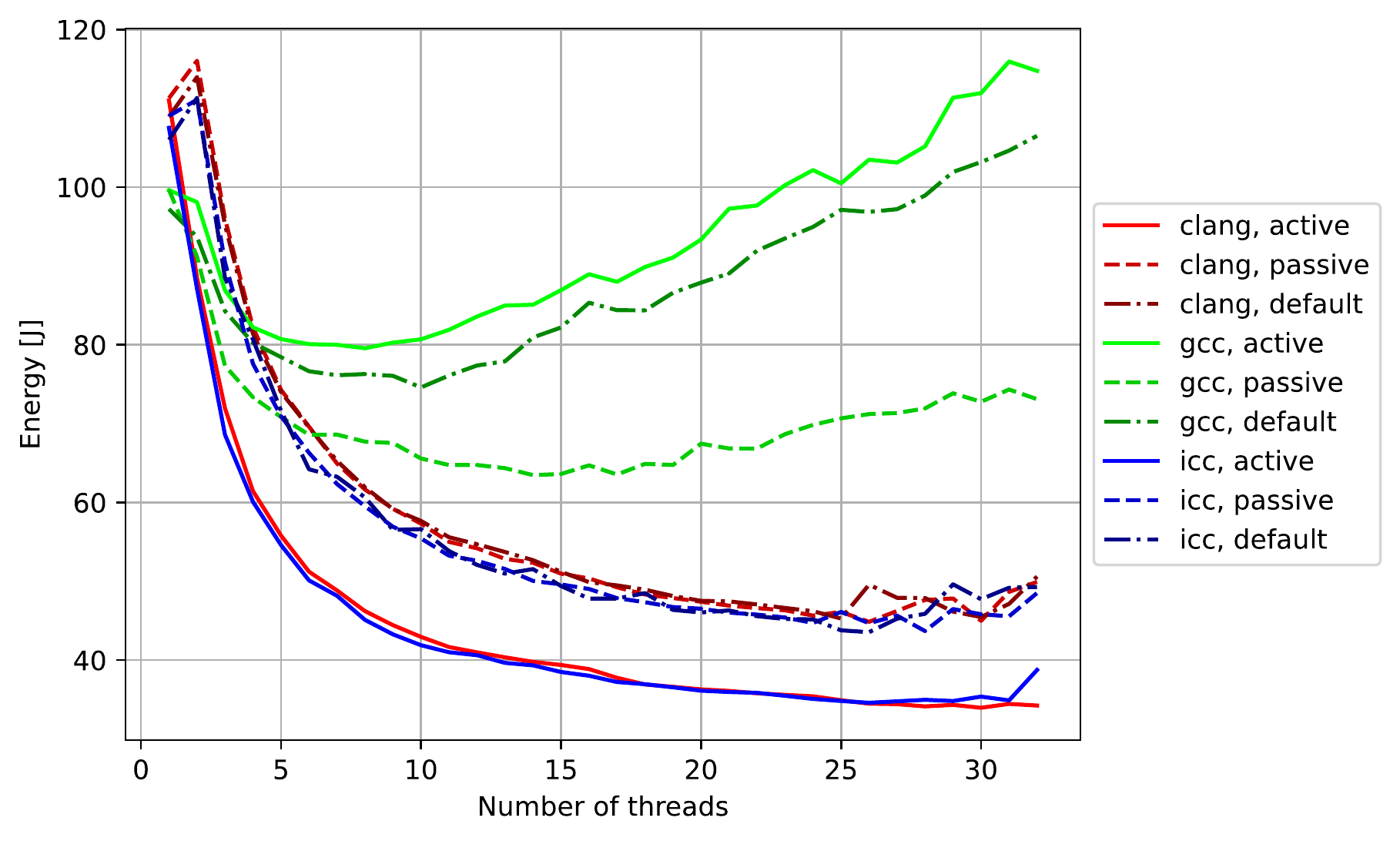}
    \caption{Average energy consumption of all BOTS programs for the three compilers and waiting policies.}
    \label{fig:bots_energy_32}
\end{figure}

The mean energy consumption can be seen in figure~\ref{fig:bots_energy_32}. The most obvious result is that GCC performs poorly with high numbers of threads. Since neither of the other compilers do so, this indicates some kind of inefficiency of tasking in GCC. The best results are produced by ICC and Clang, both with active waiting. Conversely, for GCC, active waiting performs the worst.

The corresponding graphs for execution time and power consumption can be found in the appendix, in figures~\ref{fig:bots_exec_32} and~\ref{fig:bots_power_32}. The curve for execution time is very similar to that of energy consumption with some minor differences. 
Looking only at the two best configurations that were best in terms of energy, which were ICC and Clang with active waiting, we see that they are again best in terms of execution time. For high numbers of threads we do however see that Clang is slightly faster, but that ICC consumes less power, leading them to be roughly equal in terms of energy as seen before.

We conclude that for a task-heavy workload such as the BOTS programs, Clang or ICC are preferred and should use active waiting.

\subsubsection{Single-threaded and multi-threaded task generation}
\label{sec:results:bots_single_vs_multi_tasks}

\begin{figure}[h]
  \centering
  \includegraphics[width=\textwidth]{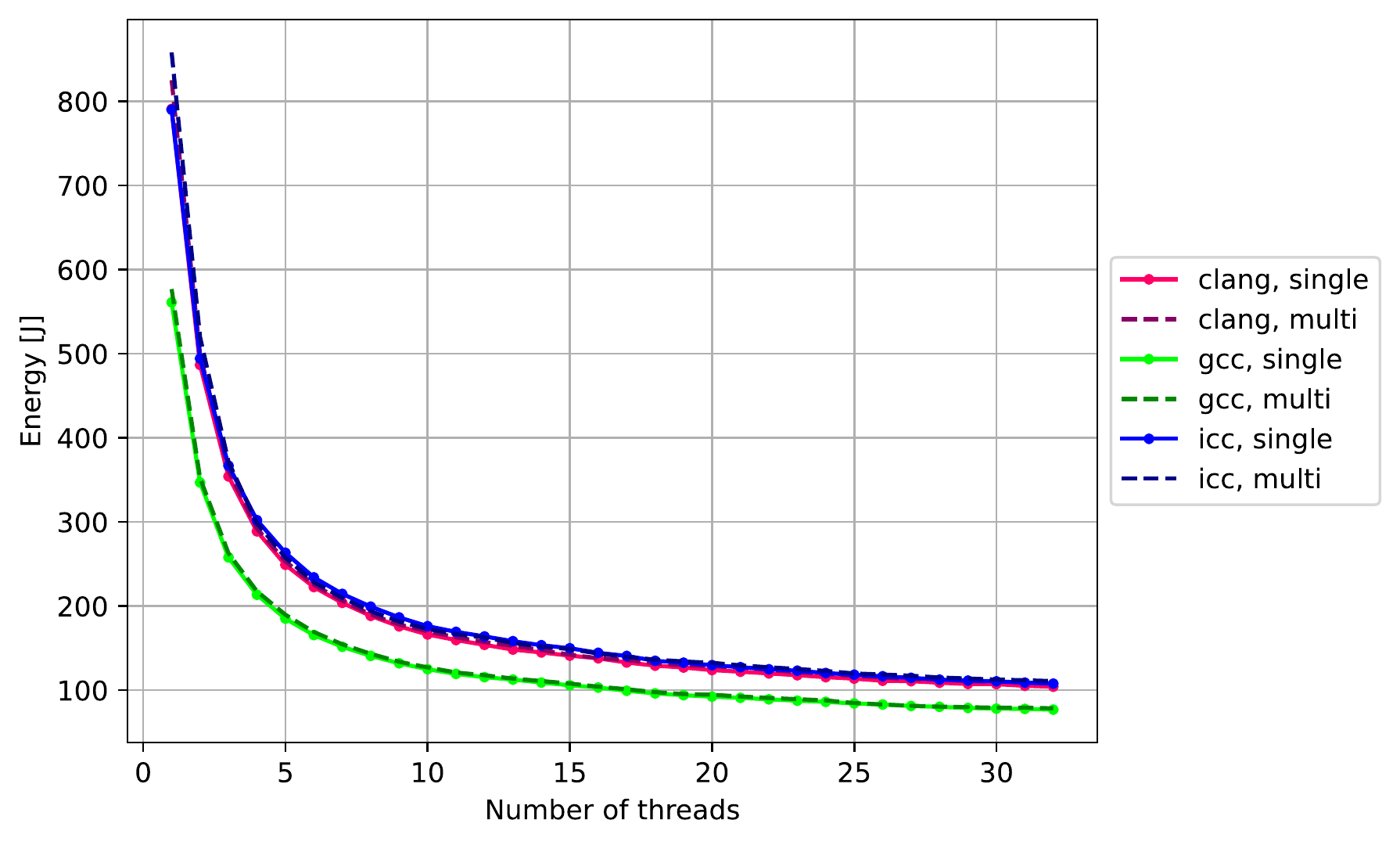}
  \caption{Energy consumption comparison between single-threaded and multi-threaded task generation, for the two programs supporting both in BOTS. We see that there is very little difference between the task creation variants.}
  \label{fig:bots_for_v_single}
\end{figure}

The \emph{alignment} and \emph{sparselu} programs include both single- and multi-threaded task generation, working similarly to the parallel construct benchmark described in section~\ref{sec:theory:parallelismconstructsmicrobenchmark}. In figures~\ref{fig:bots_for_v_single} we compare the two methods for these programs for each compiler, with the waiting policy set to default. While there is a major difference between the compilers for these programs, GCC being much faster, there is almost no difference between the tasking methods. Single-threaded generation consumes 2.4 \%, 1.6 \%, and 0.7 \% less energy for Clang, GCC and ICC, respectively. We draw the conclusion that for tasks of this granularity (which should represent most realistic workloads), there is no reason to use multi-threaded task generation when single-threaded is more energy efficient and easier to implement. The results are almost exactly the same regardless of waiting policy as well.

\subsubsection{Loop Unrolling}
\label{sec:results:bots_unrolling}

Two benchmarks from BOTS were chosen to be analysed with unrolling, \emph{strassen} and \emph{alignment}. These were chosen due to most of their execution time (from profiling) was spent in functions where unrolling could be applied.

\begin{figure}[h]
    \centering
    \includegraphics[width=0.8\textwidth]{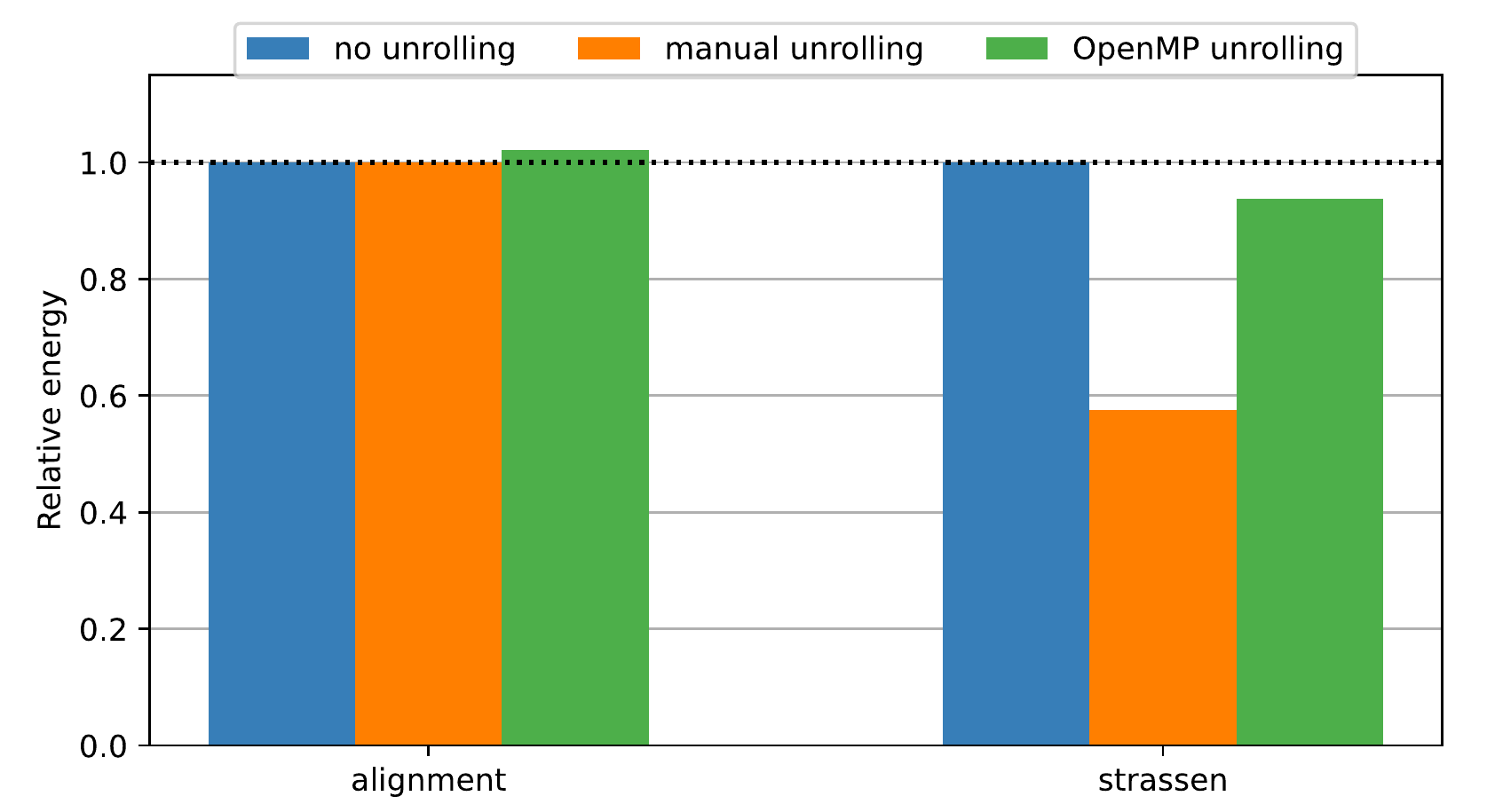}
    \caption{Relative energy used for manual and OpenMP unrolling compared to no unrolling for the alignment and strassen programs. Several unrolling factors were tested and only the best for each version is shown.}
    \label{fig:bots_unrolling}
\end{figure}

Both programs were tested with manual unrolling and the OpenMP \texttt{unroll} directive. The results for these tests can be seen in in figure \ref{fig:bots_unrolling}. The figure shows the relative amount of energy used by each configuration compared to the implementation without explicit unrolling. The results shown are the median relative energy values when running the program with 1 to 32 threads and compiled with Clang. Several unrolling factors were tested and only the best performing is shown for each program and implementation. It should also be mentioned that the execution time followed the energy usage very closely for the following tests, so any gain in energy is a result of a decrees directly proportional to the execution time.

For \emph{alignment} one very simple for loop which initialises two arrays with values was analysed. For this program neither unrolling version had any positive impact on the performance, where manual and OpenMP unrolling had with their best implementation no impact and an increase of 2.2 \% on energy respectively. These implementation had an unrolling factor of 32 for the manual and 8 for the OpenMP version. For some of the tested unrolling factors energy could increase by as much as 15 \%. 

In the case of the \emph{strassens} program the original implementation was already manually unrolled, so the \emph{no unrolling} baseline for figure \ref{fig:bots_unrolling} is instead a modified version we have created where this unrolling is removed. The code section that is being unrolled for this program is the outer loop of two nested loops, inside of which some calculations are done. This code section is almost identical to the \texttt{COMPLEX\_NESTED} program, described in section \ref{sec:background:unrolling_microbenchmark}, and was the inspiration for the microbenchmark. For the manual version the \emph{loop jamming} optimisation has also here been similarly applied. 

In contrast to the case of \emph{alignment}, unrolling for \emph{strassen} always lowered the total energy usage. With the best performing implementation, the original manual unrolling with unrolling factor 8, using only about 55 \% of the energy of having no unrolling, and the OpenMP version, with the same unrolling factor, saving around 8 \%. The reason why the manual method performed significantly better than both the default and OpenMP version is because the way the code is restructured allows for auto-vectorisation by the compiler.

\subsection{NAS Parallel Benchmarks}
\label{sec:results:nas}

In this section we present the results of experiments on the NAS Parallel benchmarks. The impact of the waiting policy is explored in subsection~\ref{sec:results:nas:waiting_policy} by executing all (unmodified) programs with the three waiting policy options.
Loop transformations are explored in subsection~\ref{sec:results:nas:transforms} by applying transformations to loops in functions where the most execution time is spent for each program. This is mainly done with loop unrolling, since it is significantly more difficult to apply loop tiling to programs.

\subsubsection{Waiting policies}
\label{sec:results:nas:waiting_policy}

To evaluate the impact of the OpenMP waiting policy we execute the otherwise unmodified programs with active, passive and default waiting with all three compilers.

\begin{figure}[h]   
    \centering
    \includegraphics[width=\textwidth]{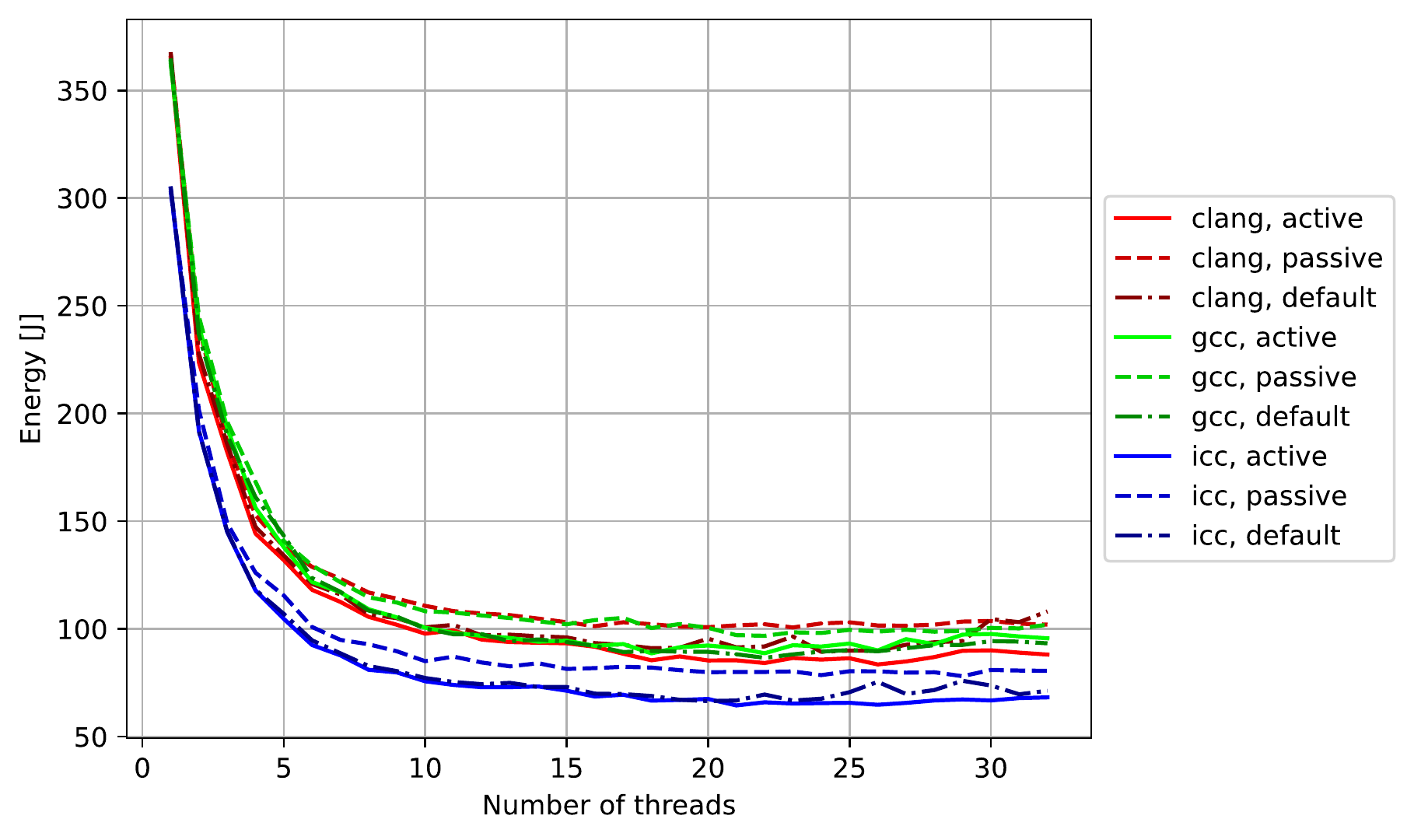}
    \caption{Average energy consumption of all NPB programs for different numbers of threads, compilers and waiting policies.}
    \label{fig:npb_waitpol_energy}
\end{figure}

In figure~\ref{fig:npb_waitpol_energy} we see the average energy results of executing the NPB programs with the different compilers and waiting policies. We see that, while the difference is small, active waiting consumes the least energy and passive consumes the most. ICC with active waiting consistently performs the best. Clang with active waiting also performs well. For BOTS, GCC scaled poorly beyond about 10 threads, which is clearly not the case here. This could further indicate that GCC performs fairly well with parallel loops, as used here, but not so well with tasking, as used in BOTS. We also note that GCC performs the best with default waiting in this case.

\begin{figure}[h]
    \centering
    \includegraphics[width=\textwidth]{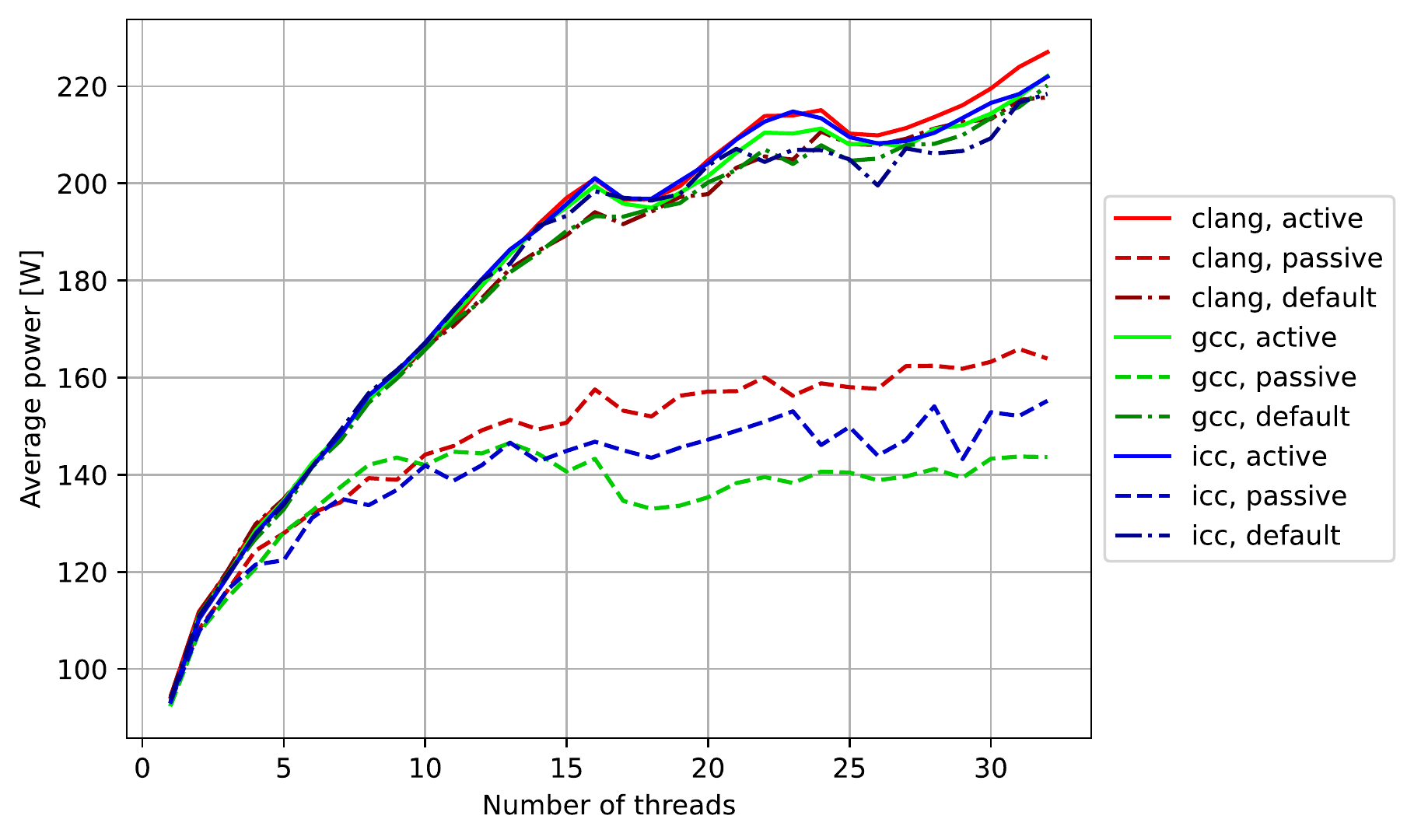}
    \caption{Average power consumption of all NPB programs for different numbers of threads, compilers and waiting policies.}
    \label{fig:npb_waitpol_power}
\end{figure}

The corresponding plot for the execution time is very similar to the energy and can be found in the appendix, in figure~\ref{fig:npb_waitpol_exec}. The main difference is that passive waiting performs worse for all compilers. The power can be seen in figure~\ref{fig:npb_waitpol_power}, where we see that active and default waiting follows roughly the same curve for all compilers, while passive waiting consumes significantly less power. 

Table~\ref{tab:npb_waitpol_summary} summarizes the results and shows the execution time, power and energy averaged across all numbers of threads. For Clang and ICC we confirm that active waiting consumes about 5 \% less energy than default waiting, while default waiting is (barely) the best for GCC. We also see that passive waiting consumes around 20 \% less power.

ICC is by far the best compiler in this case due to its comparable power consumption but much lower execution time. Clang and GCC are about equal.

\begin{table}[h]
\begin{center}
\caption{Summary of waiting policy results for NPB.}
\label{tab:npb_waitpol_summary}
\begin{tabular}{|c|c|ccc|ccc|}
\hline
Compiler & Policy & \multicolumn{3}{c|}{Absolute} & \multicolumn{3}{c|}{Relative} \\ \hline
 &  & \multicolumn{1}{c|}{T {[}s{]}} & \multicolumn{1}{c|}{P {[}W{]}} & E {[}J{]} & \multicolumn{1}{c|}{T} & \multicolumn{1}{c|}{P} & \textbf{E} \\ \hline
clang & Default & \multicolumn{1}{c|}{99.55} & \multicolumn{1}{c|}{160.92} & 218.94 & \multicolumn{1}{c|}{1} & \multicolumn{1}{c|}{1} & \textbf{1} \\ \hline
clang & Active & \multicolumn{1}{c|}{94.04} & \multicolumn{1}{c|}{163.24} & 206.6 & \multicolumn{1}{c|}{0.945} & \multicolumn{1}{c|}{1.014} & \textbf{0.944} \\ \hline
clang & Passive & \multicolumn{1}{c|}{105.9} & \multicolumn{1}{c|}{132.2} & 237.39 & \multicolumn{1}{c|}{1.064} & \multicolumn{1}{c|}{0.822} & \textbf{1.084} \\ \hline
gcc & Default & \multicolumn{1}{c|}{97.55} & \multicolumn{1}{c|}{159.84} & 214.69 & \multicolumn{1}{c|}{1} & \multicolumn{1}{c|}{1} & \textbf{1} \\ \hline
gcc & Active & \multicolumn{1}{c|}{98.83} & \multicolumn{1}{c|}{161.8} & 217.28 & \multicolumn{1}{c|}{1.013} & \multicolumn{1}{c|}{1.012} & \textbf{1.012} \\ \hline
gcc & Passive & \multicolumn{1}{c|}{104.71} & \multicolumn{1}{c|}{122.16} & 237.17 & \multicolumn{1}{c|}{1.073} & \multicolumn{1}{c|}{0.764} & \textbf{1.105} \\ \hline
icc & Default & \multicolumn{1}{c|}{75.2} & \multicolumn{1}{c|}{160.41} & 165.49 & \multicolumn{1}{c|}{1} & \multicolumn{1}{c|}{1} & \textbf{1} \\ \hline
icc & Active & \multicolumn{1}{c|}{72.74} & \multicolumn{1}{c|}{161.92} & 159.95 & \multicolumn{1}{c|}{0.9673} & \multicolumn{1}{c|}{1.009} & \textbf{0.967} \\ \hline
icc & Passive & \multicolumn{1}{c|}{83.84} & \multicolumn{1}{c|}{125.86} & 188.97 & \multicolumn{1}{c|}{1.115} & \multicolumn{1}{c|}{0.785} & \textbf{1.142} \\ \hline
\end{tabular}
\end{center}
\end{table}

\subsubsection{Loop Transformations}
\label{sec:results:nas:transforms}

To explore the impact of loop transformations we profile each program using the \emph{perf} tool. Then, we apply transformations to loops in these functions both manually and with OpenMP. In some cases the loops are already unrolled, in which case we also experiment with rolling the loop again. It is worth mentioning that the implemented transformations are to the best of the authors abilities.

In table~\ref{tab:results:nas:impl_overview} we show an overview of the implemented transformations. As shown, programs where both OpenMP and manual unrolling have been applied are BT, CG, EP and MG. In IS and LU we could not find any suitable loops.
Tiling is significantly more difficult to apply to existing programs, especially those already optimised to such a level as there. The only  program we applied it to was FT. This however only lead to performance detriments, so the results are not presented further.

\begin{table}[h]
\begin{center}
\caption{Implemented optimisations for each of the NPB programs.}
\label{tab:results:nas:impl_overview}
\begin{tabular}{|l|l|l|l|l|}
\hline
Program & Rolled & OpenMP Unrolling & Manual Unrolling & OpenMP Tiling \\ \hline
BT & Yes & Yes & Yes &  \\ \hline
CG & Yes & Yes & Yes &  \\ \hline
EP &  & Yes & Yes &  \\ \hline
FT &  & Yes &  & Yes \\ \hline
IS &  &  &  &  \\ \hline
LU &  &  &  &  \\ \hline
MG &  & Yes & Yes &  \\ \hline
\end{tabular}
\end{center}
\end{table}

\begin{figure}[h]
  \centering
  \includegraphics[width=0.8\textwidth]{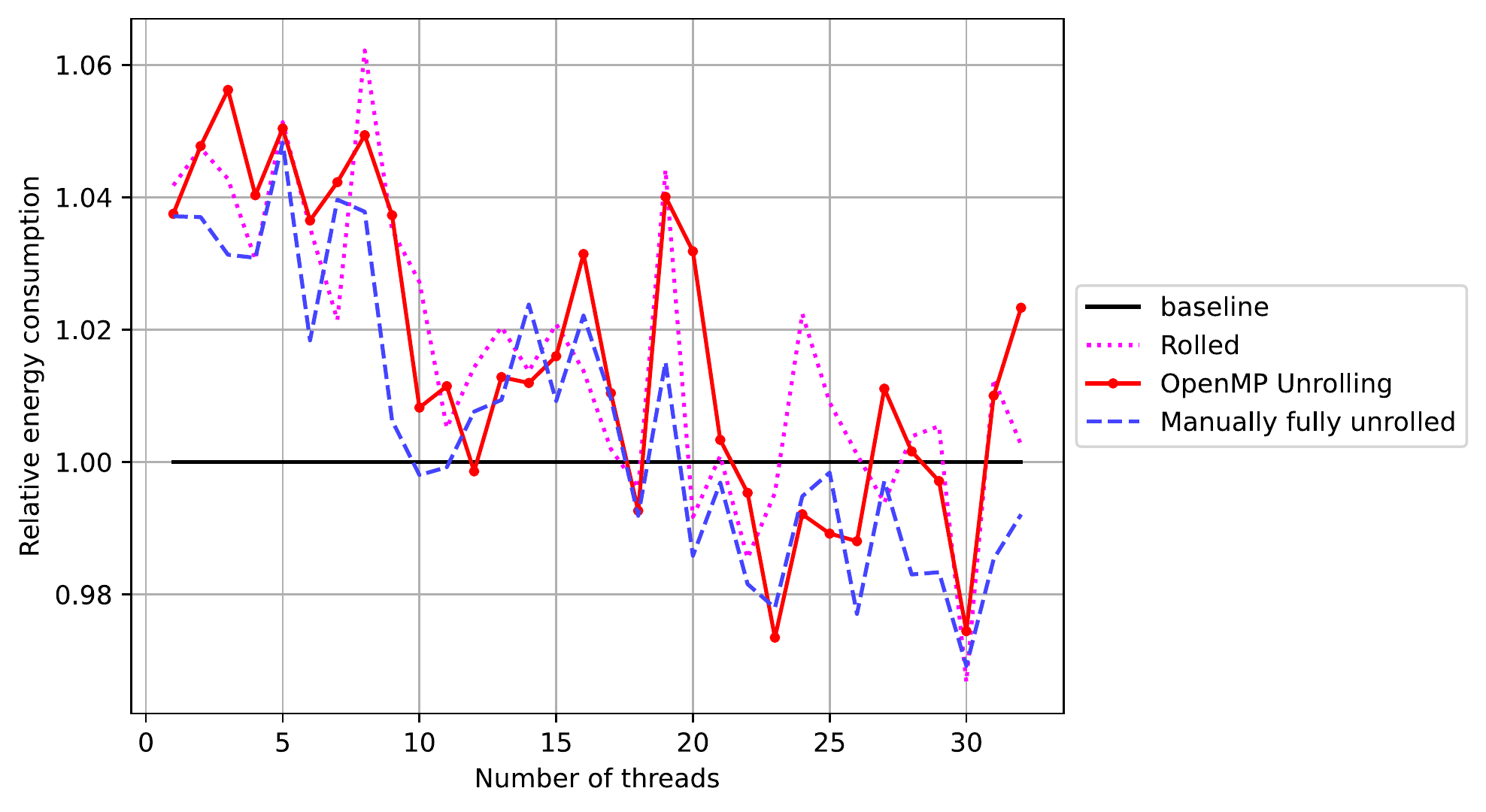}
  \caption{Relative energy consumption of unrolling the BT program.}
  \label{fig:cnas_bt_energy}
\end{figure}

\begin{figure}[h]
  \centering
  \includegraphics[width=0.8\textwidth]{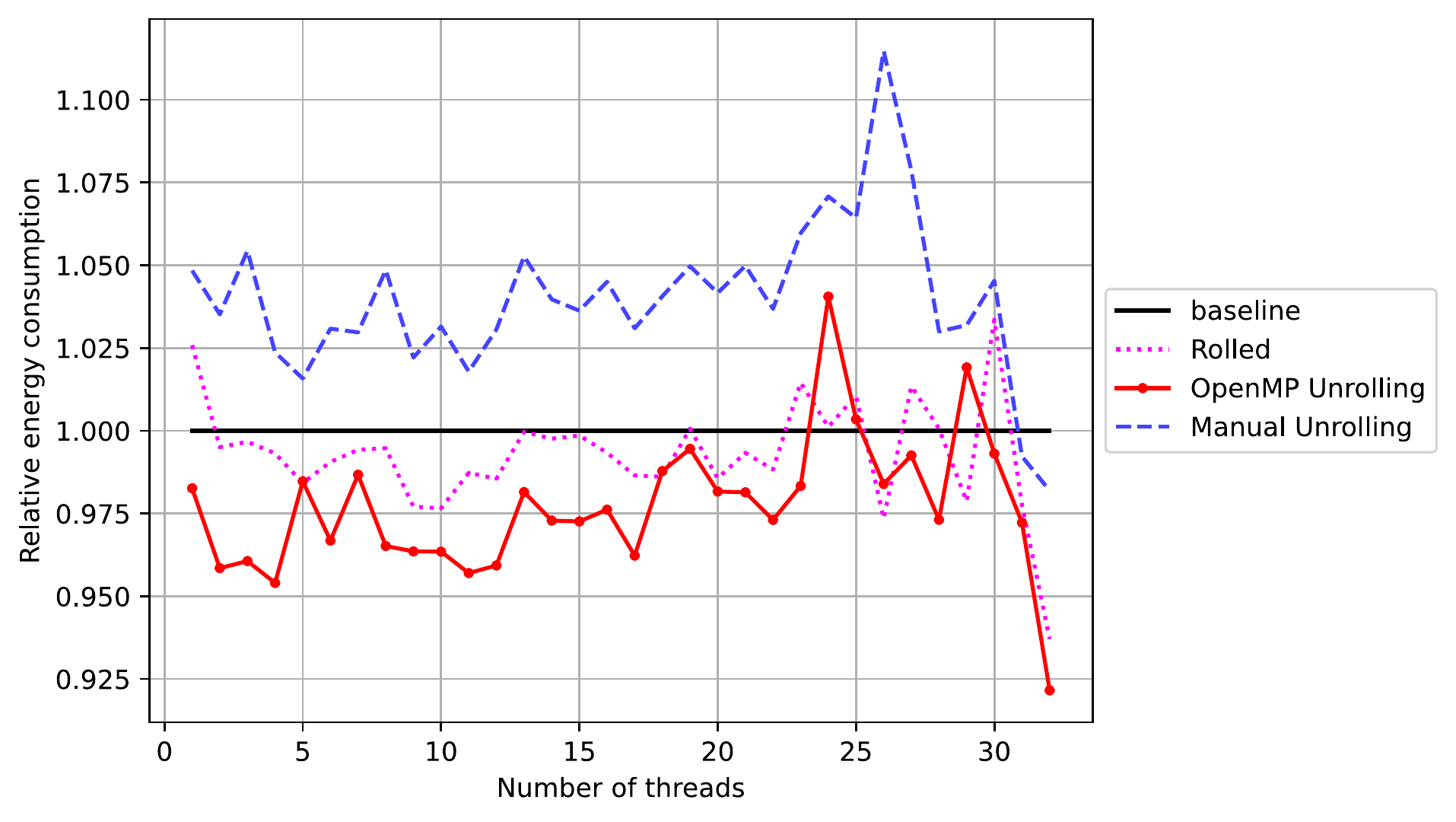}
  \caption{Relative energy consumption of unrolling the CG program.}
  \label{fig:cnas_cg_energy}
\end{figure}

\begin{figure}[h]
  \centering
  \includegraphics[width=0.8\textwidth]{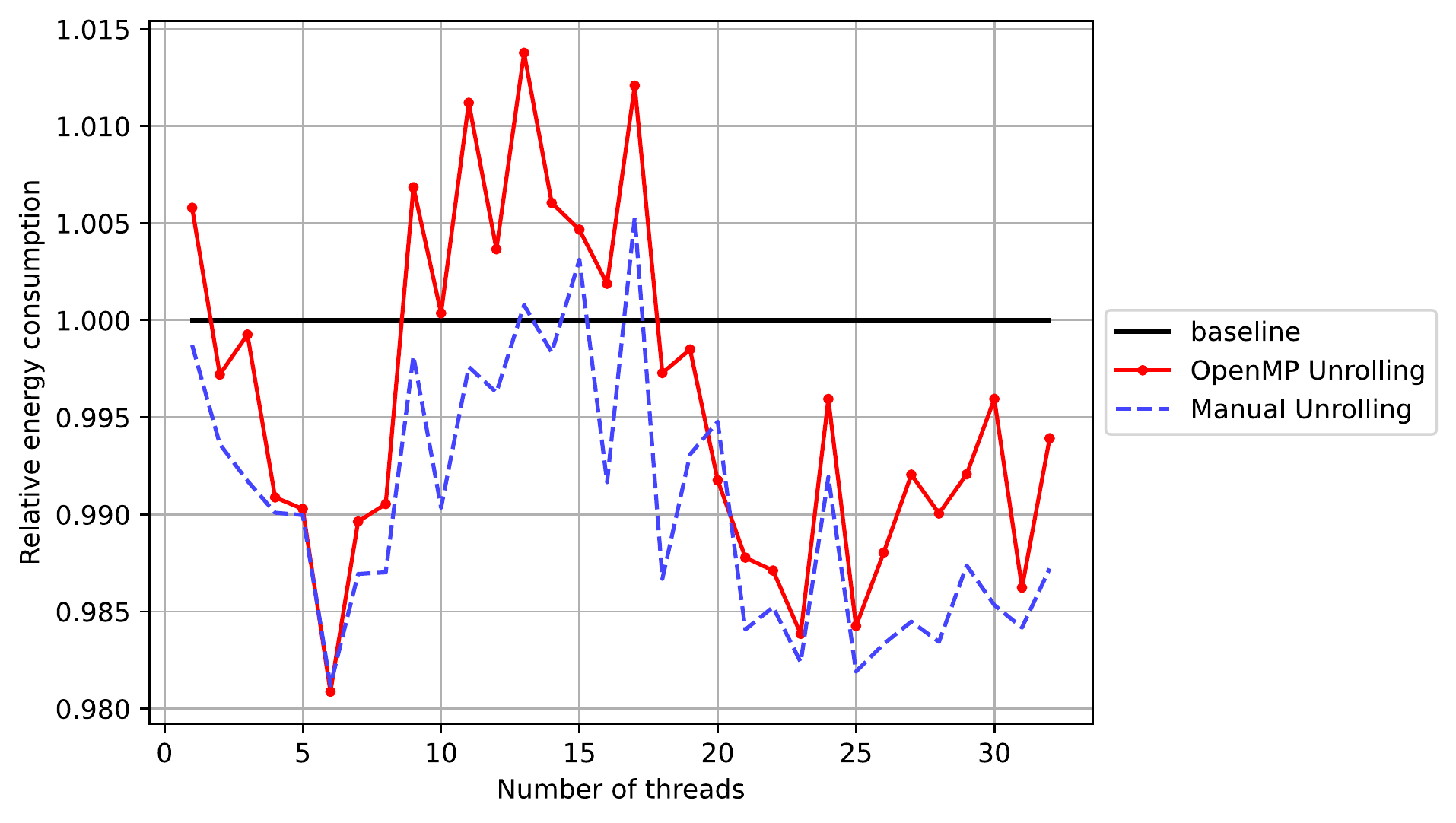}
  \caption{Relative energy consumption of unrolling the EP program.}
  \label{fig:cnas_ep_energy}
\end{figure}

\begin{figure}[h]
  \centering
  \includegraphics[width=0.8\textwidth]{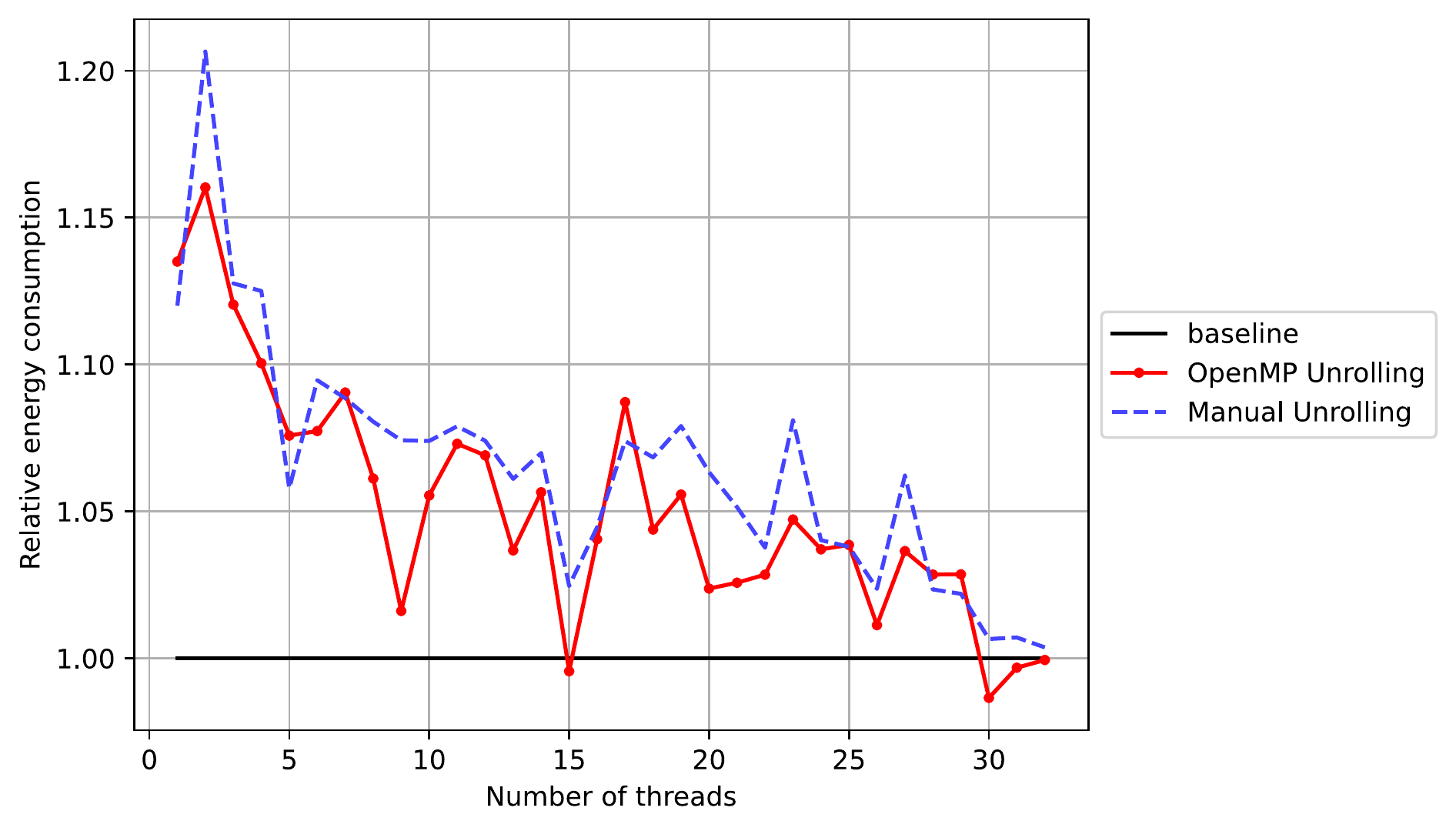}
  \caption{Relative energy consumption of unrolling the MG program.}
  \label{fig:cnas_mg_energy}
\end{figure}

First, we consider the programs where both manual and OpenMP unrolling was applied, namely BT, CG, EP and MG. Results for these are seen in figures~\ref{fig:cnas_bt_energy}, \ref{fig:cnas_cg_energy}, \ref{fig:cnas_ep_energy} and \ref{fig:cnas_mg_energy}.

In the BT program, most time is spent in the \emph{lhs\_x}, \emph{lhs\_y} and \emph{lhs\_z} functions. They contain innermost loops that are fully unrolled by default, which we roll into loops. After this, we apply full loop unrolling both with OpenMP and manually. While the baseline is best for low numbers of threads, unrolling becomes better with more threads.
The CG program, which computes a conjugate gradient, spends most of its execution time in the \emph{conj\_grad} function. Here we apply manual and OpenMP unrolling. Unrolling using OpenMP yields the best results, especially for low numbers of threads, even in comparison to the manual unrolling.
In the EP program we apply both manual and OpenMP unrolling. We see that manual unrolling is better than OpenMP unrolling in general, but that the difference is very small.
In the MG program we apply unrolling to the \emph{resid} function. Results show that unrolling yields worse results than the baseline.

These results are summarized in table~\ref{tab:nas:unroll_res}. We observe that the differences between any of the configurations are very small, usually about just one percent. We also observe that on average, unrolling in this way yields worse energy results of about a percent for OpenMP unrolling and two percent for manual unrolling.

\begin{table}[h]
\begin{center}
\caption{The improvement of unrolling each program in NPB compared to the baseline, averaged across all numbers of threads.}
\label{tab:nas:unroll_res}
\begin{tabular}{|c|c|ccc|ccc|}
\hline
Program & Unrolling Method & \multicolumn{3}{c|}{Absolute} & \multicolumn{3}{c|}{Relative} \\ \hline
 &  & \multicolumn{1}{c|}{T {[}s{]}} & \multicolumn{1}{c|}{P {[}W{]}} & E {[}J{]} & \multicolumn{1}{c|}{T} & \multicolumn{1}{c|}{P} & \textbf{E} \\ \hline
BT & Baseline & \multicolumn{1}{c|}{8.812} & \multicolumn{1}{c|}{191.2} & 1685 & \multicolumn{1}{c|}{1} & \multicolumn{1}{c|}{1} & \textbf{1} \\ \hline
BT & Rolled & \multicolumn{1}{c|}{9.058} & \multicolumn{1}{c|}{189.0} & 1712 & \multicolumn{1}{c|}{1.028} & \multicolumn{1}{c|}{0.988} & \textbf{1.016} \\ \hline
BT & OpenMP & \multicolumn{1}{c|}{9.060} & \multicolumn{1}{c|}{188.9} & 1712 & \multicolumn{1}{c|}{1.028} & \multicolumn{1}{c|}{0.988} & \textbf{1.016} \\ \hline
BT & Manual & \multicolumn{1}{c|}{9.002} & \multicolumn{1}{c|}{188.3} & 1695 & \multicolumn{1}{c|}{1.021} & \multicolumn{1}{c|}{0.985} & \textbf{1.006} \\ \hline
CG & Baseline & \multicolumn{1}{c|}{0.112} & \multicolumn{1}{c|}{178.3} & 19.94 & \multicolumn{1}{c|}{1} & \multicolumn{1}{c|}{1} & \textbf{1} \\ \hline
CG & Rolled & \multicolumn{1}{c|}{0.111} & \multicolumn{1}{c|}{178.7} & 19.80 & \multicolumn{1}{c|}{0.991} & \multicolumn{1}{c|}{1.002} & \textbf{0.993} \\ \hline
CG & OpenMP & \multicolumn{1}{c|}{0.109} & \multicolumn{1}{c|}{178.5} & 19.49 & \multicolumn{1}{c|}{0.976} & \multicolumn{1}{c|}{1.001} & \textbf{0.977} \\ \hline
CG & Manual & \multicolumn{1}{c|}{0.110} & \multicolumn{1}{c|}{188.5} & 20.74 & \multicolumn{1}{c|}{0.984} & \multicolumn{1}{c|}{1.057} & \textbf{1.040} \\ \hline
EP & Baseline & \multicolumn{1}{c|}{1.946} & \multicolumn{1}{c|}{148.3} & 288.5 & \multicolumn{1}{c|}{1} & \multicolumn{1}{c|}{1} & \textbf{1} \\ \hline
EP & OpenMP & \multicolumn{1}{c|}{1.945} & \multicolumn{1}{c|}{147.7} & 287.3 & \multicolumn{1}{c|}{1.000} & \multicolumn{1}{c|}{0.996} & \textbf{0.996} \\ \hline
EP & Manual & \multicolumn{1}{c|}{1.936} & \multicolumn{1}{c|}{147.6} & 285.8 & \multicolumn{1}{c|}{0.995} & \multicolumn{1}{c|}{0.995} & \textbf{0.990} \\ \hline
MG & Baseline & \multicolumn{1}{c|}{1.946} & \multicolumn{1}{c|}{148.3} & 288.5 & \multicolumn{1}{c|}{1} & \multicolumn{1}{c|}{1} & \textbf{1} \\ \hline
MG & OpenMP & \multicolumn{1}{c|}{1.945} & \multicolumn{1}{c|}{147.7} & 287.3 & \multicolumn{1}{c|}{1.000} & \multicolumn{1}{c|}{0.996} & \textbf{0.996} \\ \hline
MG & Manual & \multicolumn{1}{c|}{1.936} & \multicolumn{1}{c|}{147.6} & 285.8 & \multicolumn{1}{c|}{0.995} & \multicolumn{1}{c|}{0.995} & \textbf{0.990} \\ \hline
\end{tabular}
\end{center}
\end{table}

\subsection{PARSEC benchmark}
\label{sec:results:parsec}

In this section we present the results from the two programs tested for the PARSEC benchmark suit, \emph{blackscholes} and \emph{fluidanimate}. The programs were tested using the \emph{parallel for} and \emph{tasking} implementations as well as with all available compilers and waiting policies. The parallel for versions were tested using the default \emph{static} scheduling policy, unless otherwise stated.

\begin{figure}
    \centering
    \begin{subfigure}[b]{0.45\textwidth}    
        \centering
        \includegraphics[width=\textwidth]{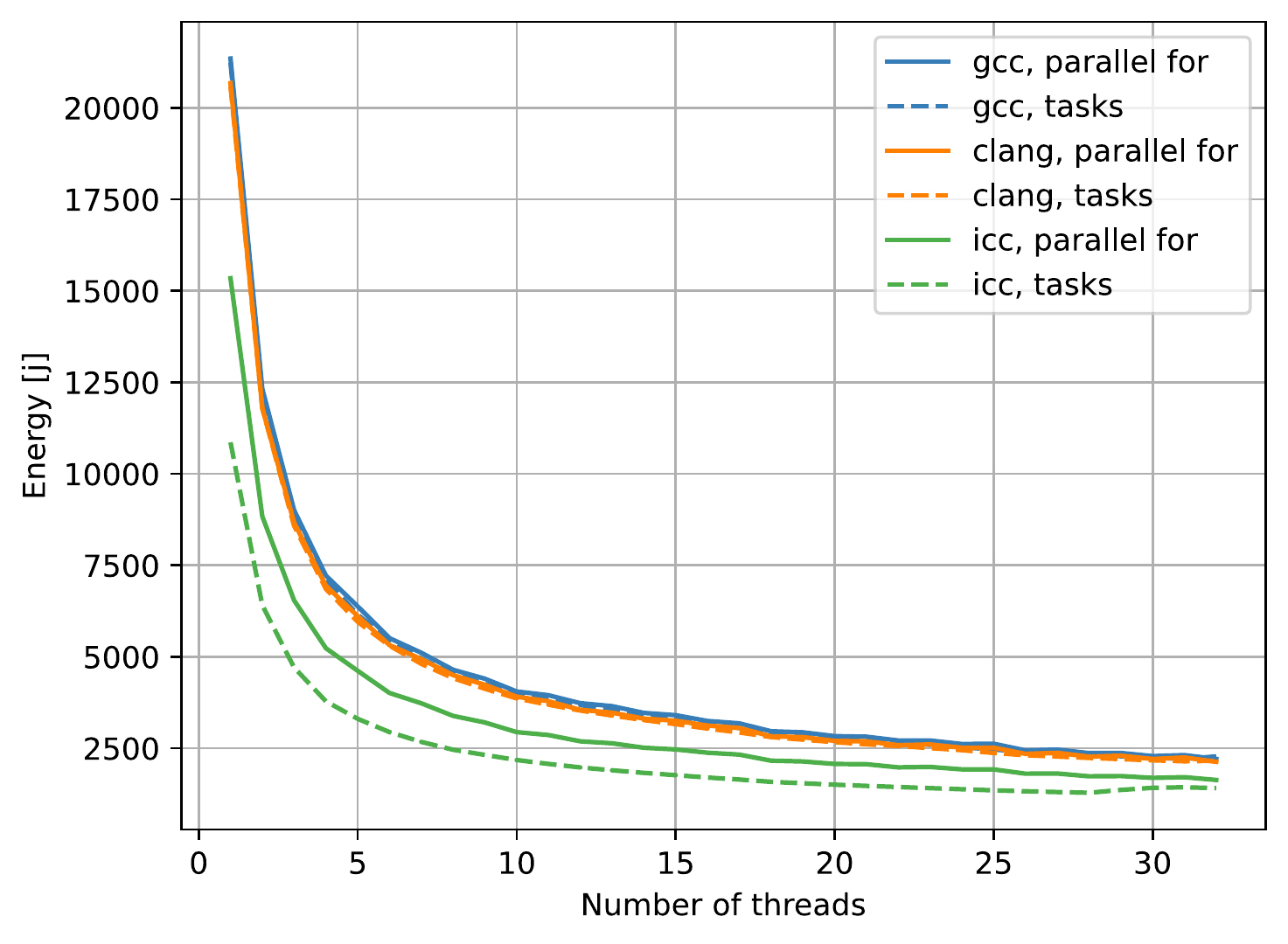}
        \caption{Blackscholes.}
        \label{fig:PARSEC_black_versions}
    \end{subfigure}\hfill
    \begin{subfigure}[b]{0.45\textwidth}    
        \centering
        \includegraphics[width=\textwidth]{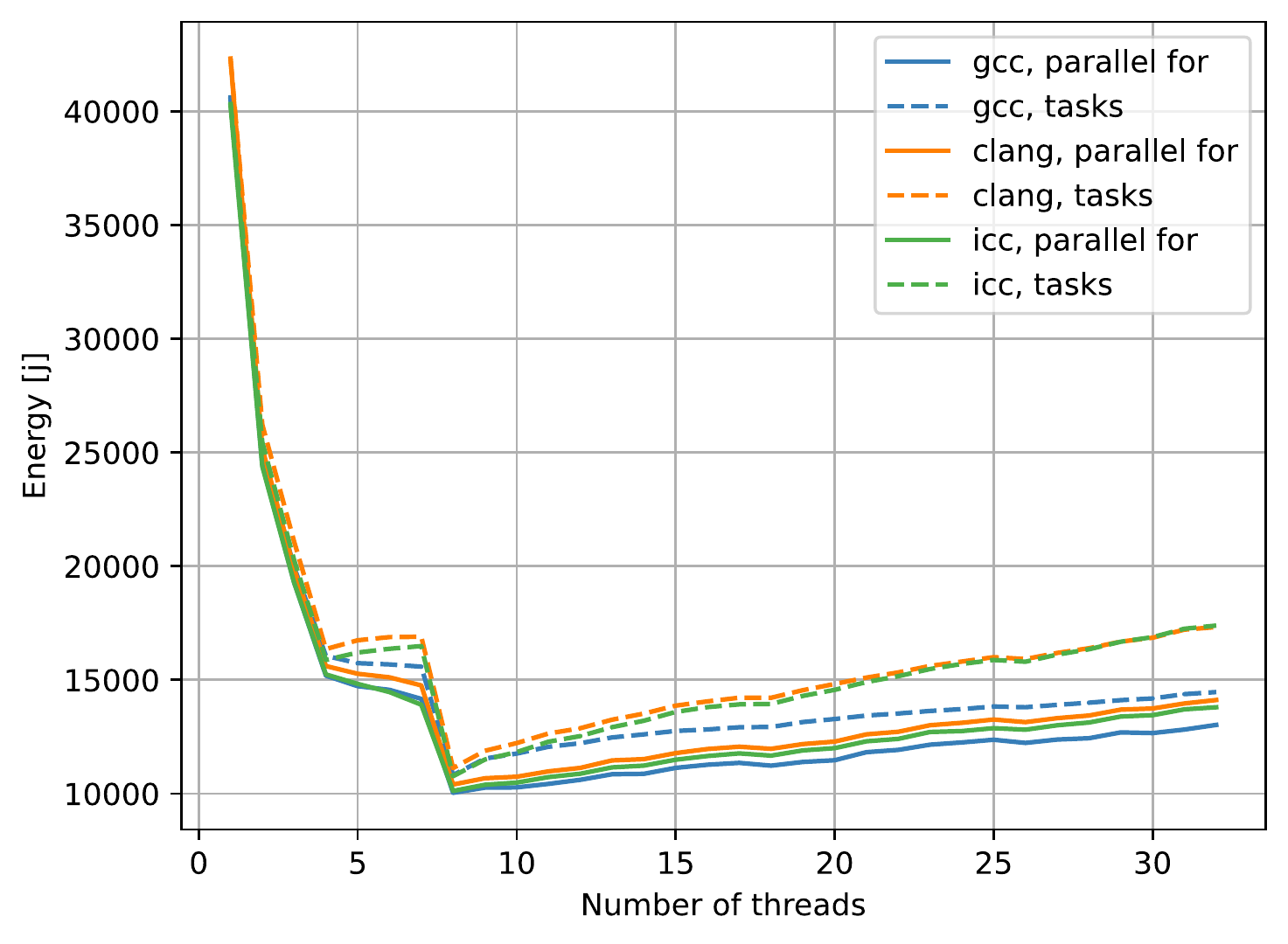}
        \caption{Fluidanimate.}
        \label{fig:PARSEC_fluid_versions}
    \end{subfigure}
    \caption{Energy used for the two PARSEC programs, for the \emph{parallel for} and \emph{tasking} implementations, using the three different compilers.}
    \label{fig:PARSEC_versions}
\end{figure}

In figure \ref{fig:PARSEC_versions} the energy usage for the two programs can be seen while using the two parallelisation implementation. The results shown are obtained by taking the geometric mean using the three different waiting policies. For blackscholes, figure \ref{fig:PARSEC_black_versions}, the tasking version performs better on average for all three compilers. For both implementations, using the ICC compiler performed significantly better than their GCC and Clang counterparts. When taking the geometric mean of the energy saved by using tasks compared to the parallel for implementation over all threads, the tasking version for GCC uses 2.5 \% less energy, Clang uses 2.0 \% less and for ICC it is 27 \%. For this program some additional scheduling policies were tested, specifically the \emph{guided} setting, which resulted in \emph{parallel for} being at best roughly 5 \% better than their tasking version while using GCC and Clang. For ICC the \emph{tasking} version was, however, consistently the best. For fluidanimate, figure \ref{fig:PARSEC_fluid_versions}, the opposite relation is true where instead \emph{parallel for} is the best option. The energy saved for the three compilers is between 10 - 15 \% compared to their tasking versions. Overall this program has some unusual performance characteristics. From one to four threads the performance improves linearly, then little improvement is seen up until eight threads, where the performance improves substantially. After eight threads the programs execution time stays the same, leading the energy used to increase when more threads are added.

To figure out why ICC performed so much better than the two other compilers for the blackscholes program, further analysis was conducted using the perf tool. There we noticed that GCC and Clang spent the majority of their execution time in the \texttt{exp} function, which computes the Euler number raised to some given input, included in the \texttt{math.h} library, something the ICC didn't spend nearly the same amount of time in. The \texttt{exp} function was therefore tested in isolation for the different compilers. The results from these test was that the ICC implementation of the function was 4x faster than for the two other compilers. This finding explains why ICC had the best performance for the blackscholes program.

\begin{figure}
    \centering
    \begin{subfigure}[b]{0.45\textwidth}    
        \centering
        \includegraphics[width=\textwidth]{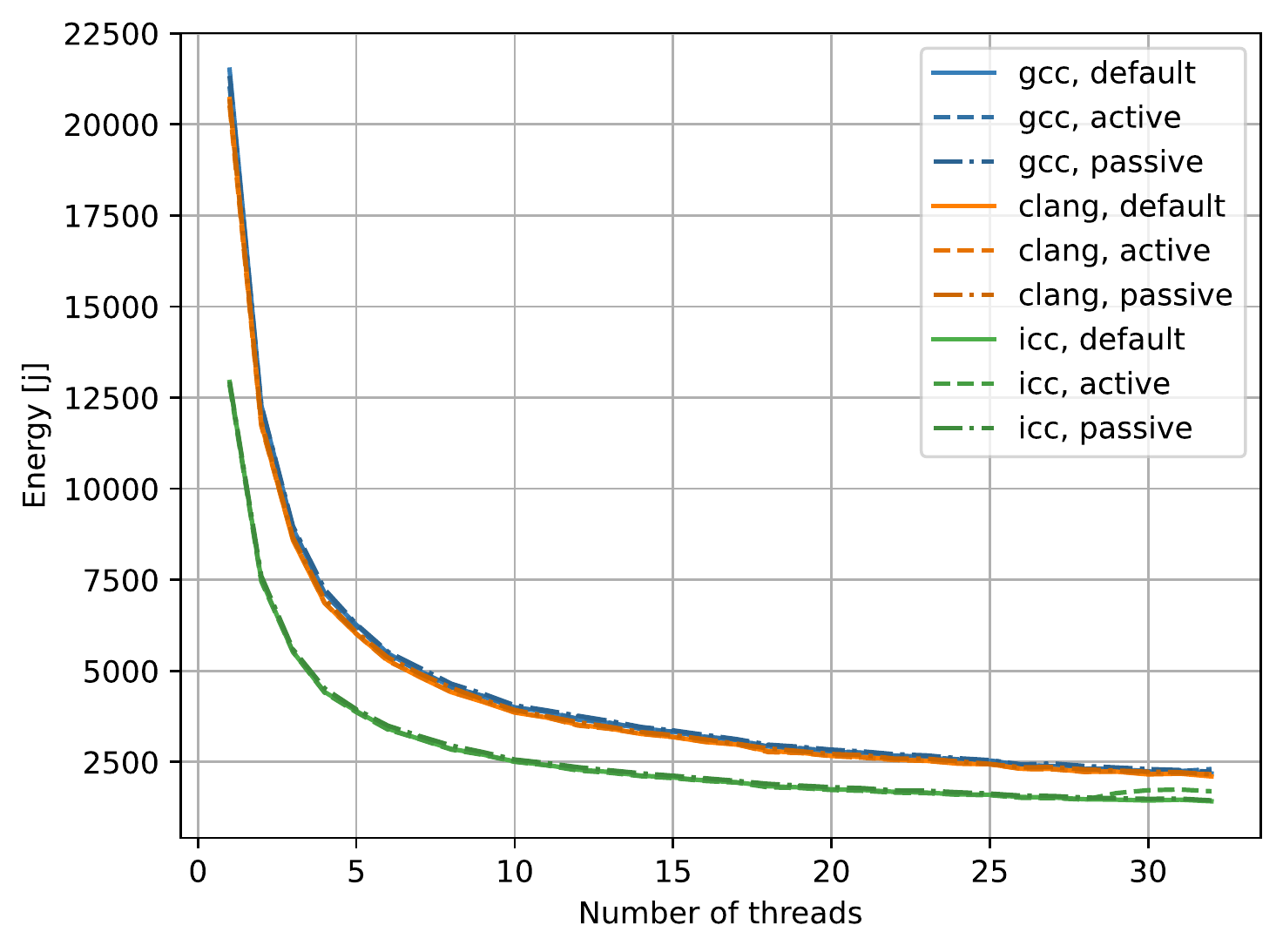}
        \caption{Blackscholes.}
        \label{fig:PARSEC_black_waiting}
    \end{subfigure}\hfill
    \begin{subfigure}[b]{0.45\textwidth}    
        \centering
        \includegraphics[width=\textwidth]{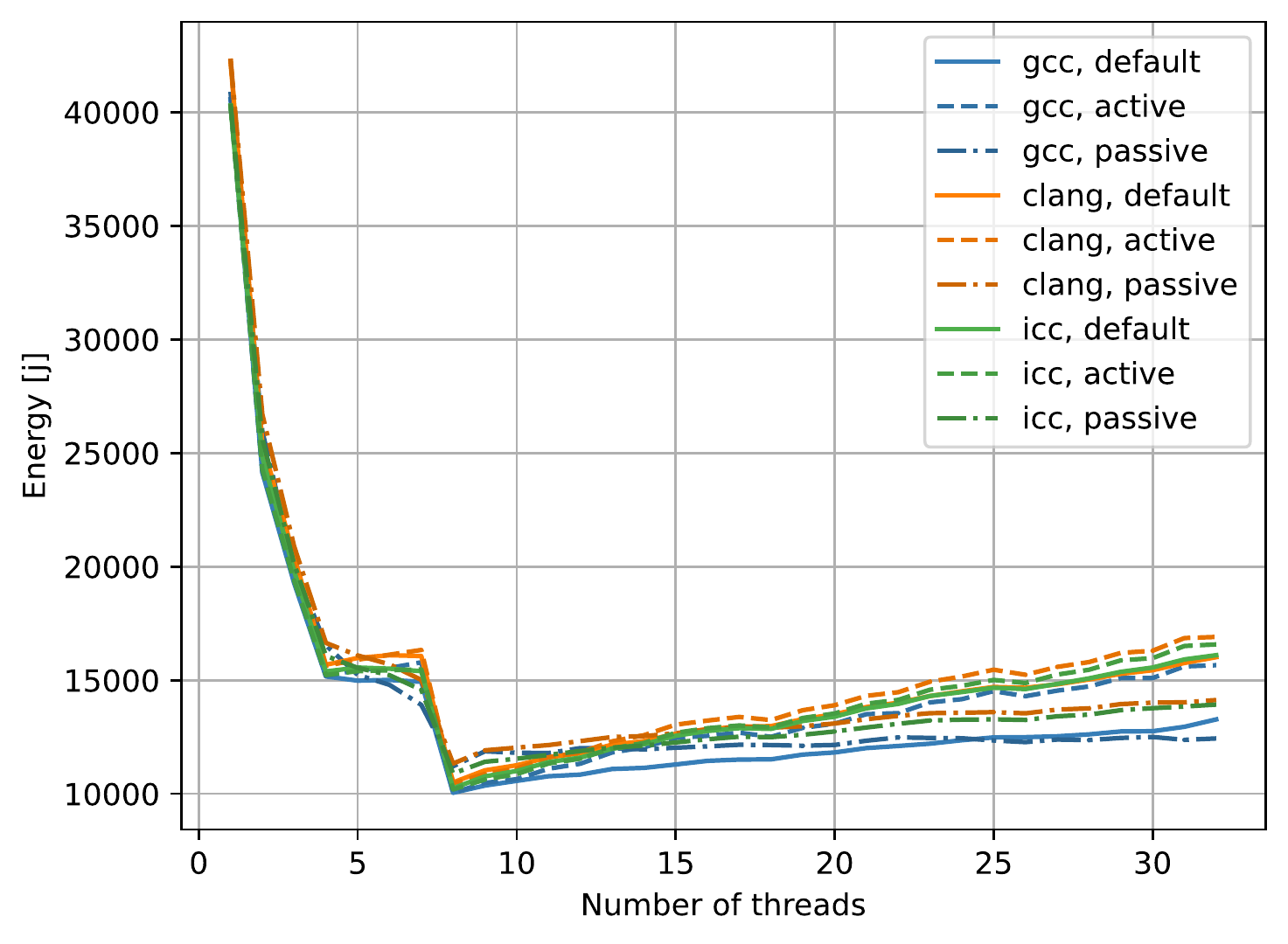}
        \caption{Fluidanimate}
        \label{fig:PARSEC_fluid_waiting}
    \end{subfigure}
    \caption{Energy used for the two PARSEC programs, for the three different waiting policies, using the three different compilers.}
    \label{fig:PARSEC_waiting}
\end{figure}

In figure \ref{fig:PARSEC_waiting} the energy usage for the two programs can be seen while using the three waiting policies. The results shown are obtained by taking the geometric mean using the two parallelisation implementation. For blackscholes, figure \ref{fig:PARSEC_black_waiting}, the default option performs best for all three compilers, but the difference compared to the other waiting policies are very slight, only 1 - 2 \%. An outlier is the active option for ICC, where from 28 to 32 threads it gets noticeable slower. For fluidanimate, figure \ref{fig:PARSEC_fluid_waiting}, what waiting policy performed best heavily depends on how many threads was used. For eight threads, which was the thread count that overall used the least amount of energy, the active and default setting performed the best, with passive being 5 - 10 \% worse depending on the compiler. As more threads are used this trend reverses and the active and default setting gets significantly worse while the passive just increasing slightly in energy. For 32 threads the passive setting uses between 10 - 20 \% less energy compared to the active setting for the different compilers.

What the results from two PARSEC benchmarks have demonstrated is that there is not one size fit all when it comes to optimising for energy usage for parallel programs. The two programs tested shared no parameter when it came to their overall best configurations. For blackscholes the overall best configuration was with active waiting policiy, tasking implementation and ICC as the compiler, while for fluidanimate the best configuration was instead passive waiting policy, parallel for implementation and GCC as the compiler. 

\subsection{Summary in context of research questions}
\label{sec:results:summary}

The results presented thus far have been in the context of each program. 
This section aims to summarise the results and draw possible conclusions in terms of each research question. The final question, which concerns novel directives for energy consumption, is covered on its own in chapter~\ref{chap:directives}.


\subsubsection*{\textbf{RQ1}: How much impact do code transformations such the recently introduced unrolling and tiling have on energy consumption?}

This question has been evaluated using our own matrix multiplication (both naive and reordered) and 2D stencil programs, as well as our efforts of applying the transformations to the common benchmark suites NPB and BOTS. Unrolling specifically has also been evaluated using the unrolling microbenchmark.

We have seen that tiling can provide a significant energy reduction (50-60 \%) in the naive matrix multiplication, with similar results for the 2D stencil. However, we have also seen that tiling with OpenMP has consistently provided worse results than manual unrolling. In the benchmark suites, we have not managed to apply tiling with much success due to the strict requirements for doing so. We have also seen how tiling can drastically deteriorate performance in the reordered matrix multiplication. In general, we have also seen that the effectiveness of tiling slightly decreases with higher numbers of threads.

To conclude, loop tiling can potentially provide huge improvements to both execution time and power, and in turn energy consumption. However, it is often not applicable to real programmes, because it requires nested loops where memory is accessed in a certain way. The programmer can not blindly apply tiling and expect their programs to be faster, but must instead empirically profile their program with and without tiling to not risk a major performance degradation, likely due to preventing other optimisations.

The effects of explicit unrolling, either through the OpenMP directive or manually modifying the code, differs heavily between programs, but overall it tends to have either an insignificant or negative impact when done naively. This can be seen for the unrolling results for the Alignment program in BOTS, the four benchmarks tested in NAS, or the \texttt{SIMPLE\_COMP\_DEPEND} program in the unrolling microbenchmark. In some rare cases, performance can also significantly degrade, like for the \texttt{SIMPLE\_MEM} program, where the unrolling disturbed automatic optimisations by the compiler, increasing total energy usage by 70 \%. Lastly there were some programs that had an noticeable positive effect on overall performance, like the 2D stencil benchmark, with an improvement of about 5 \%.

An area which showed more promising results was manual unrolling where some additional optimisation had been applied, especially if that optimisation exposes some opportunity for ILP by the compiler. Examples of this can be seen for the Strassen benchmark from BOTS, or the \texttt{COMPLEX\_NESTED} in the unrolling microbenchmark. In those cases \emph{loop jamming} was used, which groups the logic from several loops into one, making it so SIMD instructions could be utilised, significantly reducing execution time and energy usage by 45 \% and 80 \% respectively.

In conclusion, it is best to let the compiler handle unrolling optimisations except in cases where the programmer can  expose opportunities for ILP to the compiler by manually unrolling, like for the \emph{loop jamming} optimisation. Naively adding unrolling has in many cases at best a very small positive impact, and at worst a big performance downgrade, making it so the safest option is to not unroll at all. 

\subsubsection*{\textbf{RQ2}: How do the runtime parallelism constructs of OpenMP (tasks and parallel loops) compare in terms of energy?}

This has been examined using our parallel construct microbenchmarks at a fine-grained level. Single- and multithreaded task generation has been analysed using the BOTS programs where both methods are available. Parallel loops versus tasking has also been tested using PARSEC.

We first consider the different variants of tasking. From the microbenchmark we have seen that single-threaded task generation is ineffective for very small tasks (tens of microseconds), but that the difference becomes negligible for larger task granularities. This is confirmed by the BOTS programs, where we see that single-threaded task generation is about 1.5 \% faster and consumes 1.6 \% less energy. We can conclude that in most cases, there is no major difference between the task generation methods. Single-threaded is, however, easier to implement most of the time and should be preferred for tasks of granularities beyond tens of microseconds.

In the parallel constructs microbenchmark we have compared the best version of tasking versus parallel loops. Here, we conclude that parallel for loops provide the best energy reduction given that the correct scheduling is used. In the case where parallel loops use static scheduling, task generation techniques can provide better results due to uneven work sizes. However, we have shown that parallel loops will perform better than tasking in that case given that dynamic scheduling is used.
 
For the PARSEC benchmark suite, while for one program the most energy efficient implementation ultimately used tasking, in general when looking at the performance across all three compilers, the parallel for was on average better when using an appropriate scheduling policy.

We conclude that in cases where both parallel loops and tasking can be used, parallel loops should be preferred.

\subsubsection*{\textbf{RQ3}: How does the waiting policy of OpenMP affect power and energy?}

This is perhaps the aspect we have seen the biggest improvements from using the different configurations, and has been examined at a fine level using our inactivity microbenchmark and on a coarse level using the BOTS, NPB and PARSEC benchmark suites, as well as the parallel constructs microbenchmark to a degree.

From the inactivity benchmark we have seen that active waiting is better than passive waiting for very small task granularities (tens of microseconds) while passive waiting becomes better for larger tasks.
Threads must be inactive for fairly long periods of time (30-40 microseconds for Clang and ICC, and around 60-70 microseconds for GCC) for passive waiting to yield lower energy consumption.
Furthermore, in the parallel constructs microbenchmark, we have seen active waiting outperform passive waiting as long as threads have a reasonable amount of work to perform. In situations where this is not the case, such as with single-threaded task generation with more threads than can be utilized, energy is saved by using passive waiting. Passive waiting also tends to be favored with parallel for loops, especially when using GCC.

In a more realistic context, represented by BOTS and NPB, we have seen that active waiting is favorable in general, as this consumes the least energy and has the best execution times. This is however only true for Clang and ICC, as GCC performs the best with passive waiting in this case due to active and default waiting scaling poorly with many threads.
From PARSEC we have seen that active waiting is preferred when using tasking while passive waiting is preferred with parallel for loops, which confirms the results from the parallel constructs microbenchmark.

To conclude, the OpenMP waiting policy often has a major impact on both power and energy in programs where threads are blocked by each other. Passive waiting generally provides lower power at the cost of higher execution time and active waiting does the opposite. The used compiler has also a major impact.
The choice of waiting policy should be a major configuration parameter for the programmer or data center system administrator.

\subsubsection*{\textbf{RQ4}: How do different implementations of OpenMP impact the research questions above?}

In our analysis we have used three different compilers which includes its own unique implementation of OpenMP. The differences we have observed between them is a results of both in the standalone compiler and their respective OpenMP implementation. In the following discussion we won't make any attempts to isolate the contribution from the specific OpenMP implementation but instead compare them as a whole.

For GCC and Clang, the results for loop transformations have been very similar. Where a transformation has been beneficial for one, it has been for the other as well and vice versa. ICC has however shown more varied results, which has been the most prominent for matrix multiplication. Here, where loop tiling has been beneficial for GCC and Clang, it has been a major detriment to ICC. This is due to ICC performing loop reordering by itself, which it fails to do if tiling is applied. We have also seen ICC benefit the most from assuming that the loop iterations is divisible by tile size or unrolling factor.

For the 2D stencil program, we have seen that both tiling and unrolling consistently provide some decreased energy consumption. ICC has however seen the most benefit, followed by GCC, and then Clang.
Also note that for both matrix multiplication and the 2D stencil, the best results have been from some version with ICC.

Next is the question of waiting policies and parallel constructs. 
In BOTS, we have seen that ICC and Clang have fairly similar results regardless of waiting policy, but that clang performs somewhat better for high numbers of threads. GCC however, show very poor results, only scaling well up to about 10 threads before becoming worse with more threads regardless of waiting policy, and also performing the best with passive waiting.
Strangely, it simultaneously performs the best with low numbers of threads.
We draw the conclusion that GCC (with passive waiting) should be used with low numbers of threads to optimise for power, while ICC or Clang with active waiting should be used in all other cases.

For NPB, while there is significantly less variance in the results, about the same conclusions can be drawn. The main difference here is that GCC performs much better, about equal to Clang. ICC outperforms them both by far at any numbers of threads.

%% file: include/directives.tex
In this chapter we further interpret the results and suggest how to apply the findings. Some findings are more coarse-grained and general and, therefore, fit better as a recommendation for the programmer, while others are more suitable to be applied in the form of OpenMP directives. 
We begin with the general programmer recommendations, found in section~\ref{sec:directives:programmerrecs}. The purpose of this section is not to present new data or conclusions, but to concretise the findings for the common programmer.
After that, section~\ref{sec:directives:lengthcheck}, we discuss how a descriptive extension to the unrolling and tiling directives could be added to alleviate compiler optimisations and estimate the effects of such an extension. Finally, section~\ref{sec:directives:unrollreduction} covers how unrolling can be used to remove dependencies between variables leading to faster reduction operations.

\subsection{Programmer recommendations}
\label{sec:directives:programmerrecs}

In the following list, we cover broad programmer recommendations aimed at increasing the energy efficiency for parallel programs. These recommendations have been created by condensing the key finding of our results into easily applicable guidelines that do not require deep understanding of the inner workings of OpenMP by the programmer. We list the recommendations roughly in the order of greatest to lowest significance in the context of reducing energy consumption.

\begin{itemize}
    \item \textbf{Compiler choice} has shown to be one of the main factors in reducing energy consumption. For single-threaded programs, GCC tends to perform the best, while ICC and Clang tend to perform better with many threads. GCC should be avoided in tasking-heavy programs as it tends to scale poorly. Using the active waiting policy tends to be the best choice when using Clang or ICC due to shortening the execution time, while GCC often performs the best with passive waiting due to a lower power consumption.
    
    \item The \textbf{OpenMP waiting policy} has also a major impact on energy consumption. Using \emph{active} waiting tends to be the best as it shortens execution time. The waiting policy is especially important in tasking-heavy programs.
    
    \item In cases where \textbf{tasking and parallel for loops} can be used, they tend to perform about the same when the granularity of loop iterations are larger than about 100 microseconds. This is the case in most programs. Smaller than that, we have seen that parallel for loops tend to perform best in terms of energy consumption as well as execution time as long as the correct scheduling policy is used.
    
    \item When applying parallelism to a loop using tasking, it can be done in a \textbf{single-threaded or multi-threaded task generation} manner. For realistic workloads, there is almost no difference. Only when the tasks are very small, in the order of tens of microseconds, single-threaded task generation should be avoided as it throttles parallelism.
    
    \item \textbf{Loop tiling} is a complex subject. On the one hand, it can improve performance considerably in programs such as matrix multiplication and stencil applications, which act on multidimensional data. In these programs, we have seen energy reductions of the scale 40-50 \%, mainly due to shortened execution time. On the other hand, it can be greatly detrimental to performance if it causes other optimisations not to be applied by the compiler, if these would otherwise be applied if tiling was not used. In these cases, we have seen loop tiling increase energy consumption by a factor 2-3$\times$. It should also be mentioned that loop tiling is very situational due to being limited to nested loops, and often nontrivial to apply to existing programs.
    
    \item \textbf{Loop unrolling} is equally complex, but for other reasons. Explicit loop unrolling by the programmer should be considered only in very specific cases. Altering source code and applying unrolling naively will in the best case have a very minor benefit (1-2 \% energy reduction), and in the worst case worsen performance drastically (2-3X energy consumption) due to preventing other optimisations. It does have a potential use if it can be combined with additional clever optimisations and allow for techniques such as auto-vectorisation, in which case it can be highly beneficial. However, as this benefit is not due to the unrolling itself as well as being a highly advanced optimisation technique, we conclude that the common programmer should almost always just leave unrolling to the compiler.
    
    \item In general, when minimising energy consumption on a homogeneous platform such as the one used in this project, \emph{minimising energy consumption is most easily done by minimising execution time rather than power consumption}.
    
\end{itemize} 
As a final note, we emphasise that there are always exceptions to these guidelines. When applying an optimisation, it should be empirically evaluated to see if the program performs better or worse with the optimisation compared to before. This is especially important for tiling and unrolling, which we have seen both drastically improve and impair performance. OpenMP-applied tiling and unrolling can be helpful in this regard, as it requires less effort from the programmer to apply than doing so manually.

\subsection{Loop transformation length check clauses}
\label{sec:directives:lengthcheck}

Throughout our testing we have used two variants of manual tiling and unrolling: one where the tile size or unrolling factor is assumed to evenly divide the number of loop iterations, and one where this is not assumed.
In the case where passing this additional information would make a significant difference, it would make for a good optional modifier to the \emph{tile} and \emph{unroll} directives, exemplified by the listings~\ref{lst:directives:unroll_extension} and~\ref{lst:directives:tiling_extension}.
The potential gain from such an extension would be that overhead logic from uneven tile sizes and unroll factors could be avoided. 

\noindent\begin{minipage}{\textwidth}
\begin{lstlisting}[caption=Possible extension to the unrolling directive.,frame=tlrb,label={lst:directives:unroll_extension}]{Name}
int i;
int n[N];
#pragma omp unroll partial(5) nocheck
for(i = 0; i < N; i++)
{
  n[i] = 10 * i
}
\end{lstlisting}
\end{minipage}
\begin{minipage}{\textwidth}
\begin{lstlisting}[caption=Possible extension to the tiling directive.,frame=tlrb, label={lst:directives:tiling_extension}]{Name}
#pragma omp tile sizes(8, 8, 8) nocheck
for (int row = 0; row < N; row++) {
  for (int col = 0; col < N; col++) {
    for (int k = 0; k < N; k++) {
      C[row*N + col] += A[row*N + k] * B[k*N + col];
    }
  }
}
\end{lstlisting}
\end{minipage}

To see how much effect this could have in practise, we first consider the data gathered from applying tiling to matrix multiplication, which is best summarised in table~\ref{tab:matmul_tiling_summary}. By comparing the tiling versions (Manual v1 and v2), we can estimate the improvement of this optimisation. For Clang, we see that the possible energy reduction is $1 - \frac{22.47}{29.66} \approx 0.24 = 24 \%$, which is a fairly significant amount. For GCC, the improvement is even higher at 65 \%! Such high numbers are, however, unlikely to be just due to this optimisation, but rather because the logic is simplified for the compiler, which then can optimise better. The corresponding numbers for the 2D stencil are much more conservative, with only a 2.4 \% improvement for Clang and no difference for GCC. This is more in line with what is expected from such a small change.
For unrolling, we see that the change would have next to no effect in practice, which is confirmed by the unrolling microbenchmark.

A compiler could choose to generate two versions of an unrolled or tiled loop, one where the property is assumed, and one where it is not. 

\subsection{Loop unrolling reduction clause}
\label{sec:directives:unrollreduction}

In OpenMP there is a clause called \texttt{reduction} which can be used together with the \texttt{parallel}, \texttt{for} or \texttt{sections} directives. It specifies that at the end of that parallel region, a specified variable should be subject to a reduction using some specified operation. The clause works by first making a private copy of the specified variable for each thread, then the threads do their calculations saving the results to their own local copy. Lastly, at the end of the parallel region, all the private copies are reduced using the specified operation, such as addition.

What we propose is to extend the OpenMP unrolling directive so that it can also use the reduction clause, but instead of a tool to allow for TLP, it enables ILP. An example of how the directive would work can be seen for the program \texttt{SIMPLE\_COMP}, with the modified implementation in listing \ref{lst:SIMPLE_COMP_OPT}, and the unmodified version found in listing \ref{lst:SIMPLE_COMP}. In this example the specified directive would look like this: \texttt{\#pragma omp unroll partial(2) reduction(sum:+)}. The directive splits up the unrolling so several independent variables are used for the calculations, then does a reduction on these variables after the loop.

The goal of adding this clause is to allow the compiler to optimise for the use of vector instructions, which can significantly improve performance. When separating the calculation in this way into several independent parts, dependencies that would otherwise prevent the use of vectorisation are removed. For the previously mentioned \texttt{SIMPLE\_COMP} program, execution time and energy were reduced by about 80 \% using this method.

To provide some additional examples of situations where the directive is, and is not useful, another program was tested using two slightly different versions. The program calculates the sum of all elements in an array. The first version took the sum of all the elements in ascending order, while the second version accessed the array based on another array that contained all the indices shuffled in random order. The two versions were specifically chosen with the aim that the first version would be favourable for vectorisation and the second not. The code for both versions, with both the original code and with the proposed directive applied, can be seen in the listings \ref{lst:vector_unmodified}, \ref{lst:vector_modified}, \ref{lst:no_vector_unmodified} and \ref{lst:no_vector_modified}. As expected, the first version had a significant performance improvement, reducing energy usage by 85 \%, while the second version saw no improvement. Both of these results were achieved using an unrolling factor of eight, instead of the factor two shown in the listings. These results support our claim that the proposed reduction unroll clause is only effective when vectorisation is possible and give no performance improvements otherwise.

%% file: include/conclusion.tex
In this paper, we analysed aspects of OpenMP from an energy consumption perspective. This is accomplished by executing novel microbenchmarks and common benchmark suites on HPC cluster nodes and measuring the energy consumption. Three main aspects are analysed: loop tiling and unrolling, methods for generating parallelism, and the policy of handling blocked threads. 
For the first aspect, we find that tiling can yield significant energy savings for some, mostly unoptimised programs, while directive-generated unrolling yields no improvements in practice. 
For the second aspect, we find that parallel for loops in general yield better results than explicit tasking loops in cases both can be used, but that the differences are very minor for everything but the smallest task sizes.
For the third, we find that significant energy savings can be made by not descheduling waiting threads but instead have them spin, at the cost of a higher power consumption, and that this holds even for realistic workloads.
We also analyse how the choice of compiler affects the above questions by compiling programs with each of \emph{ICC}, \emph{Clang} and \emph{GCC}, and find that neither is strictly better than the others.

\subsection{Future Work}

Our finding indicate that the waiting policy of OpenMP has a major impact on execution time, power and energy. In general, an \emph{active} waiting policy increases the power consumption but lowers execution time for realistic workloads, with a \emph{passive} waiting policy doing the opposite.
This could be used as a major parameter in the previous work OpenMPE \cite{alessi_application-level_2015}, which was covered in section~\ref{sec:related_work}. In short, it introduces directives for optimising a combination of energy, power, performance and quality of service.

Enhancements can also be made to the experimental methodology. For example, the chosen method of measuring energy limits us to measuring average power only, and not peak power. Future work could consider performing energy measurements in regular intervals during execution to accommodate for this.

We also perform all empirical experiments on a single, homogeneous platform, which to an extent limit the validity of results to that platform only. Including other platforms, especially with heterogeneous architectures, would be highly interesting.

%% file: include/appendix.tex
In this appendix we show additional results, figures and tables.

\subsection*{Matrix Multiplication}

\begin{figure}[h]
  \centering
  \includegraphics[width=\textwidth]{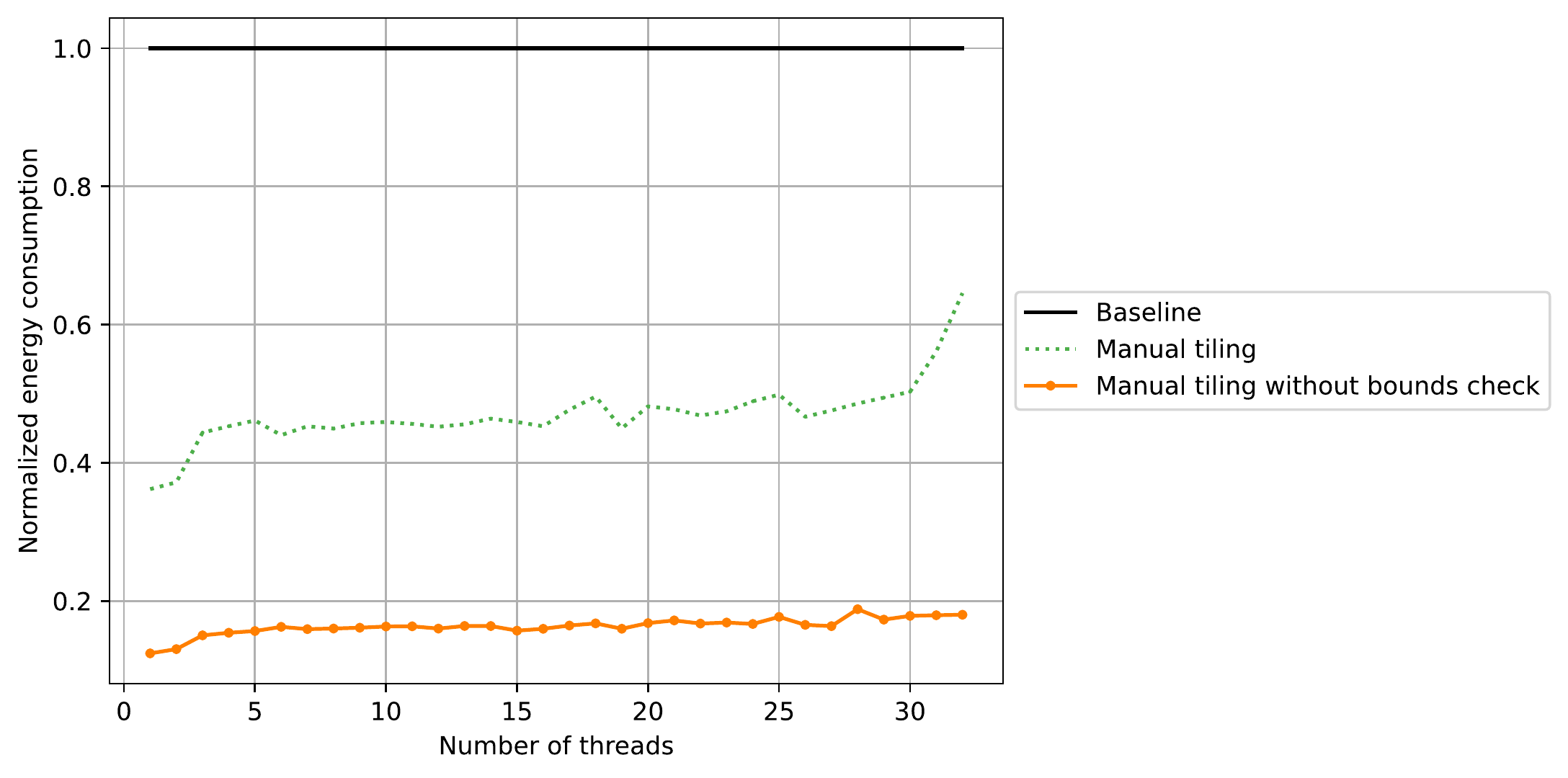}
  \caption{Relative energy consumption for the naive matrix multiplication with tiling using GCC.}
  \label{fig:matmul_gcc_tiling_naive_energy}
\end{figure}

\begin{figure}[h]
  \centering
  \includegraphics[width=\textwidth]{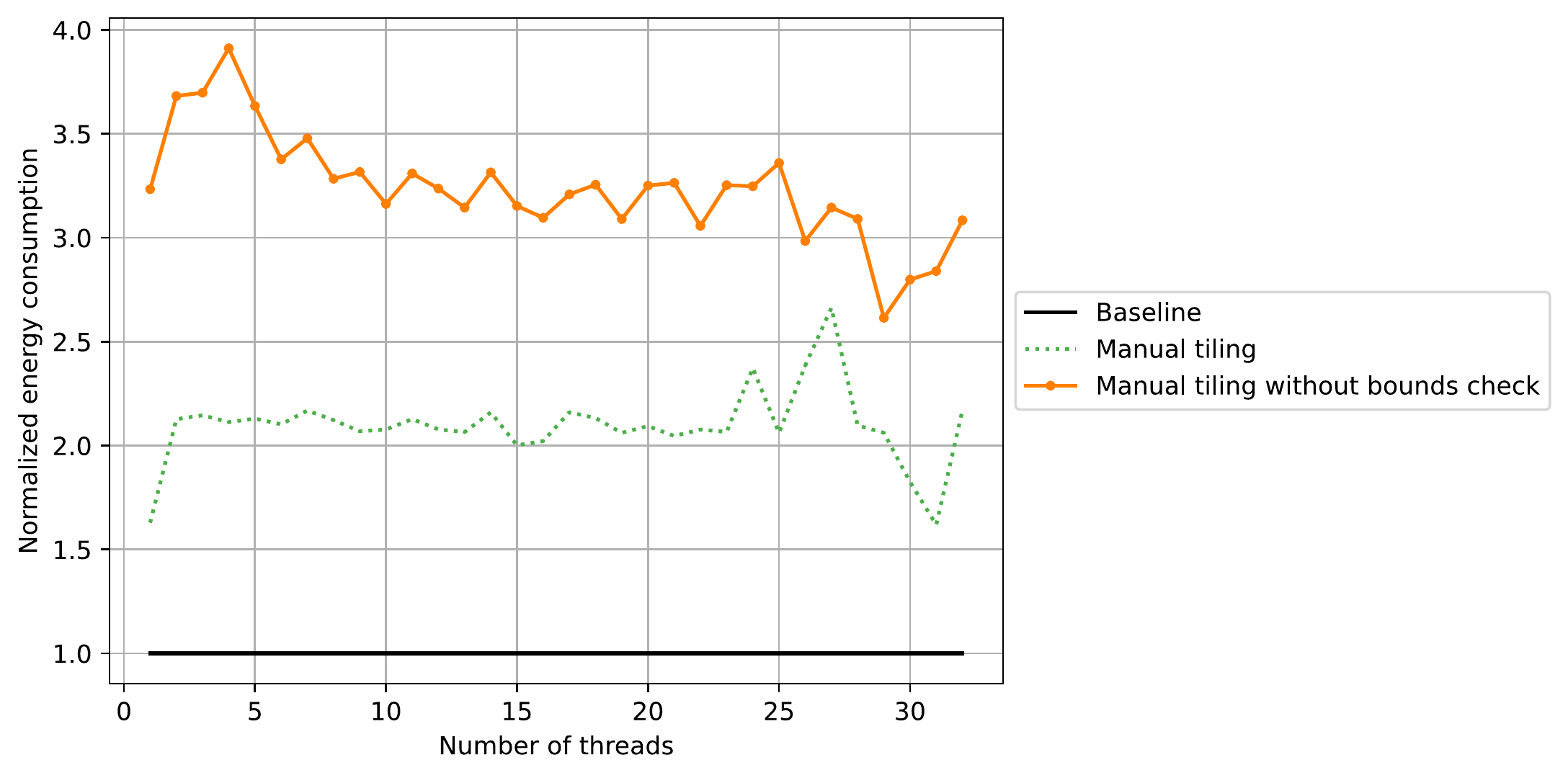}
  \caption{Relative energy consumption for the reordered matrix multiplication with tiling using GCC.}
  \label{fig:matmul_gcc_tiling_reordered_energy}
\end{figure}

\begin{figure}[h]
  \centering
  \includegraphics[width=\textwidth]{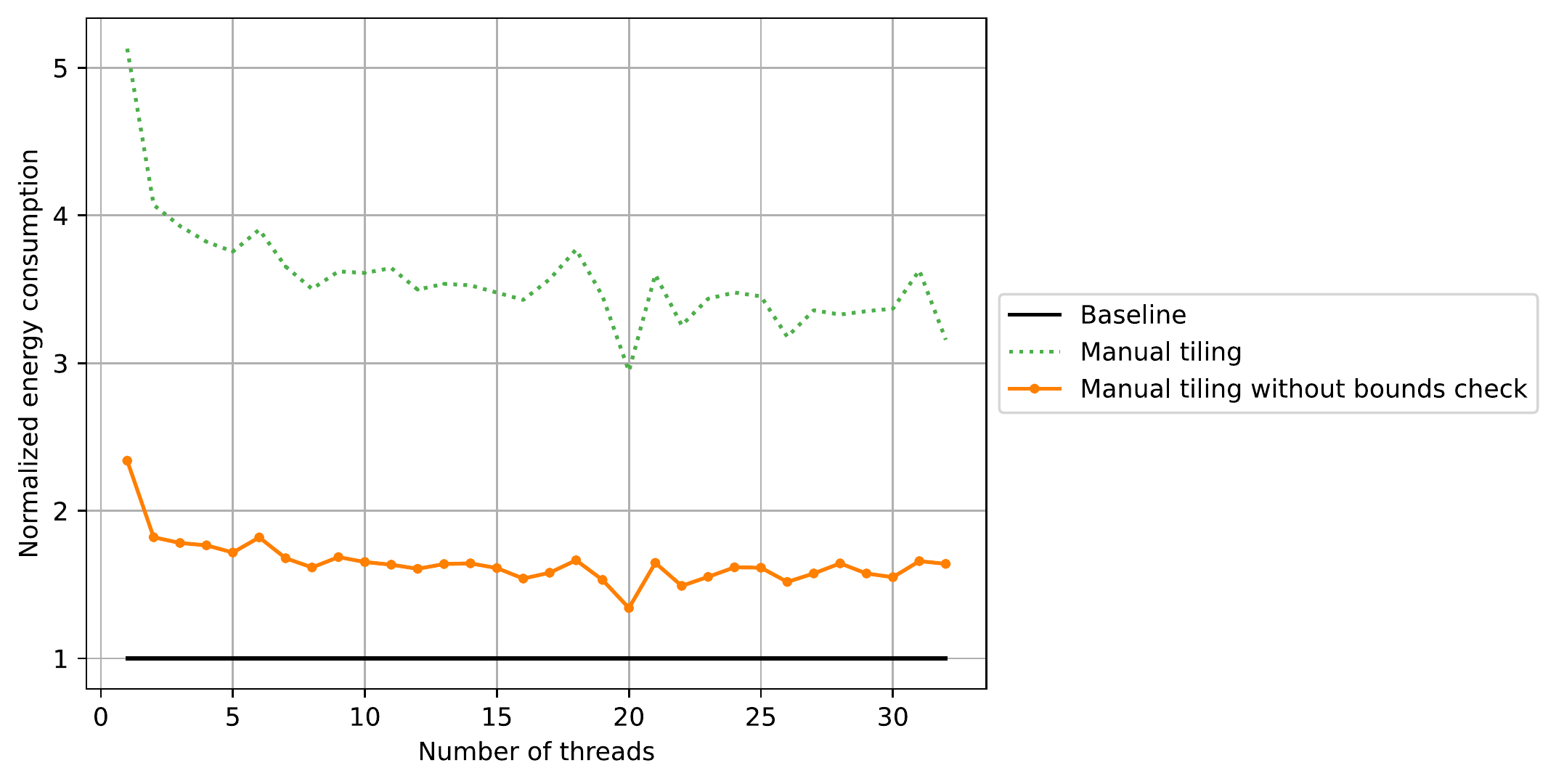}
  \caption{Relative energy consumption for the naive matrix multiplication with tiling using ICC.}
  \label{fig:matmul_icc_tiling_naive_energy}
\end{figure}

\begin{figure}[h]
  \centering
  \includegraphics[width=\textwidth]{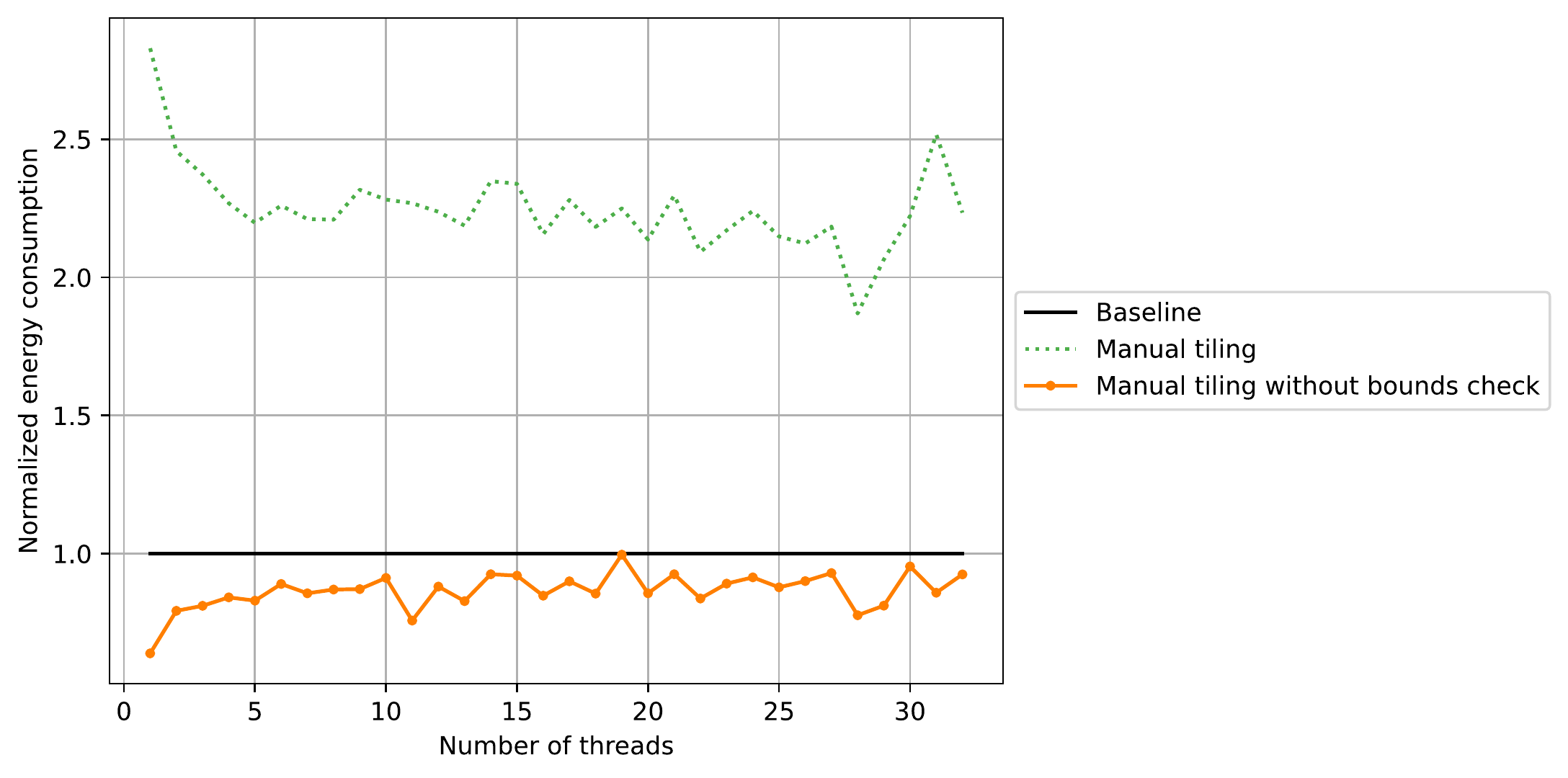}
  \caption{Relative energy consumption for the reordered matrix multiplication with tiling using ICC.}
  \label{fig:matmul_icc_tiling_reordered_energy}
\end{figure}

In figures~\ref{fig:matmul_gcc_tiling_naive_energy} and~\ref{fig:matmul_gcc_tiling_reordered_energy} we see the energy consumption of the naive and reordered matrix multiplication, respectively. The results are similar to when using Clang, with the naive algorithm heavily benefiting from tiling and the reordered algorithm becomes significantly worse. Corresponding results for ICC are seen in figures~\ref{fig:matmul_icc_tiling_naive_energy} and~\ref{fig:matmul_icc_tiling_reordered_energy}.

\begin{figure}[h]
  \centering
  \includegraphics[width=\textwidth]{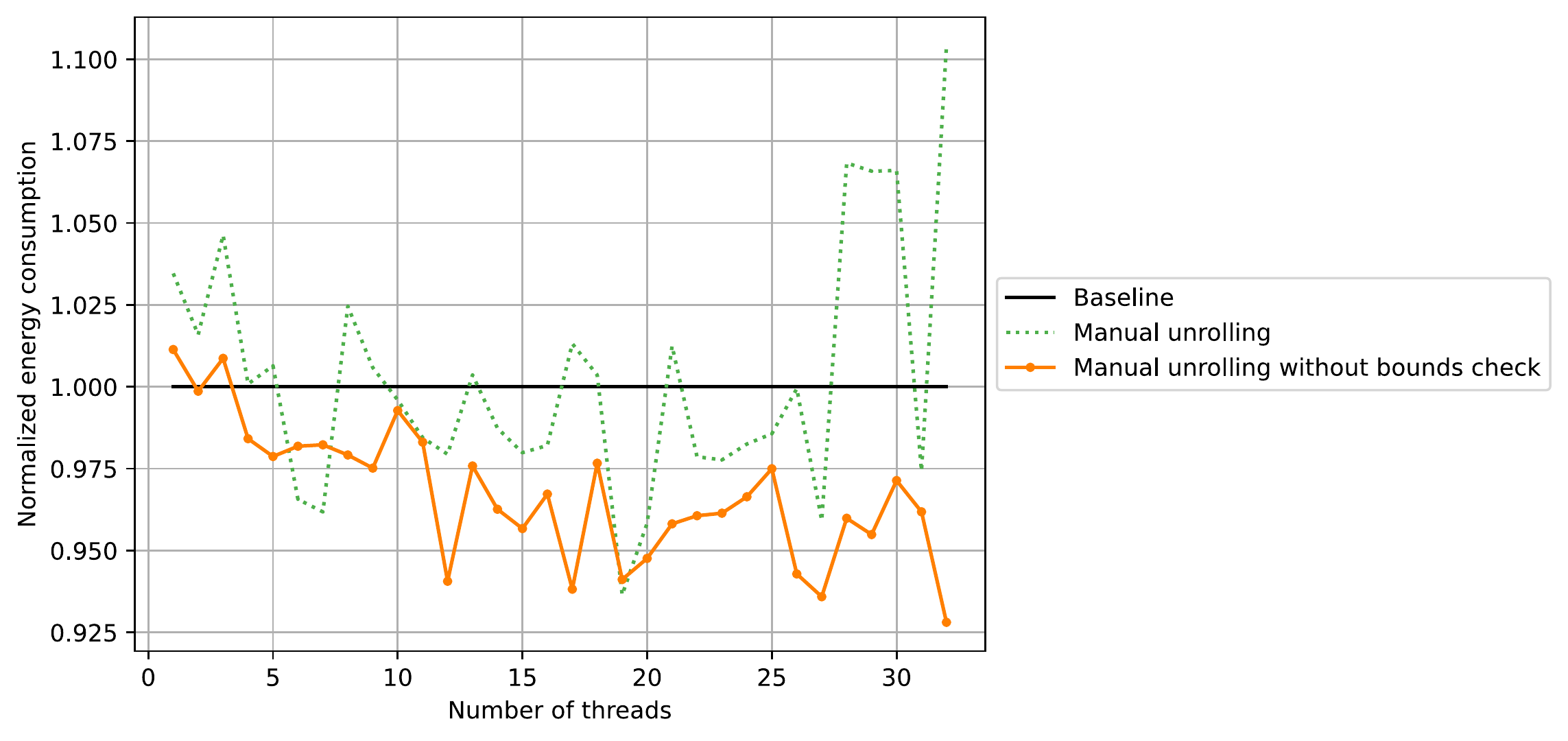}
  \caption{Relative energy consumption for the naive matrix multiplication with unrolling using GCC.}
  \label{fig:matmul_gcc_unroll_naive_energy}
\end{figure}

\begin{figure}[h]
  \centering
  \includegraphics[width=\textwidth]{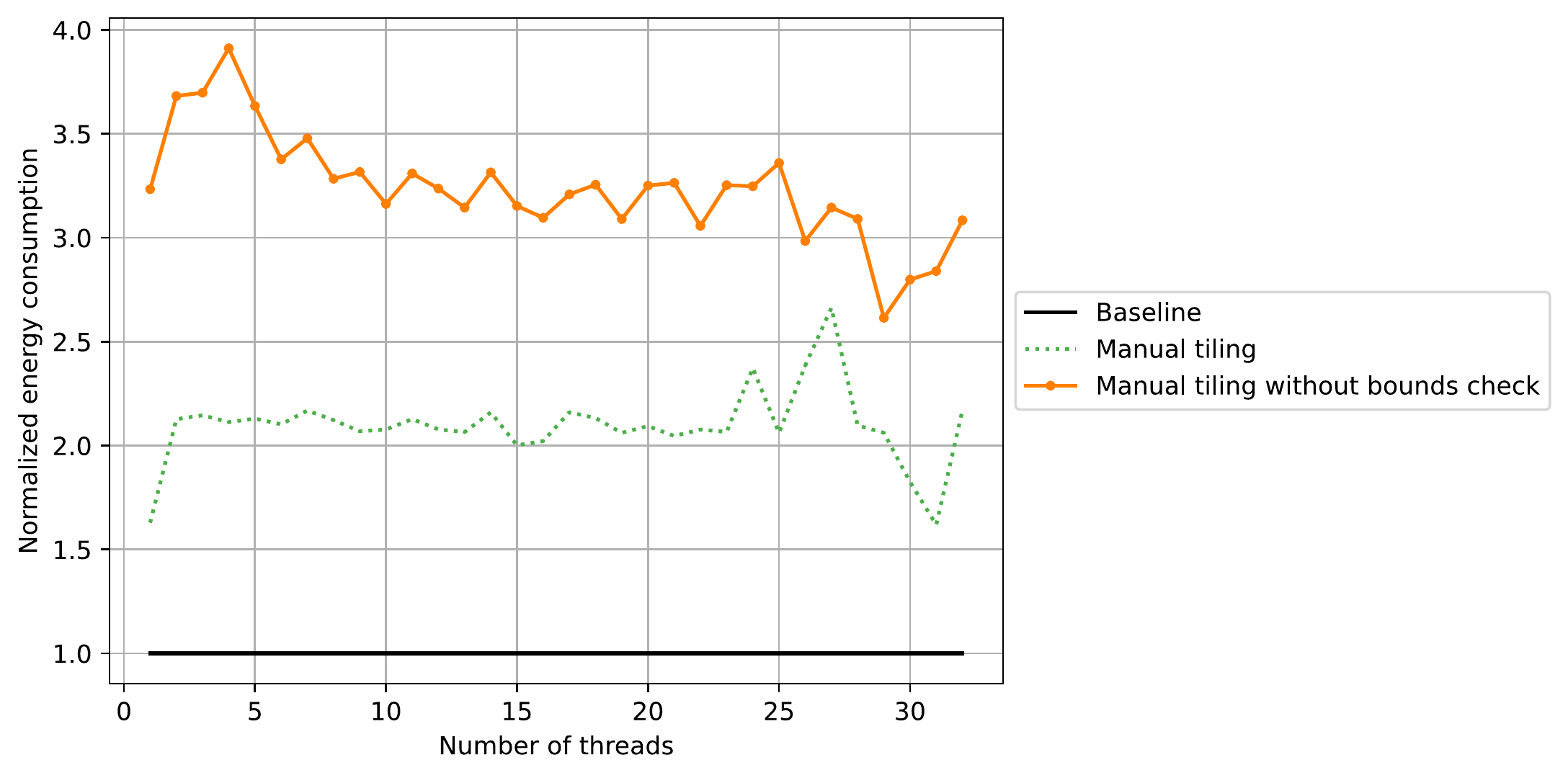}
  \caption{Relative energy consumption for the reordered matrix multiplication with unrolling using GCC.}
  \label{fig:matmul_gcc_unroll_reordered_energy}
\end{figure}

\begin{figure}[h]
  \centering
  \includegraphics[width=\textwidth]{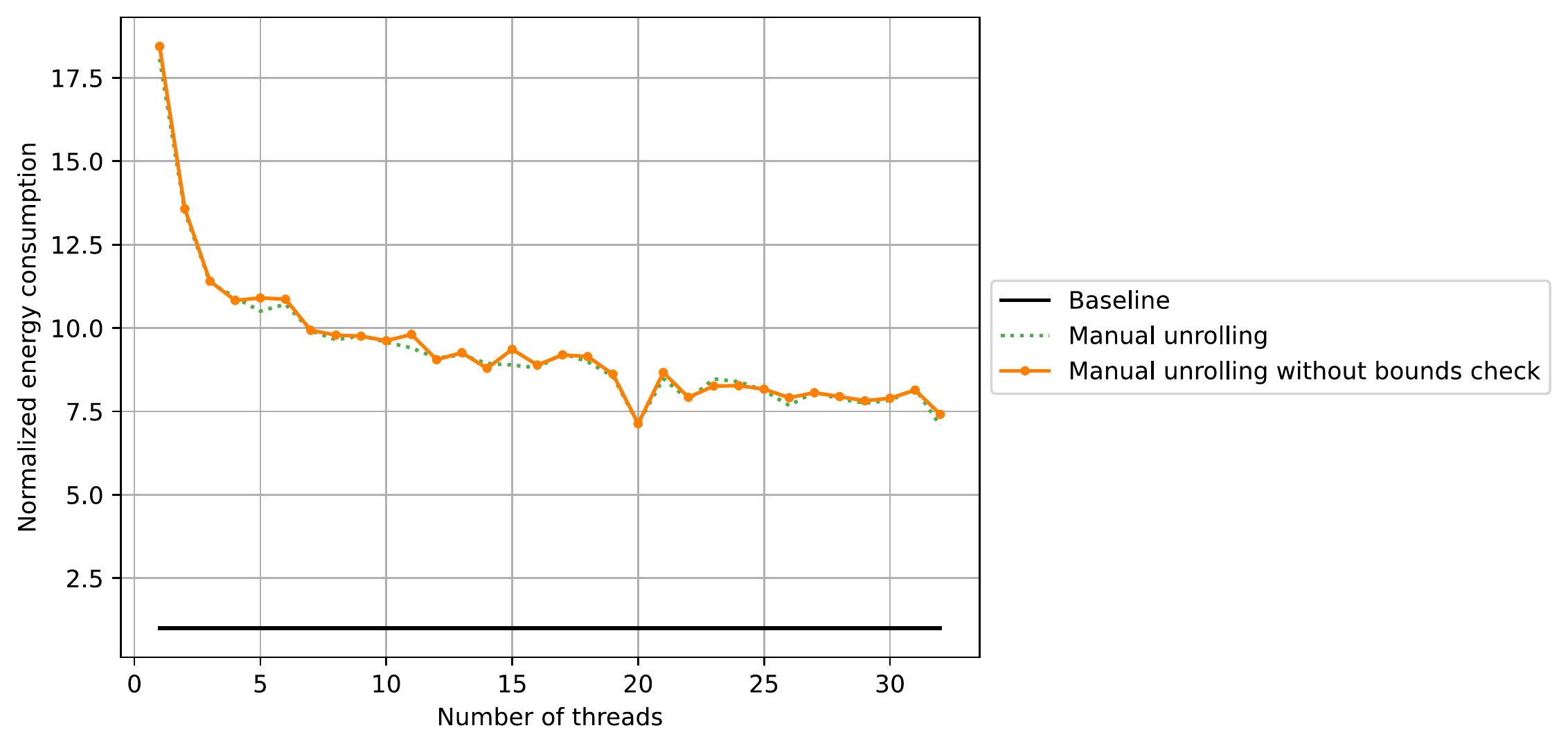}
  \caption{Relative energy consumption for the naive matrix multiplication with unrolling using ICC.}
  \label{fig:matmul_icc_unroll_naive_energy}
\end{figure}

\begin{figure}[h]
  \centering
  \includegraphics[width=\textwidth]{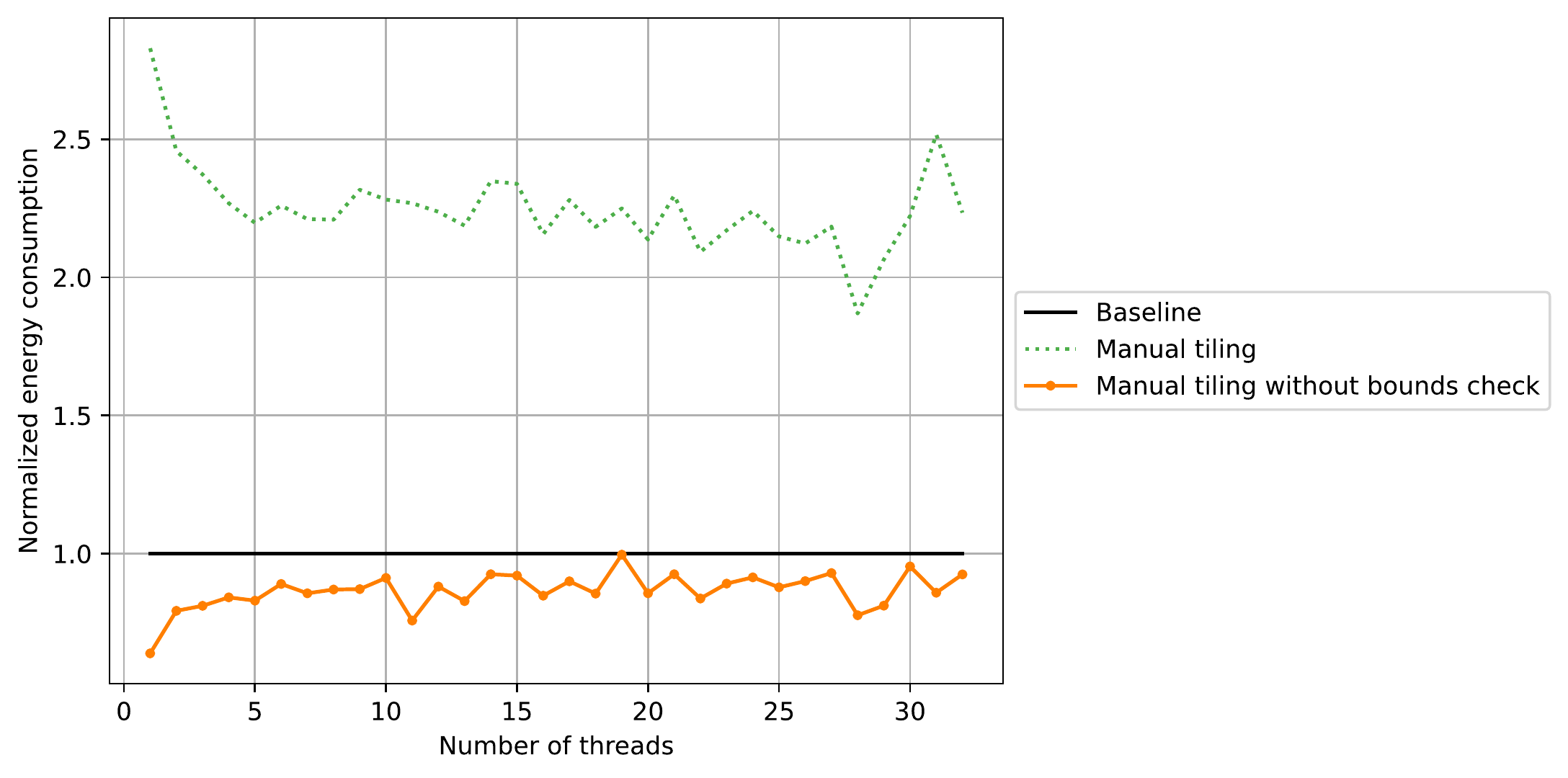}
  \caption{Relative energy consumption for the reordered matrix multiplication with unrolling using ICC.}
  \label{fig:matmul_icc_unroll_reordered_energy}
\end{figure}

Unrolling results for GCC are seen in figures~\ref{fig:matmul_gcc_unroll_naive_energy} and~\ref{fig:matmul_gcc_unroll_reordered_energy}. Corresponding results for ICC are seen in figures~\ref{fig:matmul_icc_unroll_naive_energy} and~\ref{fig:matmul_icc_unroll_reordered_energy}.

\subsection*{2D stencil}

\begin{figure}[h]
  \centering
  \includegraphics[width=\textwidth]{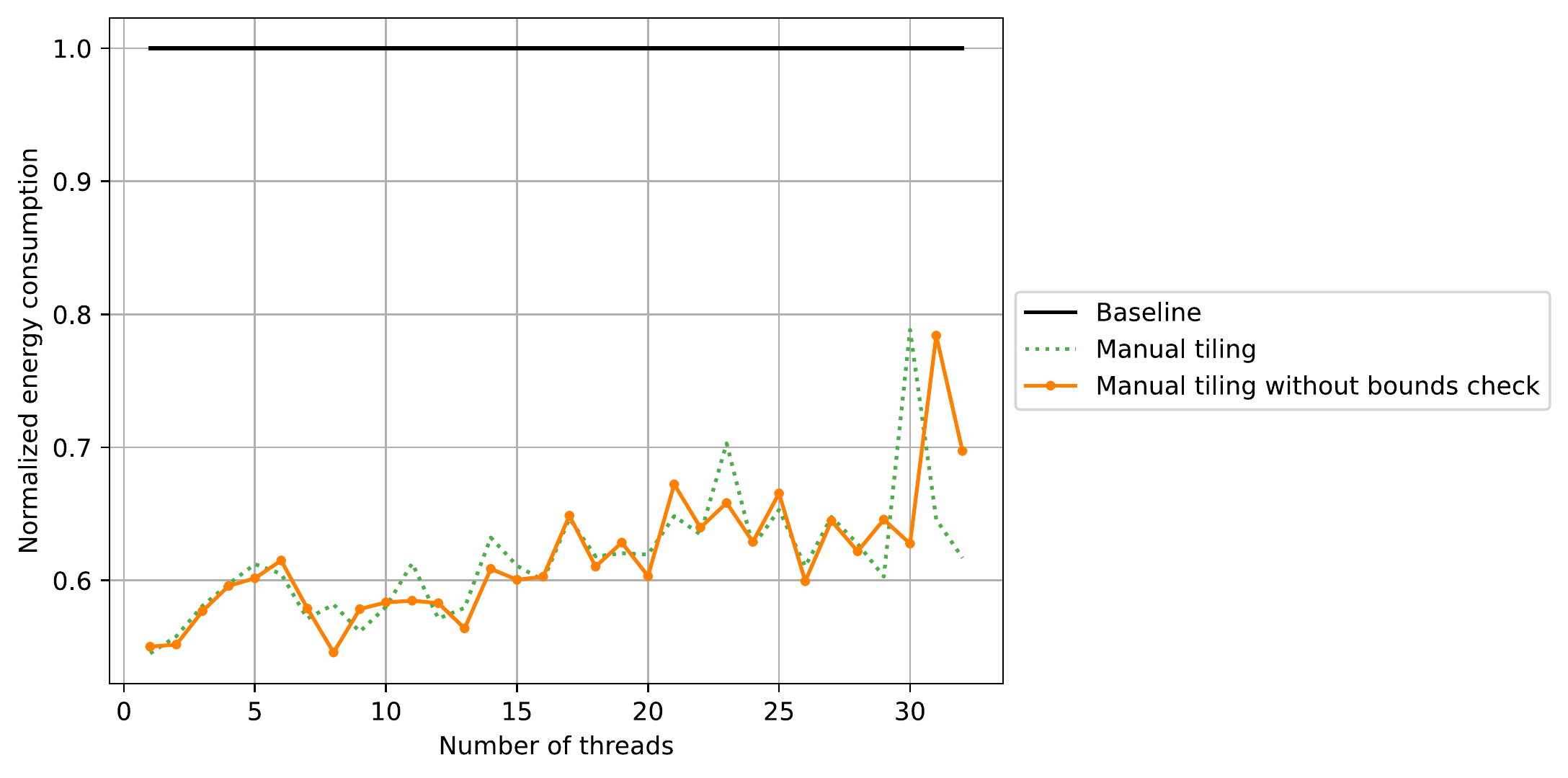}
  \caption{Relative energy consumption from applying tiling to the 2D stencil program, compiled with GCC.}
  \label{fig:stencil_gcc_energy_tiling}
\end{figure}

\begin{figure}[h]
  \centering
  \includegraphics[width=\textwidth]{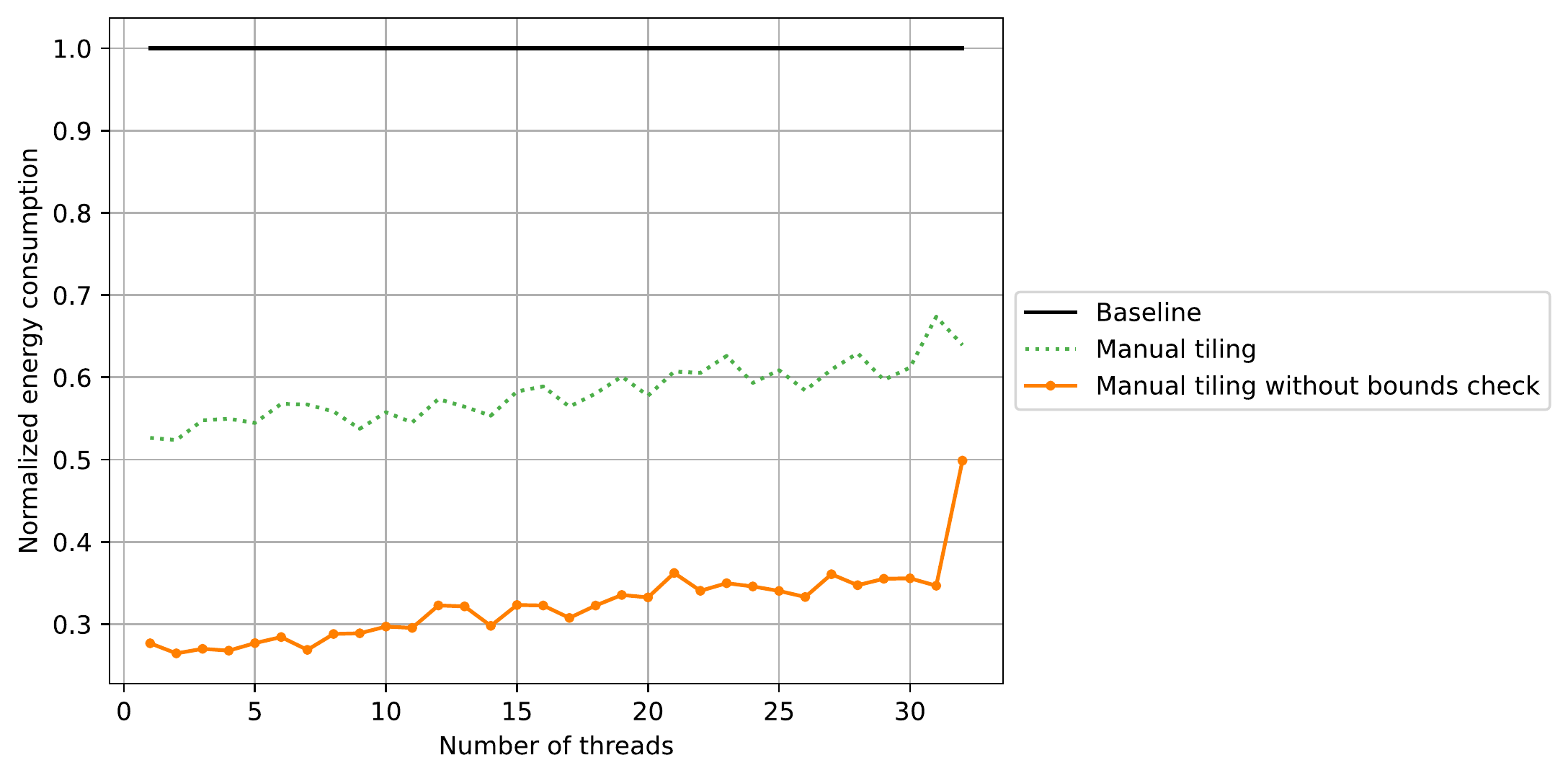}
  \caption{Relative energy consumption from applying tiling to the 2D stencil program, compiled with ICC.}
  \label{fig:stencil_icc_energy_tiling}
\end{figure}

\begin{figure}[h]
  \centering
  \includegraphics[width=\textwidth]{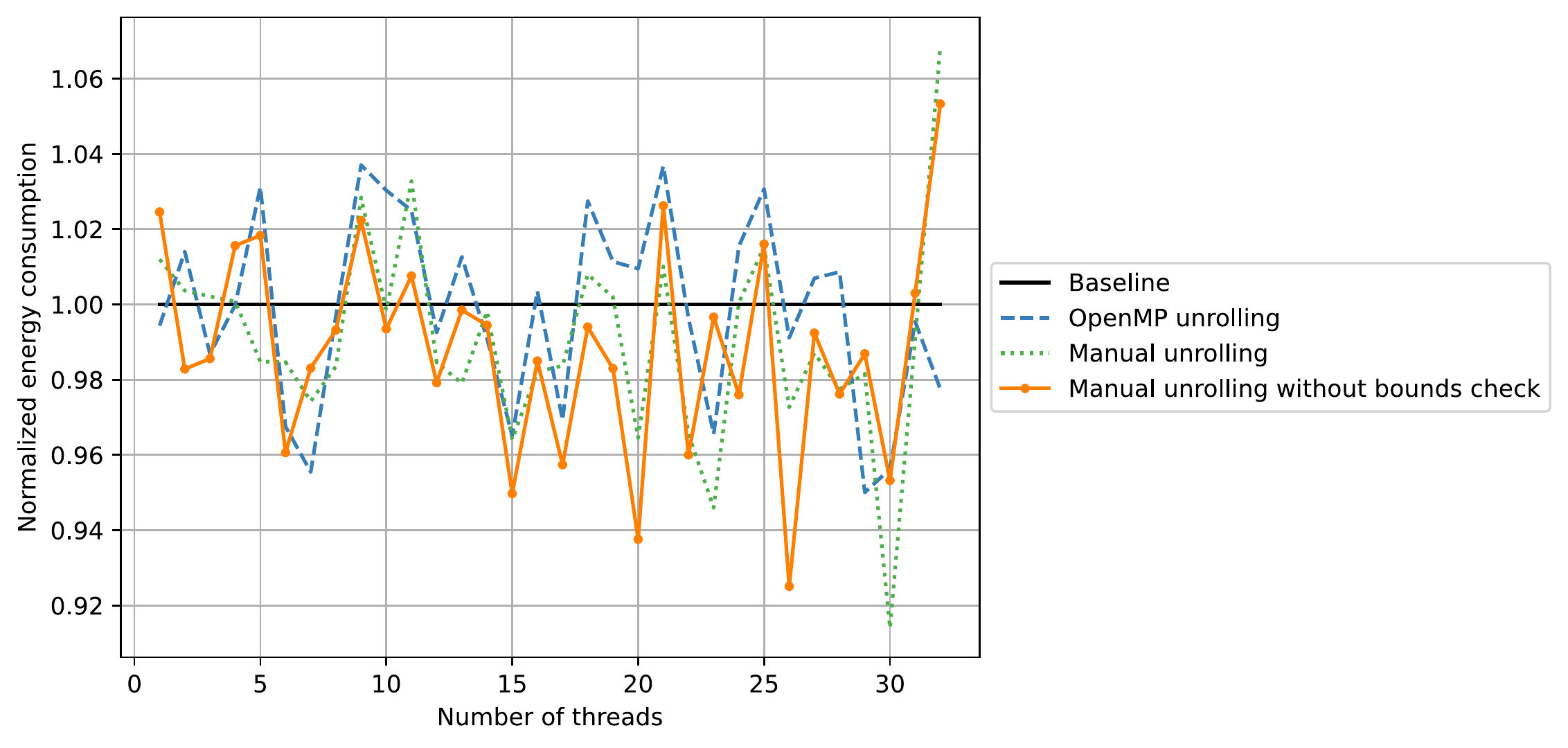}
  \caption{Energy consumption for the 2D stencil program with different versions of unrolling, compiled with Clang.}
  \label{fig:stencil_clang_energy_unroll}
\end{figure}

In figures~\ref{fig:stencil_gcc_energy_tiling} and~\ref{fig:stencil_icc_energy_tiling} we observe the energy consumption results for applying loop tiling to the 2D stencil program, compiled with GCC and ICC respectively. 
Figure~\ref{fig:stencil_clang_energy_unroll} shows the results of unrolling for Clang.

\subsection*{Parallel constructs microbenchmark}
\label{sec:appendix:tasksize}

In the main text only results for Clang and GCC is presented. Here, we present the corresponding data for ICC.
Table~\ref{tab:parconst_tasking_icc} shows the results for the different tasking variants: single-threaded task generation, multi-threaded task generation, and using the \emph{taskloop} clause of OpenMP.
Table~\ref{tab:parconst_parfor_vs_tasking_icc} shows results for parallel for loops versus the multi-threaded tasking generation.

\begin{table}[h]
\begin{center}
\caption{Characteristics of the tasking methods for the parallelism constructs microbenchmark using ICC.}
\label{tab:parconst_tasking_icc}
\begin{tabular}{|c|c|ccc|ccc|ccc|}
\hline
            &         & \multicolumn{3}{c|}{Single-threaded}                                  & \multicolumn{3}{c|}{Multi-threaded}                                   & \multicolumn{3}{c|}{Taskloop}                                         \\ \hline
Granularity & Waiting & \multicolumn{1}{c|}{T}    & \multicolumn{1}{c|}{P}   & \textbf{E}     & \multicolumn{1}{c|}{T}    & \multicolumn{1}{c|}{P}   & \textbf{E}     & \multicolumn{1}{c|}{T}    & \multicolumn{1}{c|}{P}   & \textbf{E}     \\ \hline
20          & Active  & \multicolumn{1}{c|}{0.34} & \multicolumn{1}{c|}{144} & \textbf{49.39} & \multicolumn{1}{c|}{0.09} & \multicolumn{1}{c|}{121} & \textbf{11.34} & \multicolumn{1}{c|}{0.1}  & \multicolumn{1}{c|}{124} & \textbf{11.99} \\ \hline
20          & Passive & \multicolumn{1}{c|}{0.26} & \multicolumn{1}{c|}{143} & \textbf{37.5}  & \multicolumn{1}{c|}{0.12} & \multicolumn{1}{c|}{125} & \textbf{14.71} & \multicolumn{1}{c|}{0.11} & \multicolumn{1}{c|}{121} & \textbf{13.17} \\ \hline
60          & Active  & \multicolumn{1}{c|}{0.12} & \multicolumn{1}{c|}{131} & \textbf{15.34} & \multicolumn{1}{c|}{0.09} & \multicolumn{1}{c|}{119} & \textbf{10.97} & \multicolumn{1}{c|}{0.09} & \multicolumn{1}{c|}{113} & \textbf{10.7}  \\ \hline
60          & Passive & \multicolumn{1}{c|}{0.11} & \multicolumn{1}{c|}{121} & \textbf{12.81} & \multicolumn{1}{c|}{0.1}  & \multicolumn{1}{c|}{121} & \textbf{12.65} & \multicolumn{1}{c|}{0.1}  & \multicolumn{1}{c|}{115} & \textbf{11.68} \\ \hline
100         & Active  & \multicolumn{1}{c|}{0.09} & \multicolumn{1}{c|}{118} & \textbf{11.06} & \multicolumn{1}{c|}{0.09} & \multicolumn{1}{c|}{120} & \textbf{11.13} & \multicolumn{1}{c|}{0.09} & \multicolumn{1}{c|}{117} & \textbf{11.14} \\ \hline
100         & Passive & \multicolumn{1}{c|}{0.1}  & \multicolumn{1}{c|}{114} & \textbf{11.43} & \multicolumn{1}{c|}{0.1}  & \multicolumn{1}{c|}{118} & \textbf{11.9}  & \multicolumn{1}{c|}{0.1}  & \multicolumn{1}{c|}{117} & \textbf{11.82} \\ \hline
200         & Active  & \multicolumn{1}{c|}{0.09} & \multicolumn{1}{c|}{113} & \textbf{10.57} & \multicolumn{1}{c|}{0.09} & \multicolumn{1}{c|}{120} & \textbf{11.26} & \multicolumn{1}{c|}{0.1}  & \multicolumn{1}{c|}{121} & \textbf{11.61} \\ \hline
200         & Passive & \multicolumn{1}{c|}{0.1}  & \multicolumn{1}{c|}{111} & \textbf{10.91} & \multicolumn{1}{c|}{0.1}  & \multicolumn{1}{c|}{118} & \textbf{11.66} & \multicolumn{1}{c|}{0.1}  & \multicolumn{1}{c|}{117} & \textbf{11.71} \\ \hline
\end{tabular}
\end{center}
\end{table}

\begin{table}[h]
\begin{center}
\caption{Results for the parallel for versus tasking methods, with ICC. Measurements are reported as the geometric mean of said measurement for all numbers of threads.}
\label{tab:parconst_parfor_vs_tasking_icc}
\begin{tabular}{|c|c|ccc|ccc|ccc|}
\hline
            &         & \multicolumn{3}{c|}{Tasking}                                          & \multicolumn{3}{c|}{ParFor, static}                                  & \multicolumn{3}{c|}{ParFor, dynamic}                                 \\ \hline
Granularity & Waiting & \multicolumn{1}{c|}{T}    & \multicolumn{1}{c|}{P}   & \textbf{E}     & \multicolumn{1}{c|}{T}    & \multicolumn{1}{c|}{P}   & \textbf{E}    & \multicolumn{1}{c|}{T}    & \multicolumn{1}{c|}{P}   & \textbf{E}    \\ \hline
20          & Active  & \multicolumn{1}{c|}{0.09} & \multicolumn{1}{c|}{121} & \textbf{11.34} & \multicolumn{1}{c|}{0.1}  & \multicolumn{1}{c|}{115} & 11.16         & \multicolumn{1}{c|}{0.09} & \multicolumn{1}{c|}{113} & 10.01         \\ \hline
20          & Passive & \multicolumn{1}{c|}{0.12} & \multicolumn{1}{c|}{125} & 14.71          & \multicolumn{1}{c|}{0.12} & \multicolumn{1}{c|}{81}  & \textbf{9.79} & \multicolumn{1}{c|}{0.1}  & \multicolumn{1}{c|}{96}  & \textbf{9.82} \\ \hline
60          & Active  & \multicolumn{1}{c|}{0.09} & \multicolumn{1}{c|}{119} & \textbf{10.97} & \multicolumn{1}{c|}{0.1}  & \multicolumn{1}{c|}{116} & 11.41         & \multicolumn{1}{c|}{0.09} & \multicolumn{1}{c|}{109} & 9.8           \\ \hline
60          & Passive & \multicolumn{1}{c|}{0.1}  & \multicolumn{1}{c|}{121} & 12.65          & \multicolumn{1}{c|}{0.11} & \multicolumn{1}{c|}{83}  & \textbf{8.95} & \multicolumn{1}{c|}{0.09} & \multicolumn{1}{c|}{96}  & \textbf{9.08} \\ \hline
100         & Active  & \multicolumn{1}{c|}{0.09} & \multicolumn{1}{c|}{120} & \textbf{11.13} & \multicolumn{1}{c|}{0.1}  & \multicolumn{1}{c|}{115} & 11.4          & \multicolumn{1}{c|}{0.09} & \multicolumn{1}{c|}{112} & 10.16         \\ \hline
100         & Passive & \multicolumn{1}{c|}{0.1}  & \multicolumn{1}{c|}{118} & 11.9           & \multicolumn{1}{c|}{0.11} & \multicolumn{1}{c|}{81}  & \textbf{9.05} & \multicolumn{1}{c|}{0.09} & \multicolumn{1}{c|}{102} & \textbf{9.53} \\ \hline
200         & Active  & \multicolumn{1}{c|}{0.09} & \multicolumn{1}{c|}{120} & \textbf{11.26} & \multicolumn{1}{c|}{0.1}  & \multicolumn{1}{c|}{114} & 11.43         & \multicolumn{1}{c|}{0.09} & \multicolumn{1}{c|}{110} & 10.08         \\ \hline
200         & Passive & \multicolumn{1}{c|}{0.1}  & \multicolumn{1}{c|}{118} & 11.66          & \multicolumn{1}{c|}{0.11} & \multicolumn{1}{c|}{84}  & \textbf{9.24} & \multicolumn{1}{c|}{0.09} & \multicolumn{1}{c|}{99}  & \textbf{9.33} \\ \hline
\end{tabular}
\end{center}
\end{table}

\subsection*{Inactivity microbenchmark}
\label{sec:appendix:inactivity}

\begin{figure}[h]
    \centering
    \includegraphics[width=0.75\textwidth]{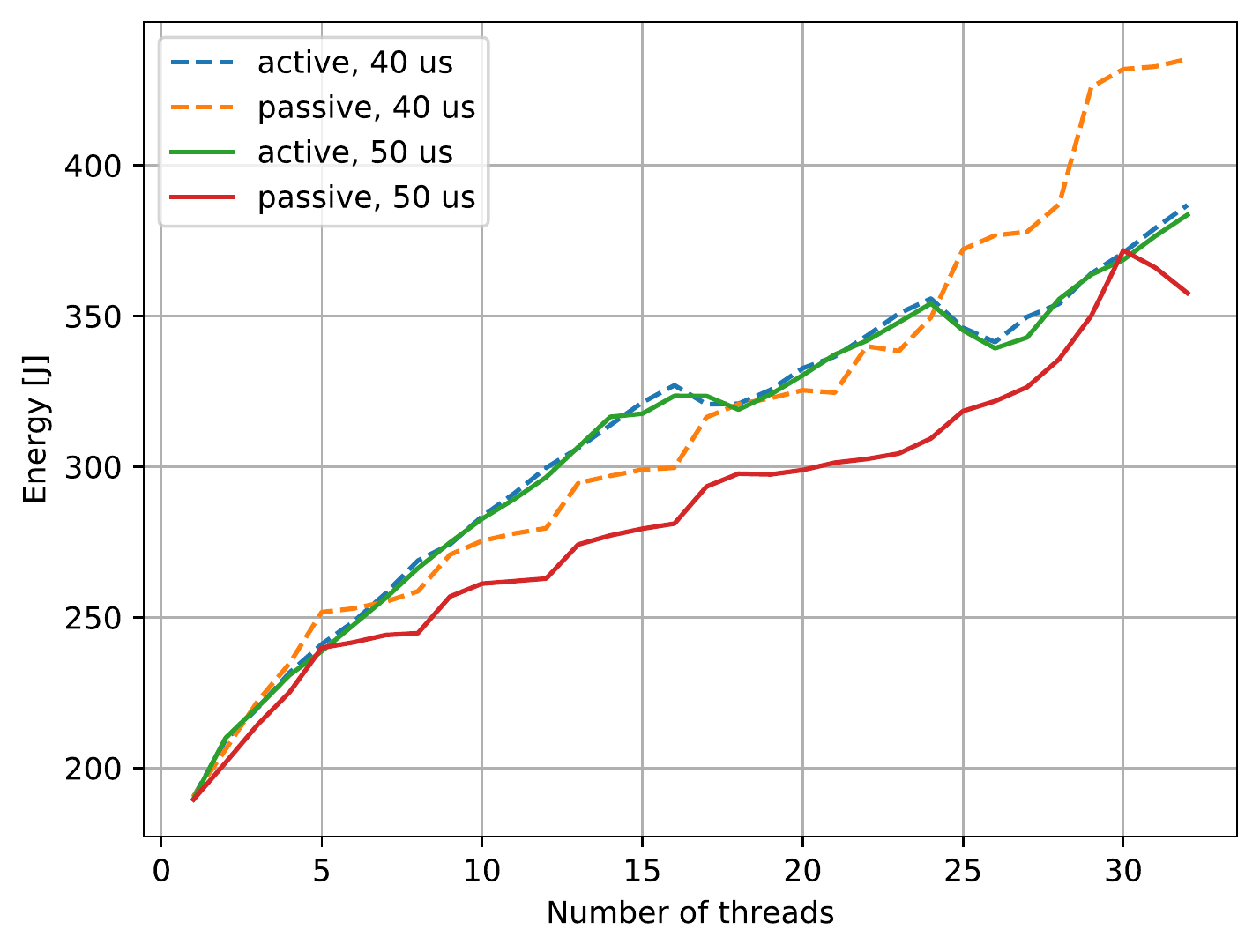}
    \caption{Energy consumption for the inactivity microbenchmark with ICC.}
    \label{fig:inactivity_icc_energy}
\end{figure}

In figure~\ref{fig:inactivity_icc_energy} we see the energy consumption for ICC around the point where passive waiting overtakes active waiting in energy efficiency. It is very similar to the figure for Clang.

\subsection*{Barcelona OpenMP Task Suite (BOTS)}
\label{sec:appendix:bots}

\begin{figure}[h]
    \centering
    \includegraphics[width=\textwidth]{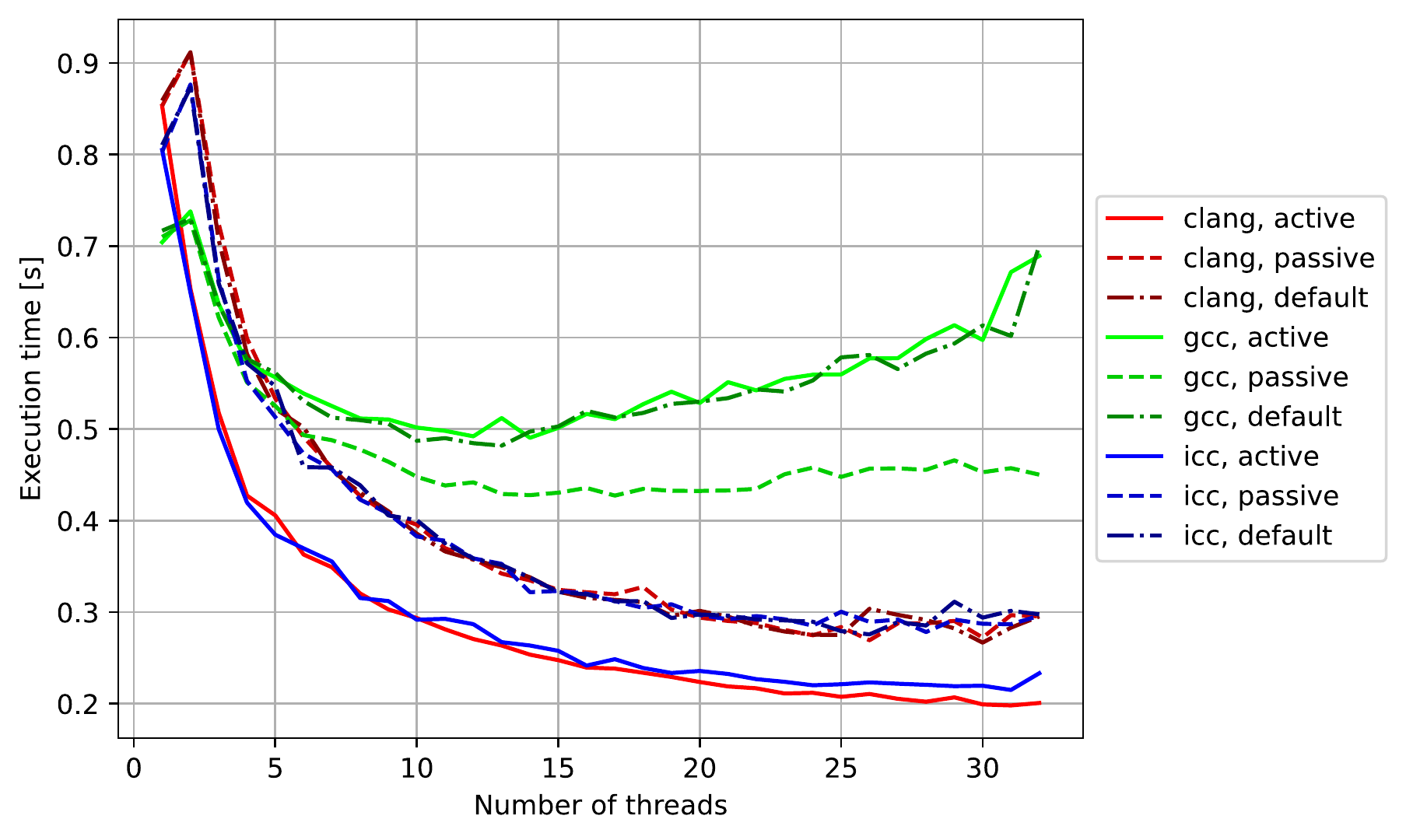}
    \caption{Average energy consumption of all BOTS programs for the three compilers and waiting policies.}
    \label{fig:bots_exec_32}
\end{figure}

\begin{figure}[h]
    \centering
    \includegraphics[width=\textwidth]{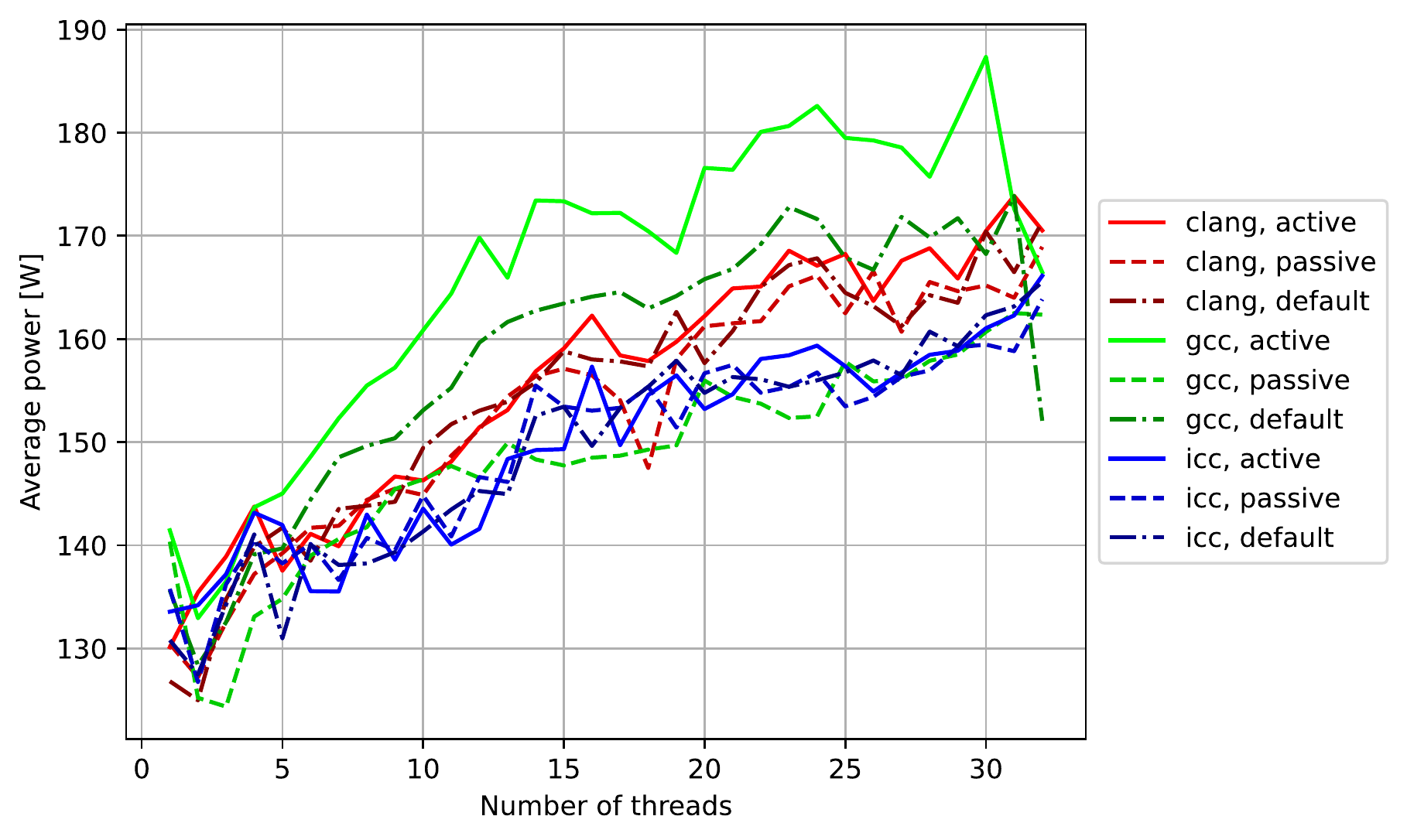}
    \caption{Average energy consumption of all BOTS programs for the three compilers and waiting policies.}
    \label{fig:bots_power_32}
\end{figure}

Figures~\ref{fig:bots_exec_32} and~\ref{fig:bots_power_32} show the execution time and power consumption of the BOTS benchmarks. The presented values are the geometric mean execution time and power of all programs.

\subsection*{NAS Parallel Benchmarks (NPB)}

\begin{figure}[h]
    \centering
    \includegraphics[width=\textwidth]{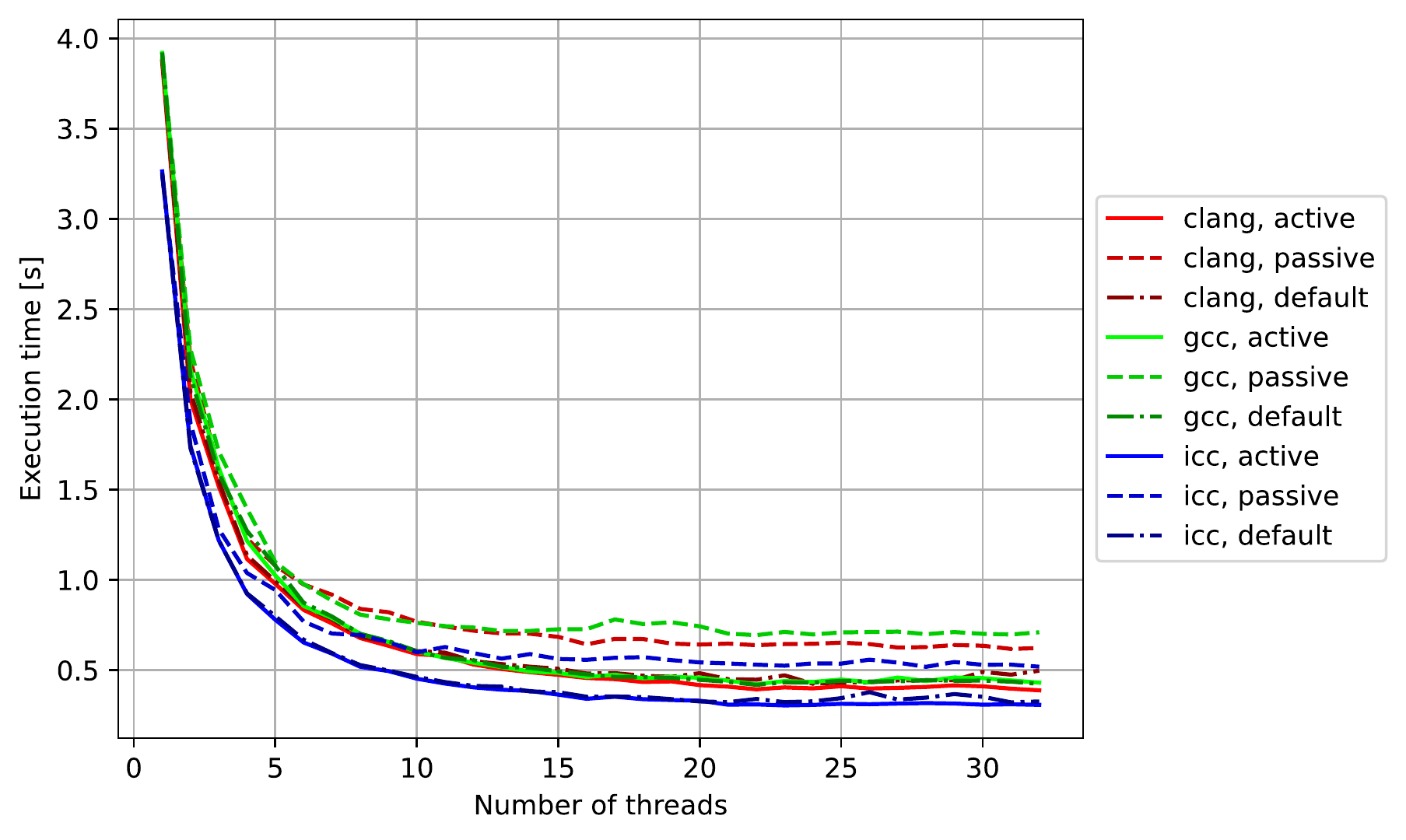}
    \caption{Execution time of the NPB programs with different waiting policies.}
    \label{fig:npb_waitpol_exec}
\end{figure}

Figure~\ref{fig:npb_waitpol_exec} shows the execution times for different compilers and waiting policies for the NPB programs.

\subsection*{Directives}

In listings \ref{lst:vector_unmodified}, \ref{lst:vector_modified}, \ref{lst:no_vector_unmodified}, and \ref{lst:no_vector_modified} the two different versions of the summation programs described in section \ref{sec:directives:unrollreduction} with their unmodified, and modified implementations are shown. The first version accesses the array in ascending order, while the second does so in a random order, with indexes precalculated and stored in the array \texttt{ind}.

\noindent\begin{minipage}{.45\textwidth}
\begin{lstlisting}[caption={First version, unmodified implementation.} ,frame=tlrb,label={lst:vector_unmodified},language=C]{Name}
float sum = 0;
for(int i=0; i<len; i++)
  sum += A[i];

\end{lstlisting}
\end{minipage}\hfill
\begin{minipage}{.45\textwidth}
\begin{lstlisting}[caption={First version, modified implementation.},frame=tlrb, label={lst:vector_modified}]{Name}
float sum0 = 0, sum1 = 0;
for(int i=0; i<len; i+=2) 
{
  sum0 += A[(i+0)];
  sum1 += A[(i+1)];
}
float sum = sum0+sum1;
\end{lstlisting}
\end{minipage}

\noindent\begin{minipage}{.45\textwidth}
\begin{lstlisting}[caption={Second version, unmodified implementation.} ,frame=tlrb,label={lst:no_vector_unmodified},language=C]{Name}
float sum = 0;
for(int i=0; i<len; i++)
  sum += A[ind[i]];

\end{lstlisting}
\end{minipage}\hfill
\begin{minipage}{.45\textwidth}
\begin{lstlisting}[caption={Second version, modified implementation.},frame=tlrb, label={lst:no_vector_modified}]{Name}
float sum0 = 0, sum1 = 0;
for(int i=0; i<len; i+=2) 
{
  sum0 += A[ind[(i+0)]];
  sum1 += A[ind[(i+1)]];
}
float sum = sum0+sum1;
\end{lstlisting}
\end{minipage}